\documentclass[onecolumn,11pt]{report}
\usepackage{amsmath}
\usepackage{dsfont}
\usepackage{amssymb}
\usepackage[utf8]{inputenc}
\usepackage[polish,english]{babel}
\usepackage[lmargin=3.5cm,rmargin=3.5cm,tmargin=3.5cm,bmargin=4.5cm]{geometry}
\usepackage{placeins}
\usepackage{caption}
\usepackage{listings}
\usepackage{subcaption}
\usepackage{dblfloatfix}
\usepackage{braket}
\usepackage{xcolor}
\usepackage{pdfpages}
\usepackage{multicol}
\usepackage{multirow}
\usepackage{dsfont}
\usepackage{bm}
\usepackage{graphicx}
\usepackage{titletoc}
\usepackage{diagbox}
\usepackage[toc,page]{appendix}
\usepackage{hyperref}
\hypersetup{
    colorlinks=true,
    linkcolor=black,
    filecolor=black,      
    urlcolor=black,
	 citecolor=black,
    pdfpagemode=FullScreen,
    }
\usepackage{fancyhdr}

\numberwithin{equation}{chapter}
\title{Local spin description of fermions on a lattice}
\author{Adam Wyrzykowski}
\date{\today}
\begin{document}
\lstset{language=Mathematica}
\begin{center}
\large{Jagiellonian University in Kraków\\
Faculty of Physics, Astronomy and Applied Computer Science,\\
Institute of Theoretical Physics, Particle Theory Department}\\
\vspace{2cm}
\huge{\textbf{Local spin description of fermions\\ on a lattice}}\\
\vspace{1cm}
\Large{\textbf{Adam Wyrzykowski}}\\
\end{center}
\vspace{2cm}
\begin{flushright}
\begin{Large}
PhD thesis written under the supervision of\\
\textbf{prof. dr hab. Jacek Wosiek} and\\
\textbf{dr hab. Piotr Korcyl}
\end{Large}
\end{flushright}
\vspace{1.5cm}
\begin{center}
\includegraphics[width=3.7cm]{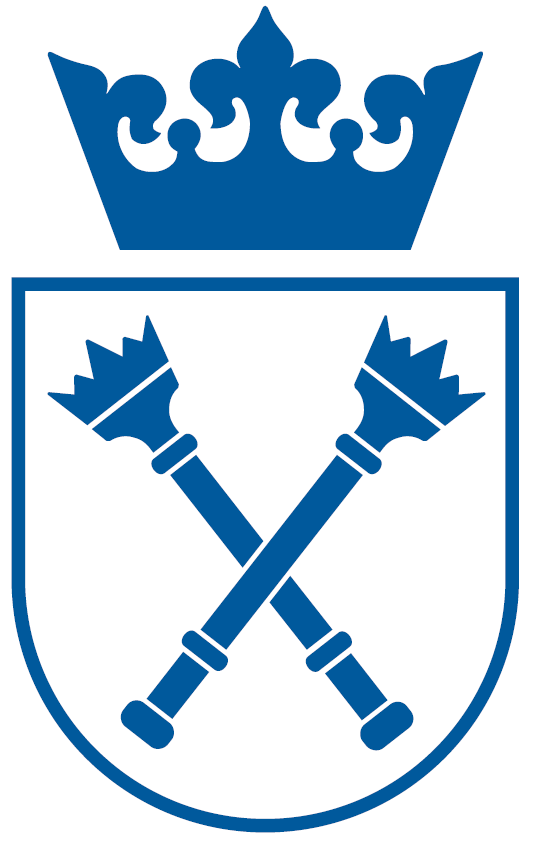}\\
\vspace{0.5 cm}
\large{Kraków 2022}
\end{center}
\thispagestyle{empty}
\pagebreak
\setcounter{page}{1}
\chapter*{Abstract}
A local transformation from fermionic operators to spin matrices is proposed and studied in this work. For this purpose, a system of fermions on a lattice is considered and one applies the scheme to replace the fermionic variables with spin matrices, while the transformation relates only those fermionic/spin operators which are assigned to nearby lattice sites. In one dimension, this proposal yields the same result as the well-known Jordan-Wigner transformation, while not being restricted to $d=1$ dimension, but having a straightforward generalization to $d>1$.

To obtain the equivalent description in the spin picture, one needs to impose constraints on the spin space. Since finding the reduced spin Hilbert space constitutes a substantial stage of the whole procedure, the constraints are paid particular attention. The full set of necessary constraints is determined in both representations. The proper boundary conditions for the fermionic operators and the spin-like variables on finite-size lattices are established for 1-dimensional and 2-dimensional lattices.

Apart from the free fermions case, the method to introduce the interaction with the external $\mathbb{Z}_2$ field is also proposed. In particular, the case of fermions in a constant external field is studied, for which the Hamiltonians are derived and their eigenvalues are found.

To approach the task to solve the constraints and find the exact form of the basis vectors of the reduced spin Hilbert space, a suitable basis is constructed. A detailed scheme to implement this basis in Wolfram Mathematica programs is explained. The introduction of the basis in the spin representation along with the construction of the constraints and the Hamiltonian in this basis show how the transformation proposed in this work can be applied to obtain observables in the spin picture. Explicit construction of the constraints in the basis allows one to solve them and, once the basis vectors of the reduced spin Hilbert space are found, the spin Hamiltonian is expressed in this basis and diagonalized. This goal -- finding the exact solution of the constraints -- is the main motivation and achievement of this work. Application of the algorithms and programs proposed in this thesis enables this task, which used to be unaccesible due to its complexity. This step makes the prescription discussed in this thesis complete and suitable for actual physics problems.

The constraints are constructed in the basis as discussed above and analyzed with the Wolfram Mathematica programs for lattice sizes $3\times3$, $4\times3$ and $4\times4$. Their mutual relations are determined and the reduced spin Hilbert space is specified. The Hamiltonian is constructed in this representation and diagonalized. It is verified that the eigenenergies obtained in the spin picture agree with the analytic formulas from the fermionic representation, which provides an additional check of the equivalence between the two descriptions.
\newpage
\tableofcontents
\newpage
\chapter{Introduction}
\section{Background}
\begin{figure}[!h]
\begin{center}
\includegraphics[width=10cm]{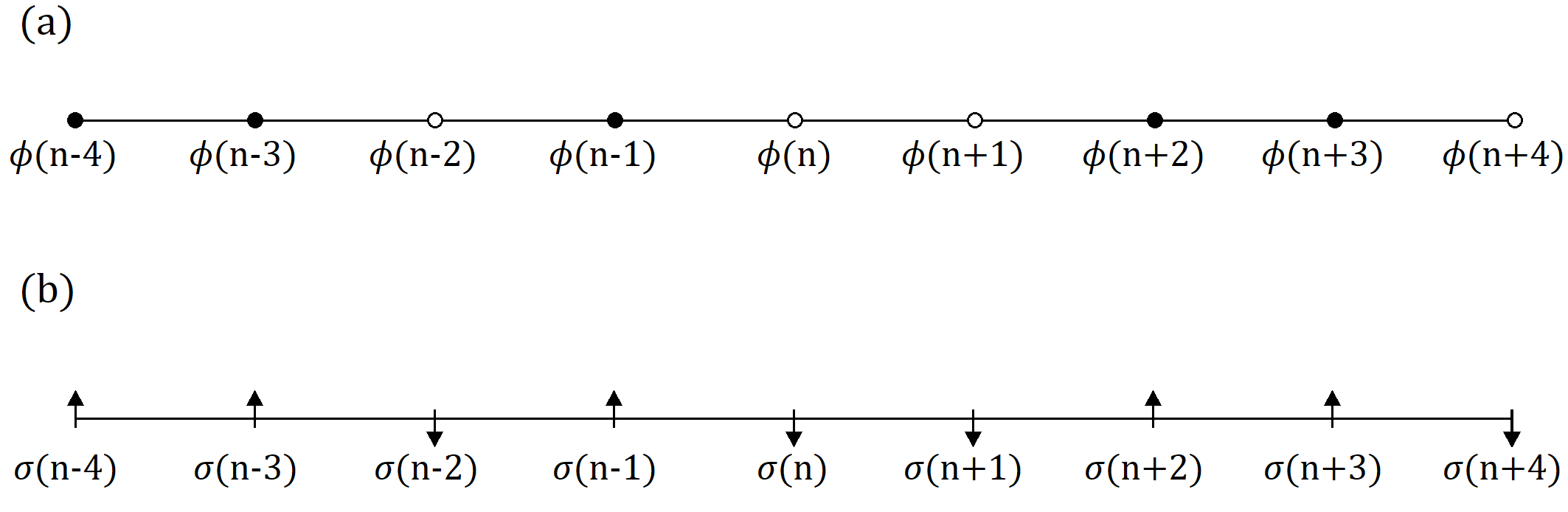}
\end{center}
\caption{The two dual systems related by Jordan-Wigner transformation (\ref{intr1})--(\ref{intr2}):\\ (a) complex fermions on a 1-dimensional lattice, (b) 1/2 spins on a 1-dimensional lattice.}
\label{fig000}
\end{figure}

The central interest of this work is the transformation from Grassmann to spin variables in the description of fermionic systems on lattices. In one dimension a map between fermionic creation/annihilation operators and spin operators is widely known as Jordan-Wigner transformation \cite{jw}. It is applied to replace fermionic operators $\phi(n)^\dag$ and $\phi(n)$ with spin matrices $\sigma^+(n)$ and $\sigma^-(n)$ according to the rules:
\begin{equation}
\phi(n)^\dag\ \longrightarrow\ \sigma^+(n)\prod_{j=1}^{n-1}e^{-i\pi\sigma^+(j)\sigma^-(j)}
\label{intr1}
\end{equation}
\begin{equation}
\phi(n)\ \longrightarrow\ \left(\prod_{j=1}^{n-1}e^{i\pi\sigma^+(j)\sigma^-(j)}\right)\sigma^-(n),
\label{intr2}
\end{equation}
where $n$ and $j$ number the lattice sites to which the fermionic or spin operators are assigned. In the fermionic picture, fermions live on the sites of a 1-dimensional lattice and are described by creation $\phi(n)^\dag$ and annihilation $\phi(n)$ operators, while on the spin side there are Pauli matrices $\sigma^k(n)$ ($k=x,y,z$) located at corresponding lattice sites -- as shown in Fig. \ref{fig000}. The main limitation of this transformation stems from the fact that it does not have a straightforward generalization to $d>1$. Indeed, the products of the exponents in these equations rely on a natural 1-dimensional ordering of lattice sites. A naive way to bypass this problem, in higher dimensions, is to pick a 1-dimensional path that visits every lattice site once and apply this ordering to analogues of formulas (\ref{intr1}) and (\ref{intr2}) in $d>1$. However, in this approach some local observables in one picture are not mapped to local observables in the other. The spin chains present in eqs. (\ref{intr1}) and (\ref{intr2}) mostly cancel each other out for the operators assigned to nearby lattice sites. Yet, in $d>1$ sites adjacent on the lattice are not necessarily neighbors on the 1-dimensional path -- hence the non-locality. In $d>1$, finding a transformation from fermionic to spin variables which preserves locality of observables is therefore a non-trivial task, which has attracted much interest, also in the recent years \cite{w,n5,n6,n8}. Such mappings are sometimes called local bosonizations.

The fundamental subject under study in this work is an old proposal by Wosiek and Szczerba \cite{w,n15,brww} to locally replace Grassmann variables in description of a system of free fermions on a lattice with spin-like variables in two and three space dimensions. To illustrate the method briefly, let us begin with a 1-dimensional lattice. The idea is to introduce Clifford variables:
\begin{equation}
X(n)=\phi(n)^\dag+\phi(n)\qquad\qquad Y(n)=i(\phi(n)^\dag-\phi(n))
\label{intr3}
\end{equation}
and link operators:
\begin{equation}
S(n)=iX(n)X(n+1)\qquad\qquad \tilde{S}(n)=iY(n)Y(n+1).
\label{intr4}
\end{equation}
These operators live in a space which is a tensor product of the Hilbert spaces associated with all the lattice sites, i.e. $S(n)=i\mathds{1}\otimes\mathds{1}\otimes...\otimes X(n)\otimes X(n+1)\otimes...\otimes\mathds{1}$ and $\tilde{S}(n)=i\mathds{1}\otimes\mathds{1}\otimes...\otimes Y(n)\otimes Y(n+1)\otimes...\otimes\mathds{1}$. Geometrically, they are associated with lattice links between sites $n$ and $n+1$, which is apparent in the above definitions. Since the fermionic operators fulfill standard algebra $\{\phi(n)^\dag,\phi(m)\}=\delta_{nm}$ (while all other anticommutators vanish), a simple derivation yields the algebra of the link operators:
\begin{equation*}
[S(n),\tilde{S}(m)]=0,
\tag{1.5a}
\label{intr5a}
\end{equation*}
\begin{equation*}
[S(n),S(m)]=0,\ [\tilde{S}(n),\tilde{S}(m)]=0\qquad m\neq n-1,n+1,
\tag{1.5b}
\label{intr5b}
\end{equation*}
\begin{equation*}
\{S(n),S(m)\}=0,\ \{\tilde{S}(n),\tilde{S}(m)\}=0\qquad m= n-1,n+1.
\tag{1.5c}
\label{intr5c}
\end{equation*}
To obtain the bosonization, one substitutes the link operators with spin operators in a way which preserves this algebra. A choice which satisfies this condition is:
\setcounter{equation}{5}
\begin{equation}
S(n)\longrightarrow \sigma^1(n)\sigma^2(n+1)\qquad \tilde{S}(n)\longrightarrow -\sigma^2(n)\sigma^1(n+1),
\label{intr6}
\end{equation}
where $\sigma^j(k)$ is the Pauli matrix $\sigma^j$ assigned to the lattice site $k$ and the products between the spin matrices are tensor products (in analogy to their fermionic counterparts $S(n)$ and $\tilde{S}(n)$). Transformation defined by eqs. (\ref{intr3})--(\ref{intr6}) in one dimension is the same as the Jordan-Wigner transformation, which can be seen e.g. by application of substitution (\ref{intr1})--(\ref{intr2}) to eqs. (\ref{intr3})--(\ref{intr4}) to again obtain the rules (\ref{intr6}).\\
To anticommute, the link operators (\ref{intr4}) need to be links of the same type (either $S$ or $\tilde{S}$) \emph{and} be neighbors on the lattice, i.e. share exactly one endpoint -- as specified in eq. (\ref{intr5c}). In all other cases they commute. Therefore, the link operators and their representation by spin matrices (\ref{intr6}) establish a set of variables which commute at large distances -- contrary to the original set of fermionic operators which anticommute. Thus, they resemble bosonic degrees of freedom with local interactions.

This transformation is local, i.e. eqs. (\ref{intr3}), (\ref{intr4}) and (\ref{intr6}) provide a rule to replace the fermionic variables with the spin matrices in a way that relates only those fermionic/spin operators which are assigned to nearby lattice sites. In particular, it does not have explicitly non-local terms like the spin chains in the Jordan-Wigner approach. Its generalization to an arbitrary dimension is simple \cite{w} -- assumption that $d=1$ is not vital to this proposal. In $d$ dimensions there are $2d$ links overlapping at one end, so to satisfy condition (\ref{intr5c}) one needs matrices among which there is a set of $2d$ anticommuting ones. This can be done with generalized Euclidean Dirac matrices $\Gamma^{(d)}$ \cite{w}.

Since the fermions in the model studied in this work have 2-dimensional Hilbert spaces, while the dual spin-like variables are $2d$-dimensional, it is clear that in $d>1$ dimensions the spin Hilbert space has to be a subspace of the whole tensor product of the spin spaces associated with all the lattice sites. It must be reduced to match the dimension of its fermionic counterpart for these descriptions to be equivalent. Therefore, the representation by commuting variables requires constraints.
The origin, meaning, properties and mutual relations of these constraints form an interesting structure, which is studied in this work. It is also worth noting that similar local constraints are present in other models of local bosonization in $d>1$ dimensions \cite{n5,n6} (see eq. (\ref{intr31d})).

Details of this transformation are researched throughout the work, particularly with the aim to complete the proposal \cite{w,n15} with the construction of constraints in a basis which enables their solution. A concrete representation of the spin Hilbert spaces and the constraints allows to find the (reduced) Hilbert space in the dual picture and to apply this description to obtain actual observables.
In other words, due to the necessity to impose the constraints on the spin space and since solving them is a major task, derivation of the spin-representation formulas for the Hamiltonian and the constraints is not the final step.
This work attempts to fill this gap. It is mainly focused on systems of free fermions on rectangular lattices in $d=2$ dimensions. Particular attention is paid to small lattices as they allow direct numerical studies. Dependence of boundary conditions for the fermionic and spin operators, which arise on finite-size lattices, on the lattice size and the total number of particles is determined. The constraints are examined comprehensively and relations between the individual constraints are found in order to indicate redundant ones and establish a minimal set of independent constraints.
This construction is then applied to obtain the eigenenergies of the system in the spin description. The equivalence of the two descriptions is further tested by comparison of the energetic spectra in the two dual pictures -- the eigenvalues found from exact algebraic diagonalization of the Hamiltonian in the spin representation are compared to the analytic results known in the fermionic picture. An extension of the model with free fermions to a model with fermions interacting with an external $\mathbb{Z}_2$ field is also proposed and studied.

\section{Motivation}
There are three notable motivations to this study: possible applications in numerical calculations in lattice gauge theories \cite{n11}, interesting structure 
of such dual descriptions of a system \cite{n5,n6,n8,n7,n9,n10} and relations to challenges encountered in quantum computing \cite{n11c,n12,n13,n14}. These motivations are briefly discussed below.
\subsection{Monte Carlo lattice calculations}
Understanding of the fundamental particle physics phenomena, including the quark confinement, prediction of the hadron spectrum, chiral symmetry breaking, predictions for the parameters of the Standard Model and study of models beyond the Standard Model \cite{intro}, is inevitably linked to the analysis of quantum field theories in non-perturbative regimes. Therefore, the discretization of quantum chromodynamics (QCD) proposed by Wilson in 1974 \cite{wilson}, which provides a formulation of QCD on a space-time lattice and enables various non-perturbative techniques, has begun an important direction in the research of gauge theories. Monte Carlo methods have become popular and successful means of lattice studies, including QCD. The basic motive to this research is an attempt to get one step closer towards improving efficiency of time-consuming Monte Carlo simulations in lattice gauge theories. To familiarize the reader with this topic, a concise introduction, which follows \cite{intro,gl}, is provided below. One begins with the aim to calculate expectation values of observables, which in path integral formulation of quantum field theory are written as:
\begin{equation}
\braket{O}=\frac{1}{Z}\int\mathcal{D}U\mathcal{D}\psi\mathcal{D}\bar{\psi}\ O(U,\psi,\bar{\psi})e^{-S_G(U)-S_F(U,\psi,\bar{\psi})}.
\label{intr7}
\end{equation}
In this path integral, $S_G(U)$ is the gauge part of the action while $S_F(U,\psi,\bar{\psi})$ is the fermionic action and $Z=\int\mathcal{D}U\mathcal{D}\psi\mathcal{D}\bar{\psi}\text{exp}(-S_G-S_F)$ is the partition function. Once the fermionic degrees of freedom are integrated out or in pure gauge quantum field theories, the expectation values are:
\begin{equation}
\braket{O}=\frac{1}{Z}\int\mathcal{D}U\ O(U)e^{-S_G(U)}.
\label{intr8}
\end{equation}
When the problem is discretized and studied on a space-time lattice, these expectation values are approximated numerically by averages
\begin{equation}
\bar{O}=\frac{1}{n}\sum_{i=1}^{n}O(U^{(i)}),
\label{intr9}
\end{equation}
where the field configurations $U^{(i)}$ are generated randomly with the probability distribution $\text{exp}(-S_G)$. 
This method, called \emph{importance sampling}, ensures that the exponential factor $\text{exp}(-S_G)$ is recovered, although it is not explicitly present in eq. (\ref{intr9}). Its greatest benefit is that the path integral is probed more densely in regions of the field configuration space with highest contributions to the integral. A popular procedure to generate sequences of field configurations $\{U^{(i)}\}$ with the required probability distribution is the Metropolis algorithm and its variants.\\
Since anticommuting Grassmann variables are unsuitable for computers and Monte Carlo algorithms, a typical approach for theories with fermions is to integrate out the fermionic degrees of freedom from path integrals like (\ref{intr7}) to obtain integrals over gauge field only (like (\ref{intr8})). After that, Markov chain Monte Carlo algorithms are applied, the field configurations are generated and the path integral is approximated by statistical averages. To eliminate the Grassmann variables one uses the Matthews-Salam formula \cite{gl}, which for real fermions reads:
\begin{equation}
\int\mathcal{D}\psi\mathcal{D}\bar{\psi}e^{-\int d^4x\ \bar{\psi}(x)Q\psi(x)}=\text{det }Q,
\label{intr10}
\end{equation}
where $Q$ is any operator. This identity applies to any fermionic action bilinear in $\psi$, e.g. for a free Dirac field $S_F=\int d^4x\ \bar{\psi}(x)(\gamma_\mu\partial^\mu+m)\psi(x)$, so $Q=\gamma_\mu\partial^\mu+m$. This quantity is called the \emph{fermionic determinant}. When applied to eq. (\ref{intr7}), this formula yields:
\begin{equation}
\braket{O}=\frac{1}{Z}\int\mathcal{D}U\ O'(U)\text{det }Q(U)\ e^{-S_G(U)},
\label{intr11}
\end{equation}
where
\begin{equation}
O'(U)=\frac{\int\mathcal{D}\psi\mathcal{D}\bar{\psi}\ O(U,\psi,\bar{\psi})e^{-S_F(U,\psi,\bar{\psi})}}{\int\mathcal{D}\psi\mathcal{D}\bar{\psi}\ e^{-S_F(U,\psi,\bar{\psi})}}
\label{intr12}
\end{equation}
is a version of observable $O(U,\psi,\bar{\psi})$ with fermionic variables integrated out. Eq. (\ref{intr11}) is therefore a starting point for the Metropolis algorithm (or other importance sampling algorithms) with gauge variables only. Yet, calculation of the fermionic determinant $\text{det }Q(U)$ separately for every field configuration $U^{(i)}$ -- taking into account that $Q$ is typically a huge matrix -- greatly increases computational complexity. It is a major issue which limits the capabilities of these algorithms and increases the times of simulations. One possible way to address this problem is the quenched approximation that treats $Q$ as a constant. Due to advances in algorithm design (Hybrid Monte Carlo algorithm \cite{hmc}) and dedicated hardware development, full simulations with dynamical fermions are possible since 20 years, however they still require huge computing resources.

A transformation from Grassmann to spin variables suggests an alternative way to avoid computation of the fermionic determinants. Once such a transformation is performed, one could apply the Monte Carlo methods to both the gauge field and the new bosonized variables describing fermions. The fermionic degrees of freedom are not integrated out, but instead the importance sampling algorithm and calculation of the statistical sums are done in a larger space. Locality of the transformation from fermions to spins is essential in this application. When the field configurations $\{U^{(i)}\}$ are generated with the importance sampling algorithm, a new configuration candidate $U^{(i+1)}$ usually differs only locally from the previous configuration $U^{(i)}$, e.g. by a random increment at a single lattice site. It is accepted or rejected in the sequence of configurations $\{U^{(i)}\}$ based on its action: if $S(U^{(i+1)})\leq S(U^{(i)})$, it is accepted and otherwise it is accepted with a probability $\text{exp}(-(S(U^{(i+1)})-S(U^{(i)})))$. Since the two configurations differ only locally, one obtains $S(U^{(i+1)})$ from $S(U^{(i)})$ by a local correction. The term of the action which corresponds to the lattice site with the updated value of the field is replaced with an appropriate new value. This approach saves computing time on the step which is continually repeated in the importance sampling algorithm. If the bosonization were not local, local updates to the fermionic field would result in non-local changes of the spin variables configuration. Hence, the modification to the action is also non-local in this case and the above simplification cannot be employed.

\subsection{Dualities}

Another significant motivation to study relations between the description of the system of fermions on a 2-dimensional lattice by Grassmann variables and the dual description by spin matrices is interesting structure of such dualities, which were discovered for many other systems \cite{n5,n6,n8,n9,n10,n12,montoten,maldacena,nyscarleo,n16}. These dualities provide additional understanding of the systems they involve when description in one of the two dual pictures
makes some properties more apparent, or when the calculations are easier in this picture. For example, a more convenient formulation can reveal hidden symmetries, help discover phases or apply specific ansatz\"{e}. To illustrate this, a concise discussion of a few famous dualities follows.

\subsubsection{Electromagnetic duality 
}

A basic, well-known example is the electromagnetic duality between electric and magnetic fields. In vacuum, Maxwell's equations are given by:
\begin{equation}
\begin{split}
\vec{\nabla}\cdot\vec{E}&=0\qquad\qquad\vec{\nabla}\cdot\vec{B}=0\\
\vec{\nabla}\times\vec{E}&=-\frac{\partial\vec{B}}{\partial t}\qquad\vec{\nabla}\times\vec{B}=\frac{\partial\vec{E}}{\partial t}.
\end{split}
\label{intr13a}
\end{equation}
These equations are highly symmetric -- in particular, they are invariant 
when the electric and magnetic fields are interchanged:
\begin{equation}
\vec{E}\rightarrow\vec{B},\qquad \vec{B}\rightarrow -\vec{E}.
\label{intr14a}
\end{equation}
The electric and magnetic degrees of freedom combine when the Lorentz transformation is applied, which is made manifest by introducing the field strength tensor $F_{\mu\nu}$:
\begin{equation}
F^{0i}=-F^{i0}=-E^i\qquad F^{ij}=-\epsilon_{ijk}B^k,
\label{intr15a}
\end{equation}
so that the vacuum Maxwell's equations become:
\begin{equation}
\partial_\nu F^{\mu\nu}=0\qquad \partial_\nu^\star F^{\mu\nu}=0,
\label{intr16a}
\end{equation} 
where $^\star F^{\mu\nu}=\frac{1}{2}\epsilon^{\mu\nu\lambda\rho}F_{\lambda\rho}$. In this formulation duality (\ref{intr14a}) reads:
\begin{equation}
F^{\mu\nu}\rightarrow ^\star F^{\mu\nu},\qquad ^\star F^{\mu\nu}\rightarrow -F^{\mu\nu}.
\label{intr17a}
\end{equation}
Electromagnetic duality (\ref{intr14a}) relates phenomena which occur in electric field to dual effects in magnetic field. These include effects in static electromagnetism like, e.g. duality between Faraday's law of induction and Amp\`{e}re's circuital law, as well as quantum phenomena -- like duality between a charged particle which acquires a phase shift when traveling through a magnetic potential due to Aharonov-Bohm effect and analogous Aharonov-Casher effect \cite{casher} for magnetic dipoles in electric fields.\\
However, the electromagnetic duality is limited and its extension to the case with charges is non-trivial. When a nonzero current density $j^\mu$ is present, eqs. (\ref{intr16a}) are replaced with $\partial_\nu F^{\mu\nu}=j^\mu$ and $\partial_\nu^\star F^{\mu\nu}=0$, so they are no longer invariant under transformation (\ref{intr17a}). One way out of this issue is to introduce both electric and magnetic charges:
\begin{equation}
\partial_\nu F^{\mu\nu}=j^\mu,\qquad \partial_\nu^\star F^{\mu\nu}=k^\mu
\label{intr18a}
\end{equation}
and supplement duality (\ref{intr17a}) with additional rules $j^\mu\rightarrow k^\mu$ and $k^\mu\rightarrow -j^\mu$. This leads to consideration of theories with magnetic monopoles \cite{figureoa}, which is a topic broad enough not to be discussed here. Yet, the final moral of this part is that even such a common duality as the one between electric and magnetic fields has far reaching consequences related to fundamental interactions, even string theory -- for example the duality between magnetic monopole solitons and electric charges as proposed by Olive and Montoten in \cite{montoten,olive}.

\subsubsection{Kramers-Wannier duality of the Ising model
}

Kramers-Wannier duality \cite{wannier}, being a transformation associated with a lattice system of spins, is an example closely related to the main proposal studied in this work. It is a self-duality in a sense that it links Ising model with itself, but with the ordered phase of the system in one picture mapped to the disordered phase in the dual description and vice versa. The fact that the low and high temperature regimes are projected onto each other in this duality, while the critical point is self-dual, enables one to find the temperature of the phase transition for a two-dimensional classical Ising model, which was its original application \cite{wannier}. It is also a crucial point in the famous exact solution for the partition function of $d=2$ Ising model by Onsager \cite{onsager}.

This duality can be studied either for a classical $d=2$ Ising model with $\mathbb{Z}_2$ spins on a square lattice in a magnetic field or for a quantum $d=1$ Ising model with spins in a transverse magnetic field. The latter path is taken in the following discussion. The Hamiltonian of this model reads:
\begin{equation}
H=-J\sum_i\sigma^z(i)\sigma^z(i+1)-h\sum_i\sigma^x(i),
\label{intr19b}
\end{equation}
where, similarly to the notation used in Section 1.1, $\sigma^j(i)$ is a Pauli matrix $\sigma^j$ assigned to a lattice site $i$, $J$ is the coupling of nearest-neighbor spin interactions, while $h$ is the coupling of interactions with the external transverse magnetic field. It is easily seen that introduction of new variables $\tau^j(i+1/2)$:
\begin{equation}
\begin{split}
\tau^x\left(i+\frac{1}{2}\right)&=\sigma^z(i)\sigma^z(i+1)\\
\tau^z\left(i-\frac{1}{2}\right)\tau^z\left(i+\frac{1}{2}\right)&=\sigma^x(i),
\end{split}
\label{intr20b}
\end{equation}
which are located at midpoints between neighboring sites of the original lattice, yields an identical Hamiltonian:
\begin{equation}
H=-h\sum_i\tau^z\left(i-\frac{1}{2}\right)\tau^z\left(i+\frac{1}{2}\right)-J\sum_i\tau^x\left(i+\frac{1}{2}\right).
\label{may24a}
\end{equation}
The only differences with eq. (\ref{intr19b}) is replacement of $\sigma$ matrices with $\tau$ and interchanged roles of the couplings $J$ and $h$. The second condition in eqs. (\ref{intr20b}) is solved by:
\begin{equation}
\tau^z\left(i+\frac{1}{2}\right)=\prod_{j\leq i}\sigma^x(j).
\label{intr21b}
\end{equation}
When $J>h$, the global $\mathbb{Z}_2$ symmetry ($\sigma^{z,y}(i)\to-\sigma^{z,y}(i)$) 
is spontaneously broken and the system is in the ordered (ferromagnetic) state, while for $J<h$ interactions with the transverse magnetic field prevail over nearest-neighbor interactions, which corresponds to the paramagnetic state. 
\\The dual model can be interpreted as an Ising model with kink degrees of freedom. Indeed, since $\sigma^x$ flips the eigenvalues of $\sigma^z$, operator (\ref{intr21b}) flips z-components of spin at all the spin-sites to the left from $i+1/2$. Therefore, it describes a disruption in the order parameter -- a kink. It can be understood qualitatively, why the phases in the dual model are interchanged with respect to the original one: when the spins align as a result of mutual interactions ($J>h$), the kinks are energetically costly. In the dual picture it means alignment of variables $\tau$ along z-axis is not likely, since it corresponds to many kinks (see eq. (\ref{intr21b})). 
When viewed from the perspective of the Hamiltonian (\ref{may24a}) (in the $\tau$-picture), the global symmetry $\mathbb{Z}_2$ ($\tau\to -\tau$) is unbroken for $J>h$ and the system is in the disordered state of the $\tau$ spins. Coming back to the $\sigma$ picture: in the paramagnetic state ($J<h$) the $\sigma$ spins are disordered and the kinks abundant.

Apart from Kramers-Wannier duality, the system described by Hamiltonian (\ref{intr19b}) is linked to a system of free Majorana fermions via a Jordan-Wigner transformation variant similar to (\ref{intr1})-(\ref{intr2}). Hamiltonian (\ref{intr19b}) is dual to: 
\begin{equation}
H_f=iJ\sum_{\text{even }r}\phi(r)\phi(r+1)+ih\sum_{\text{odd }r}\phi(r)\phi(r+1)
\label{intr22b}
\end{equation}
after a substitution:
\begin{equation}
\phi(r)=\begin{cases}i\sigma^z(r/2)\tau^z((r+1)/2)&\qquad\text{, even }r\\
\tau^z(r/2)\sigma^z((r+1)/2)&\qquad\text{, odd }r\end{cases}.
\label{intr23b}
\end{equation}
This duality is a good illustration of the profits that can be gained by inspection of dual theories, which were mentioned in the beginning of this subsection. Model (\ref{intr22b}) has two topologically distinct phases, which are dual to the ferromagnetic and paramagnetic phases of model (\ref{intr19b}) respectively \cite{n9}. Therefore, good knowledge of such a well-studied model as the Ising model can be applied to the fermionic one. When $J=h$, Hamiltonian (\ref{intr22b}) has translational symmetry $r\to r+1$ and it turns out this symmetry maps to the self-duality of the Ising model. Last but not least, this duality is also beneficial to the study of the spin model as its exact, analytic solution is easier found for the system of free fermions.

\subsubsection{AdS/CFT correspondence}

Another significant example of duality is the AdS/CFT correspondence. This dualism, conjectured by Maldacena \cite{maldacena}, relates type IIB string theory in anti-de Sitter background spacetime to (3+1)-dimensional $\mathcal{N}=4$ supersymmetric Yang-Mills (SYM) theory. In Maldacena's proposal, one studies conformal field theories (CFTs) on $N$ parallel, equally spaced by $r$ D-branes in low energy limit, when the brane CFTs decouple from the bulk -- i.e. a large $N$ approximation is investigated. To be more specific, let us focus on D3 branes in the product spacetime $AdS_5\times S^5$ of the five dimensional anti-de Sitter space $AdS_5$ and five-sphere $S^5$, which corresponds to compactified dimensions in this string theory. Starting from the supergravity solution carrying D3 brane charge \cite{maldacena,horowitz}:
\begin{equation}
ds^2=f^{-1/2}dx^2_{||}+f^{1/2}(dr^2+r^2d\Omega_5^2),
\label{intr13}
\end{equation}
\begin{equation}
f=1+\frac{4\pi gN\alpha'^2}{r^4},
\label{intr14}
\end{equation}
where $x_{||}$ are the coordinates along the worldvolume of the brane, $d\Omega_5$ is the metric of five-sphere and $g$ is the string coupling, one takes the decoupling limit $\alpha'\to0$, $U=r/\alpha'=\text{const}$ and finally arrives at the conclusion that (3+1)-dimensional $\mathcal{N}=4$ $U(N)$ SYM includes, for large $N$, in its Hilbert space states of type IIB supergravity on $AdS_5\times S_5$. This observation is extended by Maldacena to a more general conjecture that, even beyond this supergravity approximation, type IIB string theory on $AdS_5\times S_5$ (with some appropriate boundary conditions) is dual to (3+1)-dimensional $\mathcal{N}=4$ $U(N)$ SYM theory. It is conjectured that also string theories on other anti-de Sitter spacetimes are dual to various conformal field theories, e.g. $d=6$ (0,2) conformal field theory is dual to M-theory on $AdS_7 \times S^4$.\\
An important corollary of this duality is its potential to yield a non-perturbative formulation of string/M-theory, since the corresponding field theories can, in principle, be defined non-perturbatively. On the other hand, it is useful for studies of quantum field theories in strong coupling regime, since their dual string theories are weakly coupled. It also allows to view string theory phenomena from gauge theory perspective -- including Hawking radiation and physics of the horizon.

\subsection{Local bosonization}

Having discussed some famous cases of dualities, we refocus on the main topic of this work -- transformations which allow description of fermionic systems by spin variables or, more generally, dualities between theories with fermionic and bosonic degrees of freedom. Such maps are often called \emph{bosonizations} and, since local transformations are of particular interest as pointed out in Section 1.2.1, it is instructive to review a few successful attempts at \emph{local bosonization}.
\begin{figure}[!h]
\begin{center}
\includegraphics[width=5cm]{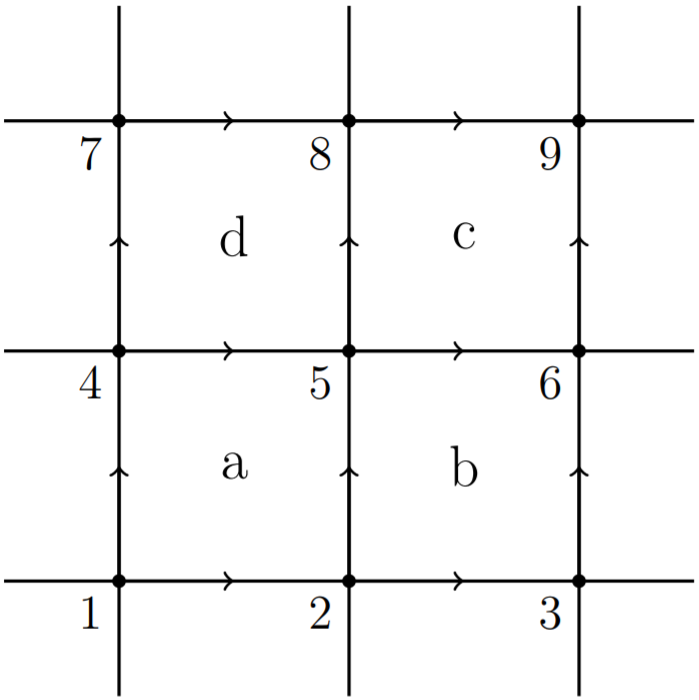}
\end{center}
\caption{The setup used in a local bosonization scheme by Chen, Kapustin and Radi\v{c}evi\'{c}. Fermions are located on the faces of the lattice, while spin matrices are associated with its edges. All horizontal edges are east-oriented and the vertical edges are oriented to the north (source of the image: \cite{n5}).}
\label{fig00}
\end{figure}
\subsubsection{Chen-Kapustin-Radi\v{c}evi\'{c} bosonization scheme
}
Let us begin with a quite recent proposal \cite{n5} by Chen, Kapustin and Radi\v{c}evi\'{c}. It bears some resemblance to the scheme studied in this work -- in particular, due to appearance of some kind of constraints on the Hilbert space in the dual picture. In this approach, fermions are located on the faces of an infinite 2-dimensional square lattice (Fig. \ref{fig00}). Thus, one starts with a set of fermionic creation $\phi(f)^\dag$ and annihilation $\phi(f)$ operators, which are assigned to lattice faces $f$ and fulfill a standard algebra $\{\phi(f)^\dag,\phi(f')\}=\delta_{f'}^f$. Similarly to eq. (\ref{intr3}), one introduces Majorana fermions:
\begin{equation}
\gamma(f)=\phi(f)^\dag+\phi(f),\qquad \gamma'(f)=i(\phi(f)^\dag-\phi(f))
\label{intr26d}
\end{equation}
and hopping operators:
\begin{equation}
S(e)=i\gamma(L(e))\gamma'(R(e)),
\label{intr27d}
\end{equation}
which are associated with lattice edges $e$. $L(e)$ and $R(e)$ are neighboring lattice faces which share the edge $e$. $L(e)$ lies to the left from this common edge, while $R(e)$ to the right -- hence the need to assign orientations to the lattice edges (this can be done arbitrarily, yet the presented formulas are valid for the choice illustrated in Fig. \ref{fig00}). Denote $F(f)=\phi(f)^\dag\phi(f)$ -- the fermion number operator at face $f$. In these new variables, the fermionic parity operator becomes:
\begin{equation}
(-1)^{F(f)}=1-i\gamma(f)\gamma'(f).
\label{intr28d}
\end{equation}
The operators (\ref{intr27d}) and (\ref{intr28d}) generate the algebra of the fermionic system. Similarly to the algebra (\ref{intr5a})-(\ref{intr5b}), they constitute a set of locally interacting variables as $S(e)$ and $(-1)^{F(f)}$ anticommute when $e$ is an edge of the face $f$, and commute otherwise. All parity operators $(-1)^{F(f)}$ commute between each other, while $S(e)$ and $S(e')$ anticommute if they are perpendicular and share a point which is the beginning to one of them and the end to the other (like $S_{25}$ and $S_{56}$ in Fig. \ref{fig00}, but not $S_{56}$ and $S_{58}$ to which point $5$ is the common beginning, but none of them ends there). In all other cases $S(e)$ and $S(e')$ commute.\\
The elementary variables of the dual, bosonic description are Pauli matrices $\sigma^j(e)$, $j\in\{x,y,z\}$ located on the lattice edges. In this picture, one defines operators:
\begin{equation}
U(e)=\sigma^x(e)\sigma^z(r(e))
\label{intr29d}
\end{equation}
and
\begin{equation}
W(f)=\prod_{e\subset f}\sigma^z(e).
\label{intr30d}
\end{equation}
$r(e)$ is the edge that precedes $e$, i.e. it is an edge perpendicular to $e$ and ending where $e$ begins (e.g. $r(e_{56})=e_{25}$). It can be shown \cite{n5} that the algebra generated by the operators $(-1)^{F(f)}$ and $S(e)$ is preserved under mapping $(-1)^{F(f)}\leftrightarrow W(f)$ and $S(e)\leftrightarrow U(e)$, provided that constraints:
\begin{equation}
W(\text{NE}(v))\prod_{e\supset v}\sigma^x(e)=1
\label{intr31d}
\end{equation}
are imposed. The product in this equation runs over all edges $e$ which begin or end in a lattice vertex $v$ and $\text{NE}(v)$ is a lattice face located north-east to this vertex. These constraints, which couple the electric charge at vertex $v$ to the magnetic flux at face $\text{NE}(v)$, are interpreted as modified Gauss law for the bosonic system. Therefore, the dual theory is a $\mathbb{Z}_2$ gauge theory.\\
This duality can be applied to many meaningful physical problems like fermions on square or honeycomb lattices, Hubbard model, simple lattice gauge theories \cite{n5} and it can be generalized to arbitrary dimensions \cite{n6}. For example, in the simple case of fermions on a square lattice with nearest-neighbor hopping and on-site chemical potential:
\begin{equation}
H_f=t\sum_e(\phi(L(e))^\dag\phi(R(e))+\phi(R(e))^\dag\phi(L(e)))+\mu\sum_f\phi(f)^\dag\phi(f),
\label{intr32d}
\end{equation}
its dual, bosonic theory is described by a Hamiltonian:
\begin{equation}
H_s=\frac{t}{2}\sum_e\sigma^x(e)\sigma^z(r(e))(1-W(L(e))W(R(e)))+\frac{\mu}{2}\sum_f(1-W(f))
\label{intr33d}
\end{equation}
and constraints $(\prod_{e\supset v}\sigma^x(e))W(\text{NE}(v))=1$ on each vertex.
\begin{figure}[!h]
\begin{center}
\includegraphics[width=15cm]{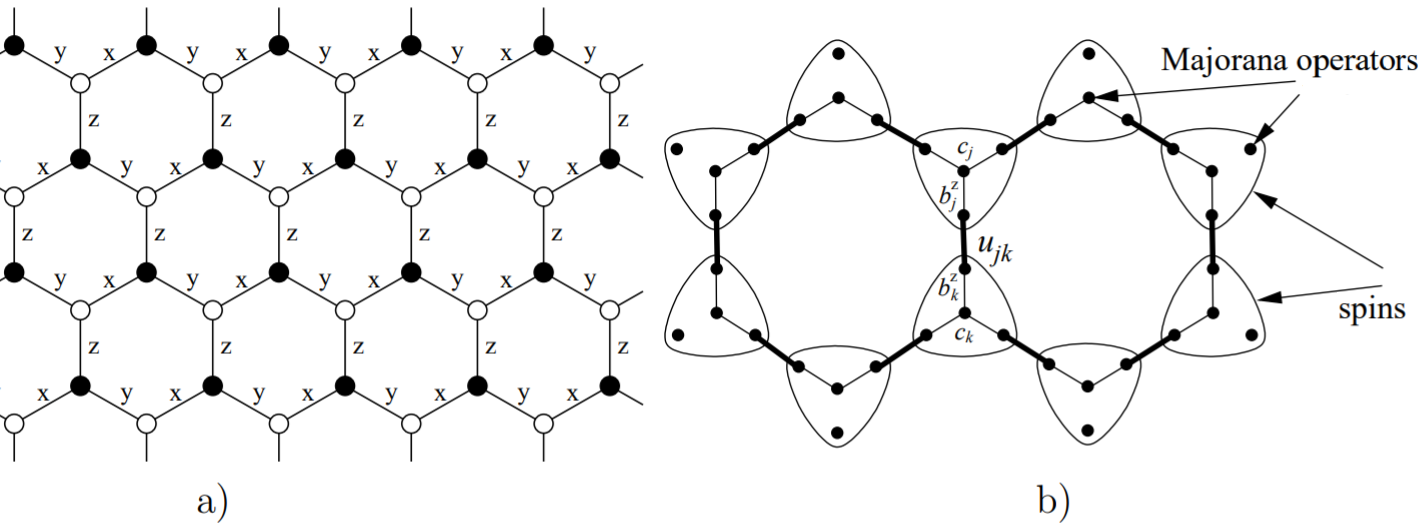}
\end{center}
\caption{The setup used in spin-fermion duality scheme by Kitaev: (a) three types of links on a honeycomb lattice, (b) graphic representation of Hamiltonian (\ref{intr38d}) (source of the images: \cite{n12}).}
\label{fig0}
\end{figure}

\subsubsection{Kitaev proposal for honeycomb lattices
}

Another, in some aspects similar approach is proposed by Kitaev \cite{n12}. In this proposal a reverse path is taken actually, i.e. a fermionic representation is designed for a spin system. One begins with a system of 1/2-spins located at the vertices of a honeycomb lattice (Fig. \ref{fig0}a) whose dynamics is governed by a Hamiltonian:
\begin{equation}
H=-J_x\sum_{x-\text{links}}\sigma^x(j)\sigma^x(k)-J_y\sum_{y-\text{links}}\sigma^y(j)\sigma^y(k)-J_z\sum_{z-\text{links}}\sigma^z(j)\sigma^z(k),
\label{intr34d}
\end{equation}
where $\sigma^i(j)$ is a Pauli matrix $\sigma^i$ assigned to a lattice site $j$ and $i$-links are classified based on their direction as shown in Fig. \ref{fig0}a ($i\in\{x,y,z\}$).\\
To represent a spin located at a single vertex two fermionic modes are used, which gives four creation/annihilation operators in total: $\phi^\dag(1)$, $\phi(1)$, $\phi^\dag(2)$ and $\phi(2)$. Introduction of Majorana fermions (\ref{intr26d}) yields 4 Majorana operators, which are denoted by $b^x$, $b^y$, $b^z$ and $c$ -- note that all operators (\ref{intr26d}) can be treated on equal basis, so such an arrangement should not be viewed as "asymmetric". Majorana operators act on the 4-dimensional Fock space $\tilde{M}$, while the 2-dimensional Hilbert space of a spin is mapped to its 2-dimensional subspace $M=\{\ket{\zeta}\in \tilde{M}:D\ket{\zeta}=\ket{\zeta}\}\subset\tilde{M}$, where $D\equiv b^xb^yb^zc$. A fermionic representation of spin operators $\sigma^x\to\tilde{\sigma}^x$, $\sigma^y\to\tilde{\sigma}^y$, $\sigma^y\to\tilde{\sigma}^y$ must be such that $M$ is an invariant subspace of these $\tilde{\sigma}^\alpha$ operators \emph{and} they must satisfy the same algebra as Pauli matrices when restricted to $M$. It is easily seen that a choice:
\begin{equation}
\tilde{\sigma}^x=ib^xc,\qquad\tilde{\sigma}^y=ib^yc,\qquad\tilde{\sigma}^z=ib^zc
\label{intr35d}
\end{equation}
satisfies these conditions. Once the duality transformation is found for a single spin/vertex, its extension to the whole lattice is straightforward. One defines:
\begin{equation}
\tilde{\sigma}^{\alpha}(j)=ib^\alpha(j)c(j),\qquad D(j)=b^x(j)b^y(j)b^z(j)c(j)\qquad (\alpha=x,y,z)
\label{intr36d}
\end{equation}
at every lattice vertex $j$ and the subspace of the fermionic Fock space which is homomorphic to the spin Hilbert space is determined by a condition:
\begin{equation}
L=\{\ket{\zeta}\in\tilde{L}:\forall\ j\quad D(j)\ket{\zeta}=\ket{\zeta}\}\subset \tilde{L},
\label{intr37d}
\end{equation}
where $L$ and $\tilde{L}$ are full-lattice analogues of $M$ and $\tilde{M}$.\\
When applied to the model (\ref{intr34d}), this transformation yields a dual, fermionic Hamiltonian:
\begin{equation}
\tilde{H}=\frac{i}{4}\sum_{j,k}\hat{A}_{jk}c(j)c(k),
\label{intr38d}
\end{equation}
where
\begin{equation}
\hat{A}_{jk}=\begin{cases}
2J_{\alpha_{jk}}\hat{u}_{jk}&\qquad\text{if j and k are connected},\\
0&\qquad\text{otherwise}
\end{cases}
\label{intr39d}
\end{equation}
and $\hat{u}_{jk}=ib^{\alpha_{jk}}(j)b^{\alpha_{jk}}(k)$. The index $\alpha_{jk}$ takes values $x$, $y$, $z$ depending on the direction of the link $(j,k)$. The structure of Hamiltonian (\ref{intr38d}) is depicted in Fig. \ref{fig0}b: there are four Majorana operators at every vertex to represent a spin located at this site and $\hat{u}_{jk}$ can be thought of as links between neighboring sites.\\
Since $\hat{u}_{jk}$ commute with the Hamiltonian and among each other, diagonalization of (\ref{intr38d}) can be performed in sectors with fixed eigenvalues $u_{jk}=\pm 1$. Within these sectors, the Hamiltonian simplifies to a form with the operators $\hat{A}_{jk}$ replaced by their eigenvalues $A_{jk}$, which corresponds to a Hamiltonian of a system of free fermions. Thus, the transformation from (\ref{intr34d}) to (\ref{intr38d}) provides a way to find the exact solution for this system.\\

An interesting general feature of many bosonization proposals is their relation to flux attachment. As shown in \cite{wilczek} a composite of a charged particle and a magnetic flux tube obeys a statistics different from the statistics of this particle alone. In particular, under certain conditions, the statistics of a fermionic particle switches to bosonic when a flux tube is attached and vice versa. Thus, some kind of electric charge-magnetic flux coupling is expected in bosonization schemes. In the case of the transformation discussed at the beginning of this section \cite{n5}, electric charge-magnetic flux coupling manifests in the constraints (\ref{intr31d}). As will be shown, it applies also to the constraints in the duality which is the main topic of this work (see eq. (\ref{may2})). For the bosonizations in quantum field theories in continuum, the flux attachment is usually introduced by including Chern-Simons terms in the action \cite{n8,n9,n10}.
\subsection{Quantum computing}

The ability to locally map fermions to spins has also potential applications in quantum computing. A whole class of problems which could benefit from such a transformation is provided by an idea to use quantum computing to simulate lattice gauge theories, which has attracted much interest over the last decade \cite{n11,n11c,n11b,n11d,n11a,n11e}. In case of analog quantum simulators \cite{n11,n11c}, i.e. when degrees of freedom of the system under study and their dynamical evolutions are directly mapped into the simulating system, physical structure of the simulators introduces limitations on lattice gauge theory problems which can be addressed. In particular, a simulator built entirely of bosons cannot simulate lattice gauge theories with fermionic degrees of freedom. Ultra-cold atoms have proven to be successful candidates to simulate fermions \cite{n11b,n11a}, while other approaches include e.g. superconducting cubits \cite{n11d} and trapped ions \cite{n11c,n11e}. In any case, these attempts experience limitations such as being restricted to $1+1$ dimensions or not being able to efficiently simulate fermions. Despite the difficulties, the advantages of quantum computing approach to lattice simulations encourage the efforts to overcome them. In particular, the analog quantum simulators do not suffer from the sign problem \cite{n11}, allow study of lattice gauge theories in real time \cite{n11e} and, as systems which are quantum in their nature, are a native choice for the study of quantum theories \cite{n11c}.\\
As a final remark to this part, the works by Kitaev \cite{n12,n14}, where the two broad topics mentioned in this introduction -- dual descriptions and quantum computation -- intersect, is worth being mentioned. These works show how a precise description of anyons, particles of a neither bosonic nor fermionic statistics, can be obtained through a duality and discuss their utility in topological quantum computation.
\section{Content of the following sections}

Motivated by the above points, we focus this work on a detailed understanding of the local bosonization proposal, in particular through the algebraic numerical studies. Its dependence on the boundary conditions is studied, the constraints which appear in the spin picture are investigated to understand their mutual relations, numerical checks of the spin Hilbert space reduction are performed and the energetic spectra obtained in both representations are compared.\\
The structure of the work is the following. In Section 2 the necessary ideas, definitions and mathematical objects are introduced. Description of the system of fermions is constructed in the Grassmann and spin representations, the constraints are derived and a proposal to add the interaction of the fermions with an external magnetic field, in a way which does not break the duality, is discussed. Section 3 is devoted to construction of the constraints and the Hamiltonian in the spin representation. It is performed in a basis which is convenient for numerical calculations. Also, the implementation of these structures in the computations is described. Section 4 contains derivations of analytic formulas in the Grassmann representation, in particular the equations for the eigenenergies of the Hamiltonians of free fermions and fermions interacting with the external field. The results of numerical studies are presented in Section 5. They are compared to the theoretical predictions of Sections 2 and 4. The conclusions of this work are summarized in Section 6.
\clearpage
\chapter{Grassmann and spin representation of fermionic algebra}

In this opening section, the procedure to find the equivalent spin description for a system of free fermions on a lattice is thoroughly explained. The Hamiltonian and the particle number operator is constructed in both the Grassmann and spin representations. The link operators are defined in the Grassmann representation and the dual, spin description is determined by the requirement that their algebra is preserved by the transformation. We begin with the 1-dimensional case, which allows one to compare the result of the Jordan-Wigner transformation with the approach based on the link operators algebra preservation. Then, the proposed method is applied to the system of fermions on a 2-dimensional lattice. In $d>1$, one needs to impose the constraints on the spin space to obtain the equivalent description. The full set of constraints is constructed and their mutual relations are examined in Section 2.3. A generalization of the model to include an interaction between the particles and an external $\mathbb{Z}_2$ field, and the corresponding transformation from Grassmann to spin variables are discussed in Section 2.4.

\section{Fermions on 1-dimensional lattice}

\subsection{Jordan-Wigner transform}

It is instructive to begin with a system of free fermions which occupy sites of a 1-dimensional lattice of size $L$. Its Hamiltonian reads:
\begin{equation}
H_f=i\sum_{n=1}^L\left(\phi(n)^\dag\phi(n+1)-\phi(n+1)^\dag\phi(n)\right),
\label{eq1}
\end{equation}
while particle number operator of the system is
\begin{equation}
N_f=\sum_{n=1}^{L}\phi(n)^\dag\phi(n),
\label{eq1b}
\end{equation}
where $\phi(n)^\dag$/$\phi(n)$ are fermionic creation/annihilation operators associated with $n$-th site. They fulfill standard algebra $\{\phi(n)^\dag,\phi(m)\}=\delta_{nm}$. Our efforts are aimed at finding and studying a transformation to represent Hamiltonian (\ref{eq1}) in terms of spin variables. As will be shown in this section, the spin equivalent of $H_f$ is:
\begin{equation}
H_s=\frac{1}{2}\sum_{n=1}^{L}\left(\sigma^1(n)\sigma^2(n+1)-\sigma^2(n)\sigma^1(n+1)\right),
\label{eq2}
\end{equation}
where $\sigma^k$ are Pauli matrices. Matrices $\sigma^j(n)$ are associated with corresponding lattice sites, just as the fermionic operators or, in other words, they are tensor products
\begin{equation}
\sigma^j(n)=\mathds{1}_2\otimes\mathds{1}_2\otimes...\otimes\sigma^j\otimes...\otimes\mathds{1}_2,
\label{eq2b}
\end{equation}
where the Pauli matrix $\sigma^j$ is the $n$-th factor of this product. In particular, this means that spin matrices from different lattice sites commute $[\sigma^j(n),\sigma^k(m)]=0$ ($m\neq n$), while their relations on the same site are identical with the standard Pauli matrices algebra.\\
To obtain eq. (\ref{eq2}) from eq. (\ref{eq1}) one applies Jordan-Wigner transform \cite{jw}. First, one defines:
\begin{equation}
\sigma^+=\frac{1}{2}\left(\sigma^1+i\sigma^2\right),\qquad \sigma^-=\frac{1}{2}\left(\sigma^1-i\sigma^2\right).
\label{eq3}
\end{equation}
Knowing that $\sigma^j\sigma^k=\delta_{jk}\mathds{1}+i\varepsilon_{jkl}\sigma^l$, it is straightforward to see that these matrices satisfy following identities:
\begin{equation}
\sigma^+\sigma^-=\frac{1+\sigma^3}{2},\quad \sigma^-\sigma^+=\frac{1-\sigma^3}{2},\quad \sigma^3\sigma^-=-\sigma^-\sigma^3=-\sigma^-,\quad \sigma^3\sigma^+=-\sigma^+\sigma^3=\sigma^+,
\label{eq4}
\end{equation}
$\sigma^+\sigma^-$ and $\sigma^-\sigma^+$ are projection operators:
\begin{equation}
\left(\frac{1+\sigma^3}{2}\right)^2=\frac{1+\sigma^3}{2},\quad \left(\frac{1-\sigma^3}{2}\right)^2=\frac{1-\sigma^3}{2},
\label{eq5}
\end{equation}
and
\begin{equation}
e^{i\pi\sigma^+\sigma^-}=e^{-i\pi\sigma^+\sigma^-}=-\sigma^3,\quad \sigma^\pm e^{i\pi\sigma^+\sigma^-}=-e^{i\pi\sigma^+\sigma^-}\sigma^\pm.
\label{eq6}
\end{equation}
Derivations of these formulas can be found in \emph{Appendix A}. To perform the Jordan-Wigner transform one substitutes:
\begin{equation}
\phi(n)^\dag\rightarrow \sigma^+(n)\prod_{j=1}^{n-1}e^{-i\pi\sigma^+(j)\sigma^-(j)},
\label{eq7}
\end{equation}
\begin{equation}
\phi(n)\rightarrow\left(\prod_{j=1}^{n-1}e^{i\pi\sigma^+(j)\sigma^-(j)}\right)\sigma^-(n).
\label{eq8}
\end{equation}
This spin representation of the creation and annihilation operators is chosen in such a manner that their algebra is preserved:
\begin{equation}
\{\phi(n)^\dag,\phi(m)\}\rightarrow\{\sigma^+(n)\prod_{j=1}^{n-1}e^{-i\pi\sigma^+(j)\sigma^-(j)},\left(\prod_{k=1}^{m-1}e^{i\pi\sigma^+(k)\sigma^-(k)}\right)\sigma^-(m)\}=\delta^m_n
\label{eq8b}
\end{equation}
(see \emph{Appendix A}). One easily obtains the spin equivalent of the particle number operator because the products of exponential factors cancel each other out in this case:
\begin{equation}
N_s=\sum_{n=1}^{L} \sigma^+(n)\sigma^-(n)=\sum_{n=1}^{L}\frac{1+\sigma^3(n)}{2}.
\label{eq8c}
\end{equation}
To express the Hamiltonian in spin variables, one applies the Jordan-Wigner transformation to a single hopping term from eq. (\ref{eq1}):
\begin{equation}
\begin{split}
\phi(n)^\dag&\phi(n+1)\rightarrow\sigma^+(n)\left(\prod_{j=1}^{n-1}e^{-i\pi\sigma^+(j)\sigma^-(j)}\right)\left(\prod_{j=1}^{n}e^{i\pi\sigma^+(j)\sigma^-(j)}\right)\sigma^-(n+1)=\\
&=\sigma^+(n)e^{i\pi\sigma^+(n)\sigma^-(n)}\sigma^-(n+1)\stackrel{(\ref{eq6})}{=}\sigma^+(n)(-\sigma^3(n))\sigma^-(n+1)\stackrel{(\ref{eq4})}{=}\sigma^+(n)\sigma^-(n+1).
\end{split}
\label{eq9}
\end{equation}
This derivation holds for $1\leq n<L$, i.e. $n\neq L$. Under this assumption the products of exponents cancel except a single term. Special case $n=L$ is addressed in the next section. From eqs. (\ref{eq3}) and (\ref{eq9}) it follows that $\phi(n+1)^\dag\phi(n)\rightarrow\sigma^+(n+1)\sigma^-(n)$, thus:
\begin{equation}
\begin{split}
H_s&=i\sum_{n=1}^L\left(\sigma^+(n)\sigma^-(n+1)-\sigma^+(n+1)\sigma^-(n)\right)=\\
&=i\sum_{n=1}^L\left[\frac{1}{4}(\sigma^1(n)+i\sigma^2(n))(\sigma^1(n+1)-i\sigma^2(n+1))-\frac{1}{4}(\sigma^1(n+1)+i\sigma^2(n+1))(\sigma^1(n)-i\sigma^2(n))\right]=\\
&=\frac{i}{4}\sum_{n=1}^L[-2i\sigma^1(n)\sigma^2(n+1)+2i\sigma^1(n+1)\sigma^2(n)]=\frac{1}{2}\sum_{n=1}^L[\sigma^1(n)\sigma^2(n+1)-\sigma^1(n+1)\sigma^2(n)].
\end{split}
\label{eq10}
\end{equation}
Proof of eq. (\ref{eq2}) is therefore almost complete, but there still remains the special case $n=L$. It requires separate approach and this emphasizes the importance of a careful treatment of boundary conditions.

\subsection{Boundary conditions}

To deal with a finite size of the lattice, one imposes cyclic boundary conditions on fermionic variables:
\begin{equation}
\phi(L+1)=\epsilon\phi(1).
\label{eq11}
\end{equation}
Similarly for spin matrices:
\begin{equation}
\sigma^1(L+1)=\epsilon'\sigma^1(1),\qquad \sigma^2(L+1)=\epsilon'\sigma^2(1).
\label{eq12}
\end{equation}
In a general case $\epsilon=e^{2\pi i\theta}$ \cite{gl}. However, in this problem, it is sufficient to restrict to periodic ($\theta=0$) and anti-periodic ($\theta=1/2$) boundary conditions, i.e. $\epsilon,\epsilon'\in\{-1,1\}$. Thus application of the Jordan-Wigner transform to a term $\phi(n)^\dag\phi(n+1)$ in the special case $n=L$ yields:
\begin{equation}
\begin{split}
\phi(L)^\dag&\phi(L+1)=\epsilon\phi(L)^\dag\phi(1)\rightarrow\epsilon\sigma^+(L)\left(\prod_{j=1}^{L-1}e^{-i\pi\sigma^+(j)\sigma^-(j)}\right)\sigma^-(1)=\\
&\stackrel{(\ref{eq6})}{=}\epsilon\sigma^+(L)\left(\prod_{j=1}^{L-1}(-\sigma^3(j))\right)\sigma^-(1)=\epsilon\sigma^+(L)(-\sigma^3(L))\left(\prod_{j=1}^{L}(-\sigma^3(j))\right)\sigma^-(1)=\\
&\stackrel{(\ref{eq4})}{=}\epsilon\sigma^+(L)\left(\prod_{j=1}^{L}(-\sigma^3(j))\right)\sigma^-(1)\stackrel{(\ref{eq4})}{=}-\epsilon\sigma^+(L)\sigma^-(1)\left(\prod_{j=1}^{L}(-\sigma^3(j))\right)=\\&=-\frac{\epsilon}{\epsilon'}\sigma^+(L)\sigma^-(L+1)\left(\prod_{j=1}^{L}(-\sigma^3(j))\right).
\end{split}
\label{eq13}
\end{equation}
Define a $p$-particle sector as a subspace in the whole Hilbert space of eigenvectors of the particle number operator to the eigenvalue $p$. Since the eigenvalue of $\prod_{j=1}^{L} (-\sigma^3(j))=\prod_{j=1}^{L}\exp(i\pi(1+\sigma^3(j))/2)=\exp(i\pi N_s)$ when acting on a state from a $p$-particle sector is $e^{i\pi p}$, the above equation reduces to:
\begin{equation}
\phi(L)^\dag\phi(L+1)\rightarrow -\frac{\epsilon}{\epsilon'}e^{i\pi p}\sigma^+(L)\sigma^-(L+1)
\label{eq13b}
\end{equation}
within this sector. With the proper choice of boundary conditions this equation should agree with eq. (\ref{eq9}) for $n=L$. Therefore, one requires:
\begin{equation}
-\frac{\epsilon}{\epsilon'}e^{i\pi p}=1\qquad\Leftrightarrow\qquad\frac{\epsilon'}{\epsilon}=-(-1)^p.
\label{eq14}
\end{equation}
Thus the boundary conditions for fermionic variables and for spin matrices $\sigma^1$, $\sigma^2$ must be taken opposite when the number of fermions $p$ is even, while $\epsilon=\epsilon'$ for odd numbers of fermions.

\subsection{Clifford variables and link operators}
Transformation from Hamiltonian (\ref{eq1}) to (\ref{eq2}) can be also done through the introduction of Clifford variables and link operators. Since this method can be applied also in the case of a 2-dimensional lattice, it will be discussed briefly now. One defines Clifford variables:
\begin{equation}
X(n)=\phi(n)^\dag+\phi(n),\qquad Y(n)=i(\phi(n)^\dag-\phi(n))
\label{eq15}
\end{equation}
and link operators
\begin{equation}
S(n)=iX(n)X(n+1),\qquad \tilde{S}(n)=iY(n)Y(n+1).
\label{eq16}
\end{equation}
As discussed in Section 1.1, they fulfill the algebra:
\begin{equation*}
[S(n),\tilde{S}(m)]=0,
\tag{2.22a}
\label{eq16x}
\end{equation*}
\begin{equation*}
[S(n),S(m)]=0,\ [\tilde{S}(n),\tilde{S}(m)]=0\qquad m\neq n-1,n+1,
\tag{2.22b}
\label{eq16y}
\end{equation*}
\begin{equation*}
\{S(n),S(m)\}=0,\ \{\tilde{S}(n),\tilde{S}(m)\}=0\qquad m= n-1,n+1,
\tag{2.22c}
\label{eq16z}
\end{equation*}
i.e. link operators anticommute when they are two operators of the same kind ($S$ or $\tilde{S}$) and have exactly one common end, while in all other cases they commute. Hamiltonian (\ref{eq1}) can be rewritten in these variables as:
\setcounter{equation}{22}
\begin{equation}
H_f=\frac{1}{2}\sum_{n=1}^L\left(S(n)+\tilde{S}(n)\right).
\label{eq17}
\end{equation}
Thus the alternative method to replace fermionic variables with spin matrices is to find a representation of $S(n)$ and $\tilde{S}(n)$ through $\sigma^1(n)$ and $\sigma^2(n)$ matrices which preserves the algebra of these operators. A choice that satisfies this condition is:
\begin{equation}
S(n)\rightarrow\sigma^1(n)\sigma^2(n+1),\qquad \tilde{S}(n)\rightarrow -\sigma^2(n)\sigma^1(n+1).
\label{eq18}
\end{equation}
After this substitution one again obtains eq. (\ref{eq2}).

\section{Fermions on 2-dimensional lattice}

\begin{figure}
\begin{center}
\includegraphics[width=8cm]{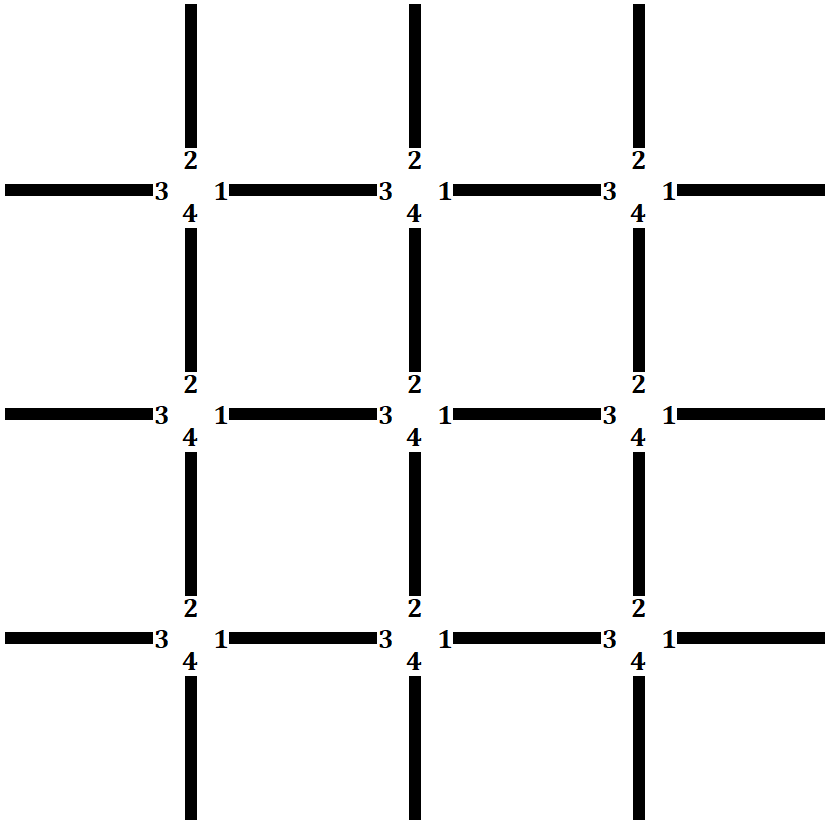}
\end{center}
\caption{The scheme to assign $\Gamma^k$ matrices to the lattice sites. A line whose ending points are marked with indices $i\in\{1,2,3,4\}$ and $j\in\{1,2,3,4\}$ depicts the link operator $S_s(\vec{n},\vec{e})=\Gamma^i(\vec{n})\Gamma^j(\vec{n}+\vec{e})$.}
\label{fig1}
\end{figure}

On a 2-dimensional $L_x\times L_y$ lattice, a natural generalization of the fermionic Hamiltonian (\ref{eq1}) is:
\begin{equation}
H_f=i\sum_{\vec{n},\vec{e}}\left(\phi(\vec{n})^\dag\phi(\vec{n}+\vec{e})-\phi(\vec{n}+\vec{e})^\dag\phi(\vec{n})\right),
\label{eq19}
\end{equation}
where the sum with respect to $\vec{n}$ goes over all the lattice sites, while $\vec{e}\in\{\hat{x},\hat{y}\}$. The lattice size is denoted as $\mathcal{N}=L_xL_y$ and the boundary conditions are:
\begin{equation}
\phi(\vec{n}+L_{e}\hat{e})=\epsilon_e\phi(\vec{n}),\qquad \Gamma^k(\vec{n}+L_e\hat{e})=\epsilon_e^\prime\Gamma^k(\vec{n}),\qquad e\in\{x,y\},
\label{eq19half}
\end{equation}
where, as in $d=1$ case, $\epsilon_e,\epsilon_e^\prime=\pm1$. New Clifford variables and link operators are also a straightforward generalization of eqs. (\ref{eq15}) and (\ref{eq16}):
\begin{equation}
X(\vec{n})=\phi(\vec{n})^\dag+\phi(\vec{n}),\qquad Y(\vec{n})=i(\phi(\vec{n})^\dag-\phi(\vec{n}))
\label{eq20}
\end{equation}
\begin{equation}
S_f(\vec{n},\vec{e})=iX(\vec{n})X(\vec{n}+\vec{e}),\qquad \tilde{S}_f(\vec{n},\vec{e})=iY(\vec{n})Y(\vec{n}+\vec{e}).
\label{eq21}
\end{equation}
They satisfy an algebra similar to (\ref{eq16x})-(\ref{eq16z}):
\begin{equation*}
[S_f(\vec{n},\vec{e}),\tilde{S}_f(\vec{m},\vec{f})]=0,
\tag{2.29a}
\label{eq21x}
\end{equation*}
\begin{equation*}
[S_f(\vec{n},\vec{e}),S_f(\vec{m},\vec{f})]=0,\ [\tilde{S}_f(\vec{n},\vec{e}),\tilde{S}_f(\vec{m},\vec{f})]=0\qquad \text{for links with no common end},
\tag{2.29b}
\label{eq21y}
\end{equation*}
\begin{equation*}
\{S_f(\vec{n},\vec{e}),S_f(\vec{m},\vec{f})\}=0,\ \{\tilde{S}_f(\vec{n},\vec{e}),\tilde{S}_f(\vec{m},\vec{f})\}=0\quad \text{for links with one common end}.
\tag{2.29c}
\label{eq21z}
\end{equation*}
With this choice of variables Hamiltonian (\ref{eq19}) can be written in a way analogous to eq. (\ref{eq17}):
\setcounter{equation}{29}
\begin{equation}
H_f=\frac{1}{2}\sum_{\vec{n},\vec{e}}\left(S_f(\vec{n},\vec{e})+\tilde{S}_f(\vec{n},\vec{e})\right)=\frac{1}{2}\sum_l\left(S_f(l)+\tilde{S}_f(l)\right),
\label{eq22}
\end{equation}
where $l=(\vec{n},\vec{e})$ are all links between neighboring sites. While in $d=1$ dimension Pauli matrices were used for a spin representation of the fermionic operators, in $d=2$ dimensions this is done through Euclidean Dirac matrices. A choice which preserves the algebra (\ref{eq21x})-(\ref{eq21z}) is:
\begin{equation}
\begin{split}
&S_s(\vec{n},\hat{x})=\Gamma^1(\vec{n})\Gamma^3(\vec{n}+\hat{x}),\qquad S_s(\vec{n},\hat{y})=\Gamma^2(\vec{n})\Gamma^4(\vec{n}+\hat{y}),\\
&\tilde{S}_s(\vec{n},\hat{x})=\tilde{\Gamma}^1(\vec{n})\tilde{\Gamma}^3(\vec{n}+\hat{x}),\qquad \tilde{S}_s(\vec{n},\hat{y})=\tilde{\Gamma}^2(\vec{n})\tilde{\Gamma}^4(\vec{n}+\hat{y}),
\end{split}
\label{eq23}
\end{equation}
where $\tilde{\Gamma}^k=i\prod_{j\neq k}\Gamma^j$. Fig. \ref{fig1} depicts how $\Gamma^k$ matrices are assigned to the lattice sites through this representation. Equivalent Hamiltonian in spin variables is thus:
\begin{equation}
H_s=\frac{1}{2}\sum_l\left(S_s(l)+\tilde{S}_s(l)\right).
\label{eq24}
\end{equation}
Similarly to eq. (\ref{eq8c}), one also obtains the spin equivalent of the fermionic particle number operator $N_f=\sum_{\vec{n}}\phi(\vec{n})^\dag\phi(\vec{n})$:
\begin{equation}
N_s=\sum_{\vec{n}}\frac{1+\Gamma^5(\vec{n})}{2}
\label{april1}
\end{equation}
or, in terms of the particle density $N_s(\vec{n})$ at a site $\vec{n}$:
\begin{equation}
N_s(\vec{n})=\frac{1+\Gamma^5(\vec{n})}{2}.
\label{april2}
\end{equation}
\section{Constraints}
The above procedure allows for calculations based on $\Gamma$ matrices instead of calculations with fermionic operators, which is the goal of this work. Yet this transformation introduces many redundant degrees of freedom. Indeed, fermionic Hilbert space has dimension $2^\mathcal{N}$ -- since there can be a fermion or a hole at every lattice site, while the dimension of spin matrices, like (\ref{eq23}) or (\ref{eq24}), is $4^\mathcal{N}$ -- since they are tensor products of $4\times4$ Dirac matrices over all the lattice sites. Thus one expects some constraints should exist which, when applied, reduce the dimension of the spin Hilbert space to the size of its fermionic counterpart. It turns out fermionic variables satisfy certain identities, while spin variables have more freedom. To make both descriptions equivalent one requires that these identities are fulfilled also by the spin matrices and these additional requirements provide necessary constraints.\\
One defines plaquette operators as:
\begin{equation}
P_f(\vec{n})=\prod_{l\in C(\vec{n})}S_f(l),\qquad P_s(\vec{n})=\prod_{l\in C(\vec{n})}S_s(l),
\label{eq25}
\end{equation}
where the fermionic and spin plaquette operators $P_f(\vec{n})$, $P_s(\vec{n})$, and a unit plaquette $C(\vec{n})$ are labeled by the position $\vec{n}$ of the lower-left corner of the plaquette.\\
Using equations (\ref{eq21}), (\ref{eq23}) and (\ref{eq25}), it is easy to see that fermionic plaquettes are equal to the identity operator:
\begin{equation}
P_f(\vec{n})=i^4X(\vec{n})X(\vec{n}+\hat{x})^2X(\vec{n}+\hat{x}+\hat{y})^2X(\vec{n}+\hat{y})^2X(\vec{n})=1,
\label{eq26}
\end{equation}
while spin plaquettes are non-trivial:
\begin{equation}
\begin{split}
P_s(\vec{n})=\Gamma^1(\vec{n})\Gamma^3(\vec{n}+\hat{x})\Gamma^2(\vec{n}+\hat{x})\Gamma^4(\vec{n}+\hat{x}&+\hat{y})\Gamma^3(\vec{n}+\hat{x}+\hat{y})\Gamma^1(\vec{n}+\hat{y})\Gamma^4(\vec{n}+\hat{y})\Gamma^2(\vec{n})=\\
&=\Gamma^{12}(\vec{n})\Gamma^{32}(\vec{n}+\hat{x})\Gamma^{43}(\vec{n}+\hat{x}+\hat{y})\Gamma^{14}(\vec{n}+\hat{y}),
\end{split}
\label{eq27}
\end{equation}
where the fact that $X^2=(\phi^\dag)^2+\phi^2+\{\phi^\dag,\phi\}=1$ is applied and a definition
\begin{equation}
\Gamma^{ij}\equiv\Gamma^i\Gamma^j
\label{eq27b}
\end{equation}
is introduced. Since $\{\Gamma^j,\Gamma^k\}=2\delta^{jk}\mathds{1}_4$, $(\Gamma^{jk})^2=-\mathds{1}_4$ and thus:
\begin{equation}
P_s(\vec{n})^2=1
\label{eq30}
\end{equation}
or $P_s(\vec{n})=\pm 1$. This, when compared to eq. (\ref{eq26}), shows that the redundant degrees of freedom can be removed by projection on the subspace of the spin Hilbert space where $P_s=1$. This is done through the projection operators:
\begin{equation}
\Sigma_{n_x,n_y}=\frac{1}{2}\left(1+P_s(\vec{n})\right)
\label{eq31}
\end{equation}
associated with plaquettes. A simple check shows that these operators are indeed projectors:
\begin{equation}
\Sigma^2_{n_x,n_y}=\frac{1}{4}\left(1+P_s(\vec{n})\right)^2=\frac{1}{4}\left(1+2P_s(\vec{n})+P_s(\vec{n})^2\right)=\frac{1}{4}\left(1+2P_s(\vec{n})+1\right)=\Sigma_{n_x,n_y}.
\label{eq31s0}
\end{equation}
The plaquettes commute with each other:
\begin{equation}
[P_s(\vec{m}),P_s(\vec{n})]=0.
\label{eq31s1}
\end{equation}
To prove this, one needs to consider three cases: a pair of disjoint plaquettes, plaquettes sharing a single vertex and plaquettes with a common side. Disjoint plaquettes commute trivially. When they share a vertex, assume without loss of generality that the common vertex is a bottom left/top right corner. Hence:
\begin{multline}
P_s(\vec{n})P_s(\vec{n}+\hat{x}+\hat{y})=\\
=\Gamma^{12}(\vec{n})\Gamma^{32}(\vec{n}+\hat{x})\Gamma^{43}(\vec{n}+\hat{x}+\hat{y})\Gamma^{14}(\vec{n}+\hat{y})\Gamma^{12}(\vec{n}+\hat{x}+\hat{y})\Gamma^{32}(\vec{n}+2\hat{x}+\hat{y})\Gamma^{43}(\vec{n}+2\hat{x}+2\hat{y})\Gamma^{14}(\vec{n}+\hat{x}+2\hat{y})\\
=\Gamma^{12}(\vec{n}+\hat{x}+\hat{y})\Gamma^{32}(\vec{n}+2\hat{x}+\hat{y})\Gamma^{43}(\vec{n}+2\hat{x}+2\hat{y})\Gamma^{14}(\vec{n}+\hat{x}+2\hat{y})\Gamma^{12}(\vec{n})\Gamma^{32}(\vec{n}+\hat{x})\Gamma^{43}(\vec{n}+\hat{x}+\hat{y})\Gamma^{14}(\vec{n}+\hat{y})\\
=P_s(\vec{n}+\hat{x}+\hat{y})P_s(\vec{n}).
\label{eq31s2}
\end{multline}
In the above expression, all matrices, except the Dirac matrices associated with the site $\vec{n}+\hat{x}+\hat{y}$, are from different vertices of the lattice, so they commute. For this common vertex, one easily checks that $\Gamma^{43}\Gamma^{12}=\Gamma^{12}\Gamma^{43}$. In the last case, for a common side, assume that this is the left/right edge of the neighboring plaquettes. Then:
\begin{multline}
P_s(\vec{n})P_s(\vec{n}+\hat{x})=\\
=\Gamma^{12}(\vec{n})\Gamma^{32}(\vec{n}+\hat{x})\Gamma^{43}(\vec{n}+\hat{x}+\hat{y})\Gamma^{14}(\vec{n}+\hat{y})\Gamma^{12}(\vec{n}+\hat{x})\Gamma^{32}(\vec{n}+2\hat{x})\Gamma^{43}(\vec{n}+2\hat{x}+\hat{y})\Gamma^{14}(\vec{n}+\hat{x}+\hat{y})\\
=\Gamma^{12}(\vec{n}+\hat{x})\Gamma^{32}(\vec{n}+2\hat{x})\Gamma^{43}(\vec{n}+2\hat{x}+\hat{y})\Gamma^{14}(\vec{n}+\hat{x}+\hat{y})\Gamma^{12}(\vec{n})\Gamma^{32}(\vec{n}+\hat{x})\Gamma^{43}(\vec{n}+\hat{x}+\hat{y})\Gamma^{14}(\vec{n}+\hat{y})\\
=P_s(\vec{n}+\hat{x})P_s(\vec{n}),
\label{eq31s3}
\end{multline}
where the reasoning is similar as in the previous case: only matrices from the common vertices $\vec{n}+\hat{x}$ and $\vec{n}+\hat{x}+\hat{y}$ may produce a non-trivial factor, but $\Gamma^{32}\Gamma^{12}=-\Gamma^{12}\Gamma^{32}$ and $\Gamma^{43}\Gamma^{14}=-\Gamma^{14}\Gamma^{43}$, so the overall factor is $+1$. As a straightforward consequence of eq. (\ref{eq31s1}) the projection operators also commute:
\begin{equation}
[\Sigma_{n_x,n_y},\Sigma_{m_x,m_y}]=0.
\label{eq31s4}
\end{equation}
Products of those projection operators are then projection operators too:
\begin{equation}
(\Sigma_{m_x,m_y}\Sigma_{n_x,n_y})^2=\Sigma_{m_x,m_y}\Sigma_{n_x,n_y}\Sigma_{m_x,m_y}\Sigma_{n_x,n_y}\stackrel{(\ref{eq31s4})}{=}\Sigma_{m_x,m_y}^2\Sigma_{n_x,n_y}^2\stackrel{(\ref{eq31s0})}{=}\Sigma_{m_x,m_y}\Sigma_{n_x,n_y},
\label{eq31s5}
\end{equation}
which has a simple generalization to any number of projection operators in the product.

When a fixed basis is needed, the following representation of gamma matrices is used:
\begin{equation}
\begin{split}
&\Gamma^1=\begin{pmatrix}
 0 & 0 & -1 & 0 \\
 0 & 0 & 0 & 1 \\
 -1 & 0 & 0 & 0 \\
 0 & 1 & 0 & 0 \\
\end{pmatrix},\quad \Gamma^2=\begin{pmatrix}
 0 & 1 & 0 & 0 \\
 1 & 0 & 0 & 0 \\
 0 & 0 & 0 & 1 \\
 0 & 0 & 1 & 0 \\
\end{pmatrix},\quad \Gamma^3=\begin{pmatrix}
 0 & -i & 0 & 0 \\
 i & 0 & 0 & 0 \\
 0 & 0 & 0 & -i \\
 0 & 0 & i & 0 \\
\end{pmatrix},\\
&\Gamma^4=\begin{pmatrix}
 0 & 0 & -i & 0 \\
 0 & 0 & 0 & i \\
 i & 0 & 0 & 0 \\
 0 & -i & 0 & 0 \\
\end{pmatrix},\quad
\Gamma^5=\begin{pmatrix}
 1 & 0 & 0 & 0 \\
 0 & -1 & 0 & 0 \\
 0 & 0 & -1 & 0 \\
 0 & 0 & 0 & 1 \\
\end{pmatrix}.
\end{split}
\label{eq28}
\end{equation}
Let us denote the basis in which those matrices are expressed by $\{e_1,e_2,e_3,e_4\}$, i.e. $e_i$ are eigenvectors of $\Gamma^5$ and $\Gamma^5e_i=e_i$ for $i\in\{1,4\}$, while $\Gamma^5e_i=-e_i$ for $i\in\{2,3\}$. The explicit forms of $\Gamma^{ij}$ matrices which occur in eq. (\ref{eq27}) are in this basis:
\begin{equation}
\begin{split}
\Gamma^{12}=\begin{pmatrix}
 0 & 0 & 0 & -1 \\
 0 & 0 & 1 & 0 \\
 0 & -1 & 0 & 0 \\
 1 & 0 & 0 & 0 \\
\end{pmatrix},\quad
&\Gamma^{32}=\begin{pmatrix}
 -i & 0 & 0 & 0 \\
 0 & i & 0 & 0 \\
 0 & 0 & -i & 0 \\
 0 & 0 & 0 & i \\
\end{pmatrix},\quad
\Gamma^{43}=\begin{pmatrix}
 0 & 0 & 0 & -1 \\
 0 & 0 & -1 & 0 \\
 0 & 1 & 0 & 0 \\
 1 & 0 & 0 & 0 \\
\end{pmatrix},\\
&\Gamma^{14}=\begin{pmatrix}
 -i & 0 & 0 & 0 \\
 0 & -i & 0 & 0 \\
 0 & 0 & i & 0 \\
 0 & 0 & 0 & i \\
\end{pmatrix}.
\end{split}
\label{eq29}
\end{equation}
Every plaquette projection operator provides reduction of the dimension of the Hilbert space by a factor two, thus it seems that $\mathcal{N}$ plaquettes associated with all vertices of the lattice provide exactly the amount of constraints needed to reduce the space from dimension $4^\mathcal{N}$ to $2^\mathcal{N}$. However, not all of these projectors are independent and two additional operators are necessary:
\begin{equation}
\Sigma_x=\frac{1}{2}\left(1+(-i)^{L_x}\epsilon_x\mathcal{L}_x\right),\qquad \Sigma_y=\frac{1}{2}\left(1+(-i)^{L_y}\epsilon_y\mathcal{L}_y\right),
\label{eq32}
\end{equation}
where $\mathcal{L}_x$ and $\mathcal{L}_y$ are "Polyakov line" operators:
\begin{equation}
\mathcal{L}_x(n_y)=\prod_{n_x=1}^{L_x}S_s(\vec{n},\hat{x}),\qquad \mathcal{L}_y(n_x)=\prod_{n_y=1}^{L_y}S_s(\vec{n},\hat{y}).
\label{eq33}
\end{equation}
The explicit form of these operators reads:
\begin{equation}
\begin{split}
\mathcal{L}_x(n_y)=\epsilon_x^\prime\Gamma^1(1,n_y)\Gamma^3(2,n_y)\Gamma^1(2,n&_y)\Gamma^3(3,n_y)\Gamma^1(3,n_y)...\Gamma^1(L_x,n_y)\Gamma^3(1,n_y)=\\
&=\epsilon_x^\prime\Gamma^{13}(1,n_y)\Gamma^{31}(2,n_y)\Gamma^{31}(3,n_y)...\Gamma^{31}(L_x,n_y),
\end{split}
\label{extra1}
\end{equation}
\begin{equation}
\mathcal{L}_y(n_x)=\epsilon_y^\prime\Gamma^{24}(n_x,1)\Gamma^{42}(n_x,2)\Gamma^{42}(n_x,3)...\Gamma^{42}(n_x,L_y),
\label{extra2}
\end{equation}
where
\begin{equation}
\Gamma^{31}=-\Gamma^{13}=\begin{pmatrix}
 0 & 0 & 0 & -i \\
 0 & 0 & -i & 0 \\
 0 & -i & 0 & 0 \\
 -i & 0 & 0 & 0 \\
\end{pmatrix},\qquad
\Gamma^{42}=-\Gamma^{24}=\begin{pmatrix}
 0 & 0 & 0 & -i \\
 0 & 0 & i & 0 \\
 0 & i & 0 & 0 \\
 -i & 0 & 0 & 0 \\
\end{pmatrix}.
\label{extra3}
\end{equation}

To see that the form of the additional projection operators (\ref{eq32}) is correct, one needs to verify that the Polyakov line operators in the fermionic picture:
\begin{equation}
\mathcal{L}_x^f(n_y)=\prod_{n_x=1}^{L_x}S_f(\vec{n},\hat{x}),\qquad \mathcal{L}_y^f(n_x)=\prod_{n_y=1}^{L_y}S_f(\vec{n},\hat{y})
\label{wombat1}
\end{equation} 
equal $i^{L_x}\epsilon_x$ and $i^{L_y}\epsilon_y$ respectively. From eqs. (\ref{eq19half})--(\ref{eq21}) one obtains:
\begin{equation}
\begin{split}
\mathcal{L}_x^f(n_y)=\prod_{n_x=1}^{L_x}S_f(\vec{n},\hat{x})&=\prod_{n_x=1}^{L_x}iX(\vec{n})X(\vec{n}+\hat{x})=\\
&=\left(\prod_{n_x=1}^{L_x-1}iX(\vec{n})X(\vec{n}+\hat{x})\right)iX(L_x,n_y)X(L_x+1,n_y)=\\
&=\left(\prod_{n_x=1}^{L_x-1}iX(\vec{n})X(\vec{n}+\hat{x})\right)iX(L_x,n_y)\epsilon_x X(1,n_y).
\end{split}
\label{wombat2}
\end{equation}
Since $X(\vec{n})^2=1$ and the pairs of $X(n_x,n_y)$ operators associated with the same lattice site are neighbors in the above product for $2\leq n_x\leq L_x$, the product of all the $X$ operators is the unity operator and only the numerical factor remains. Thus, indeed:
\begin{equation}
\mathcal{L}_x^f(n_y)=i^{L_x}\epsilon_x.
\label{wombat3}
\end{equation}
A similar reasoning shows that:
\begin{equation}
\mathcal{L}_y^f(n_x)=\prod_{n_y=1}^{L_y}S_f(\vec{n},\hat{y})=\prod_{n_y=1}^{L_y}iX(\vec{n})X(\vec{n}+\hat{y})=i^{L_y}\epsilon_y.
\label{wombat4}
\end{equation}
Evidently, projection on the subspace where the spin representation Polyakov lines $\mathcal{L}_x$ and $\mathcal{L}_y$ have these values is performed with the operators (\ref{eq32}).

Among the $L_x+L_y$ Polyakov lines, only one $\mathcal{L}_x$ and one $\mathcal{L}_y$ are independent, because all the other can be obtained by multiplication of a parallel "Polyakov line" by appropriate plaquettes. Thus, there are $\mathcal{N}+2$ projection operators, eqs. (\ref{eq31}) and (\ref{eq32}), needed to constrain $4^\mathcal{N}$-dimensional Hilbert space to $2^\mathcal{N}$ dimensions. More precise predictions on relations of these projectors are \cite{brww}:
\begin{itemize}
\item The two Polyakov lines $\mathcal{L}_x$ and $\mathcal{L}_y$ are independent when both $L_x$ and $L_y$ are even. When either $L_x$ or $L_y$ is odd, those two operators are related.
\item When both $L_x$ and $L_y$ are even, only $\mathcal{N}-2$ plaquette operators are independent. Otherwise, there are $\mathcal{N}-1$ independent plaquette operators.
\item A solution of the constraints exists only when:
\begin{equation}
(-1)^p=\eta^{L_xL_y}\left(-\frac{\epsilon_y'}{\epsilon_y}\right)^{L_x}\left(-\frac{\epsilon_x'}{\epsilon_x}\right)^{L_y},
\label{eq34}
\end{equation}
where $p$ is the number of fermions on the lattice, $\epsilon_{\hat{e}}'$ and $\epsilon_{\hat{e}}$ are boundary condition coefficients in direction $\hat{e}\in\{x,y\}$ in spin representation and for fermionic variables, respectively, and $\eta=-1$ in our case.
Value $\eta=-1$ is related to the fact that vectors from the subspace spanned by $e_1$ and $e_4$ are interpreted as states with a fermion at a site of the lattice, while $e_2$ and $e_3$ correspond to an empty site. If the interpretation of these subspaces were inverse or, speaking more physically, when the roles of particles and holes are swapped, then $\eta=1$. When the freedom of choosing either $\eta=-1$ or $\eta=1$ is taken into consideration, the formula (\ref{april2}) generalizes to:
\begin{equation}
N_s(\vec{n})=\frac{1-\eta\Gamma^5(\vec{n})}{2}.
\label{april3}
\end{equation}
In particular, the condition (\ref{eq34}) specifies which $p$-particle sectors are "good", i.e. their Hilbert spaces are not reduced to zero by application of all the projectors.
\end{itemize}

Derivations of these relations are presented in \emph{Appendix C}.

\section{Interaction with external magnetic field}
In the case of free fermions on a 2-dimensional lattice the constraints necessary to reduce the size of the Hilbert space are $P_s(\vec{n})=1$, in accordance with eq. (\ref{eq26}). However, other possible choices of the constraints $P_s(\vec{n})=\pm 1$ provide a much wider class of models to study. This suggests interesting questions what is the physical interpretation of those models and how observables, e.g. eigenvalues, are affected by the change of the constraints. The description of the system through fermionic operators is extended to allow for both $+1$ and $-1$ plaquette signs by introduction of an additional $\mathbb{Z}_2$ field $U$. One assigns variables $U(\vec{n},\vec{n}+\hat{e})\in\{-1,1\}$ to links between neighboring lattice sites, so that the sign of every plaquette $P_f(\vec{n})$ changes by $\prod_{l\in C(\vec{n})}U(l)$. Thus a general form of Hamiltonians which describe such models is:
\begin{equation}
H_f=i\sum_{\vec{n},\vec{e}}\left(U(\vec{n},\vec{n}+\vec{e})\phi(\vec{n})^\dag\phi(\vec{n}+\vec{e})-U(\vec{n},\vec{n}+\vec{e})\phi(\vec{n}+\vec{e})^\dag\phi(\vec{n})\right).
\label{s01}
\end{equation}
Every self-consistent set of constraints on the plaquette operators can be accompanied by a corresponding choice of variables $U$. In the fermionic space the interaction with an external field is introduced through a $\mathbb{Z}_2$ field $U$ assigned to the links, while in the spin space it is understood as a $\mathbb{Z}_2$ magnetic field which takes values $P_s(\vec{n})\in\{-1,1\}$ over plaquettes located at $\vec{n}$. A special case defined through conditions:
\begin{equation}
U((x,y),(x+1,y))\equiv U_x(x,y)=(-1)^y,\qquad U((x,y),(x,y+1))\equiv U_y(x,y)=1
\label{s02}
\end{equation}
is paid particular attention due to the simplicity of its interpretation and the fact that eigenvalues of the corresponding Hamiltonian can be found analytically. Since the sign of field $U$ at links directed along $x$-coordinate alternates with $y$, $U_x=(-1)^y$, this construction is valid only for even $L_y$ and only such lattices are studied in this case. This choice of field $U_{\hat{e}}(x,y)$ inverts all the plaquette operator constraints $P(x,y)=+1$ to $P(x,y)=-1$, it is therefore interpreted as a constant magnetic field. More generally, the models with the external magnetic field are described by the set of constraints:
\begin{equation}
P_s(\vec{n})=B(\vec{n}),
\label{may2}
\end{equation}
where $B(\vec{n})$ is the magnetic field on the lattice face whose lower-left corner is located at $\vec{n}$. Substitution of the values (\ref{s02}) of field $U$ in eq. (\ref{s01}) yields the Hamiltonian of fermions in a constant magnetic field:
\begin{multline}
H_f=H_++H_-=\\
=i\sum_{\vec{n}}\left(\phi(\vec{n})^\dag\phi(\vec{n}+\hat{y})-\phi(\vec{n}+\hat{y})^\dag\phi(\vec{n})\right)+i\sum_{\vec{n}}(-1)^{n_y}\left(\phi(\vec{n})^\dag\phi(\vec{n}+\hat{x})-\phi(\vec{n}+\hat{x})^\dag\phi(\vec{n})\right).
\label{s1}
\end{multline}
Its first component, which is the sum of link operators along $y$-direction, is fully analogous to the free Hamiltonian (\ref{eq19}), while the latter part differs only by the factor $(-1)^{n_y}$. This simple form allows for analytic diagonalization, which is covered in Section 4.
\clearpage

\chapter{Construction of the Hamiltonian and constraints in the spin representation}
\section{Construction of the basis}

We will call a subspace of the entire Hilbert space with a fixed number $p$ of spin-particles a $p$-particle \emph{sector}, while a subspace of a $p$-particle sector which includes all states with the same spin-particles/holes positions will be called its \emph{subsector}. An important fact, which greatly simplifies our calculations, is that subsectors are invariant subspaces of the projection operators. This is easy to see from explicit forms of matrices $\Gamma^{ij}$, since their action on the subspace spanned by $e_1$ and $e_4$ (particle on site) is invariant and the same holds for the subspace spanned by $e_2$ and $e_3$ (no particle). Similarly, sectors are invariant subspaces of the Hamiltonian -- it causes hopping, but does not create nor annihilate particles. These facts suggest that a convenient basis should have sector/subsector structure. First, it is enough to construct separate bases for $p$-particle sectors and perform the calculations within those sectors. Second, basis vectors will be arranged primarily by their subsectors.\\
The Hilbert space is a tensor product of $\mathcal{N}$ 4-dimensional vector spaces of square matrices $4\times 4$. A basis vector $v_s$ of this space can be therefore represented by a $L_x\times L_y$ matrix $s$:
\begin{equation}
v_s=\bigotimes_{1\leq i\leq L_x,\ 1\leq j\leq L_y} e_{s_{ij}}\quad \leftrightarrow\quad s=(s_{ij}).
\label{eq38}
\end{equation}
An element of this matrix, $s_{ij}\in\{1,2,3,4\}$ is a label of an eigenvector of $\Gamma^5$ which contributes to the above tensor product from the lattice site $(i,j)$.

\emph{Example:}
All matrices
\begin{equation}
\begin{pmatrix}
1&2&3\\
2&2&2\\
3&3&4\\
2&2&2
\end{pmatrix},\qquad
\begin{pmatrix}
4&3&3\\
3&3&3\\
3&3&1\\
3&3&2
\end{pmatrix},\qquad
\begin{pmatrix}
4&2&2\\
2&2&3\\
3&3&4\\
2&3&2
\end{pmatrix}
\label{eq39}
\end{equation}
represent basis vectors of the Hilbert space associated with a lattice $4\times3$, which belong to the 2-particle sector and to the same subsector. The particles are located at sites $(1,1)$ and $(3,3)$.
\vspace{0.5cm}\newline
The code used to implement the spin representation basis of a $p$-particle sector for a $3\times 3$ lattice in Wolfram Mathematica is included in \emph{Appendix B}. This construction requires in particular (see lines 5--35 in \emph{Appendix B}):
\begin{itemize}
\item Building a list \texttt{stp[[]]} of all $p$-particle states, i.e. of all $L_x\times L_y$ matrices with $(\mathcal{N}-p)$ elements from $a=\{2,3\}$ and $p$ elements from $b=\{1,4\}$. This is done with a \texttt{Do[]} loop nested $(\mathcal{N}+1)$ times (lines 19--34). The loop runs over set $a$ when filling matrix elements corresponding to the empty lattice sites and over set $b$ at the occupied sites. The last, external \texttt{Do[]} loop (lines 18--35) runs over \texttt{pocc}$\in\{1,2,...,\binom{\mathcal{N}}{p}\}$ -- the index which numbers the subsectors.
\item Development of an inverse function, which assigns indices to basis vectors. For this purpose one defines a list of positions at which the basis matrices occur in \texttt{stp[[]]}: 
\begin{lstlisting}[frame=single, numbers=left, numberstyle=\scriptsize,breaklines=true,basicstyle=\ttfamily,firstnumber=27]
JJ[[l[1],l[2],l[3],l[4],l[5],l[6],l[7],l[8],l[9],pocc]]=ii;
\end{lstlisting}
where the basis matrix is encoded with the sequence of numbers \texttt{l[i]} and the subsector number \texttt{pocc}. $\text{\texttt{l[i]}}\in\{1,2\}$ specifies which, the first or the second, element of the sets $a$ or $b$ occurs at position $i$ in this basis matrix. This list is constructed parallel to $\texttt{stp[[]]}$, within the same nested \texttt{Do[]} loop.
\end{itemize}

\section{Construction of the projection operators}

As already mentioned, subsectors are invariant subspaces of sectors under the action of the projection operators associated with plaquettes and "Polyakov lines". Thus, one can study constraints within single subsectors.
Once one restricts to a subsector, every element $s_{ij}$ of a basis vector matrix has only two possible values. Either $s_{ij}\in\{1,4\}$ at lattice sites $(i,j)$ with a particle or $s_{ij}\in\{2,3\}$ at empty ones. Therefore the dimension of a subsector is $2^\mathcal{N}$. In the whole sector one takes into account also all possible locations of particles on the lattice, so its dimension is $2^\mathcal{N}\binom{\mathcal{N}}{p}$. Hence, calculations limited to subsectors significantly reduce dimension of constraint matrices from $2^\mathcal{N}\binom{\mathcal{N}}{p}\times2^\mathcal{N}\binom{\mathcal{N}}{p}$ to $2^\mathcal{N}\times2^\mathcal{N}$, which, in particular when $p$ is close to $\mathcal{N}/2$, is a serious advantage.\\
In Wolfram Mathematica, one crops \texttt{stp[[]]} to its part corresponding to the $r$-th subsector:
\begin{lstlisting}[frame=single, numbers=left, numberstyle=\scriptsize,breaklines=true,basicstyle=\ttfamily,firstnumber=51]
stR = Table[stp[[ii]], {ii, ns0*(r - 1) + 1, ns0*r}];
\end{lstlisting}
to perform this reduction. Another simplification originates from the fact that although we deal with huge matrices, most of their elements are zero. This suggests application of sparse matrices, which are lists of coordinates at which non-zero elements appear paired with the values of those non-zero entries. Wolfram Mathematica enables employment of sparse matrices through \texttt{SparseArray[]} function and has built-in tools for algebraic operations on such objects.\\
As can be seen from eqs. (\ref{eq27}), (\ref{extra1}) and (\ref{extra2}), matrices $\Gamma^{ij}$ are building blocks of constraints operators. These matrices, however, have only one non-zero entry per row/column. This can be exploited to further improve efficiency of the numerical calculations. First, one encodes those matrices with two lists, e.g. for $\Gamma^{12}$:
\begin{lstlisting}[frame=single, numbers=left, numberstyle=\scriptsize,breaklines=true,basicstyle=\ttfamily,firstnumber=55]
   JG12 = {4, 3, 2, 1}; MG12 = {1, -1, 1, -1};
\end{lstlisting}
List \texttt{JG12[[]]} provides information on positions of non-zero entries, while their values are listed in \texttt{MG12[[]]}. \texttt{JG12[[i]]} is the row number of the only non-zero element in the i-th column of $\Gamma^{12}$ and \texttt{MG12[[i]]} is the value of this entry. Then, a sparse array corresponding to a plaquette operator, say
\begin{lstlisting}[frame=single, numbers=left, numberstyle=\scriptsize,breaklines=true,basicstyle=\ttfamily,firstnumber=60]
   PMS11 = SparseArray[{{1, 1} -> 0}, ns0];
\end{lstlisting}
for $P_s(1,1)$, is filled with its elements \texttt{PMS11[[jj,ii]]}$=\braket{jj|P_s(1,1)|ii}$ within a \texttt{Do[]} loop (\emph{Appendix B}, lines 72--84), which runs over all states $\ket{ii}$ of the subsector:
\begin{lstlisting}[frame=single, numbers=left, numberstyle=\scriptsize,breaklines=true,basicstyle=\ttfamily,firstnumber=72]
Do[{  st = stR[[ii]];

     i[1, 3]=st[[1, 3]]; i[2, 3]=st[[2, 3]]; i[3, 3]=st[[3, 3]];
     i[1, 2]=st[[1, 2]]; i[2, 2]=st[[2, 2]]; i[3, 2]=st[[3, 2]];
     i[1, 1]=st[[1, 1]]; i[2, 1]=st[[2, 1]]; i[3, 1]=st[[3, 1]];
\end{lstlisting}
Keeping in mind relation (\ref{eq38}) between basis vectors of the Hilbert space understood as tensor products of $e_i$'s and their matrix representations, the numbers \texttt{i[k,l]}$\in\{1,2,3,4\}$ specify which eigenvector of $\Gamma^5$ contributes to \texttt{ii}-th basis vector from site $(k,l)$, i.e. part of $\ket{ii}$ associated with site $(k,l)$ is $e_\text{\texttt{i[k,l]}}$. To calculate a matrix element of:
\begin{equation}
P_s(1,1)=\Gamma^{12}(1,1)\Gamma^{32}(2,1)\Gamma^{43}(2,2)\Gamma^{14}(1,2)
\label{eq40}
\end{equation}
let us focus for a moment on a single site, say the one whose coordinates are $(1,1)$. When $\Gamma^{12}$ acts on $e_{\text{\texttt{i[1,1]}}}$, the result is the \texttt{i[1,1]}-th column of $\Gamma^{12}$. Yet every column of this matrix has just one non-zero entry -- the non-zero element of this column has index \texttt{JG12[[i[1,1]]]} and value \texttt{MG12[[i[1,1]]]}. Writing \texttt{j[k,l]} for the set of numbers describing $\bra{jj}$, it follows that \texttt{j[1,1]=JG12[[i[1,1]]]} is the only choice of this index matching $\ket{ii}$, i.e. such that the resulting matrix element is non-zero. Contribution to the matrix element \texttt{PMS11[[jj,ii]]} from this site is \texttt{MG12[[i[1,1]]]}. The same reasoning holds for the other sites, e.g. \texttt{j[1,2]=JG14[[i[1,2]]]} and the multiplicative contribution to the matrix element is \texttt{MG14[[i[1,2]]]}, and so forth. Thus there is only one $\bra{jj}$ corresponding to $\ket{ii}$ which yields a non-zero matrix entry and the above procedure enables finding this $\bra{jj}$ and the related matrix entry. Remaining part of the program performed within this \texttt{Do[]} loop is therefore:
\begin{lstlisting}[frame=single, numbers=left, numberstyle=\scriptsize,breaklines=true,basicstyle=\ttfamily,firstnumber=77]
j[1,3]=JId[[i[1,3]]];j[2,3]= JId[[i[2,3]]];j[3,3]=JId[[i[3,3]]];  
j[1,2]=JG14[[i[1,2]]];j[2,2]=JG43[[i[2,2]]];j[3,2]=JId[[i[3,2]]];  
j[1,1]=JG12[[i[1,1]]];j[2,1]=JG32[[i[2,1]]];j[3,1]=JId[[i[3,1]]];        
     tt = Table[j[k1, k2], {k1, 1, L}, {k2, 1, L}];
     jj = FromState[tt][[2]] - (r - 1)*ns0;          
     PMS11[[jj, ii]] = 
     MG12[[i[1,1]]] MG32[[i[2,1]]] MG43[[i[2,2]]] MG14[[i[1,2]]];
\end{lstlisting}
where \texttt{FromState[]} is a function which, when given a basis matrix \texttt{tt}, returns a two element list $\{$\texttt{listpocc,kk}$\}$ (see lines 37--48). \texttt{listpocc[[]]} is the list of indices of occupied sites of the lattice and \texttt{kk} is the index of the basis state \texttt{tt}, i.e. its position in \texttt{stp[[]]}. Since these calculations are done within a subsector ($1\leq$\texttt{ ii,jj }$\leq$\texttt{ns0}), \texttt{(r-1)*ns0} is subtracted from the index of state \texttt{tt} to obtain its index within the subsector.\\
Typically to obtain all matrix elements $\braket{jj|P_s|ii}$ of a $2^\mathcal{N}$-dimensional operator between basis states, a separate calculation for all $2^{2\mathcal{N}}$ elements is required. The main advantage of the above algorithm is that it reduces this number to $2^\mathcal{N}$ calculations. This is possible, because the loop runs over $\ket{ii}$'s, but it does not run over $\bra{jj}$'s. Instead the algorithm is designed to find the only matching $\bra{jj}$ for every $\ket{ii}$.\\

Construction of the "Polyakov line" operators is analogous. In particular, matrices $\Gamma^{31}$ and $\Gamma^{42}$ also have one non-zero element per row or column, so identical simplification is possible. The projection operators are then constructed from the plaquette operators and the "Polyakov lines". In "good" sectors, it is expected that the product of all the projectors:
\begin{equation}
\left(\prod_{1\leq n_x\leq L_x,\\ 1\leq n_y\leq L_y}\Sigma_{n_x,n_y}\right)\Sigma_x\Sigma_y
\label{eq41}
\end{equation}
reduces the dimension of subsectors from $2^\mathcal{N}$ to $1$. The contribution of individual projectors to this reduction and their relations will be studied in Section 5. For this purpose partial products of $1\leq k\leq\mathcal{N}+2$ projection operators are calculated and the trace of such products provides information on degree of reduction of the Hilbert space.\\

Another important quantity which can be obtained from numerical investigation of the projection operators, besides the Hilbert space reduction, are the basis vectors of the reduced Hilbert space, i.e. the eigenvectors of the projection operator (\ref{eq41}). A significant advantage of calculations within subsectors is that every subsector provides a single basis vector of the reduced space, which greatly simplifies the search for them. In particular, this eliminates the necessity to diagonalize eq. (\ref{eq41}) to find its eigenvectors. The simplest way to achieve this, which avoids multiplication of many $2^\mathcal{N}\times2^\mathcal{N}$ matrices, is to choose any vector $v$, i.e. any matrix $2^\mathcal{N}\times 1$, and act on it iteratively with all the $(\mathcal{N}+2)$ projection operators. The resulting vector certainly belongs to the reduced Hilbert space and, since there is only one basis vector per subsector, this is the desired vector. The only issue one can encounter is that the final vector can be zero. In numerical calculations the following procedure was applied: starting with $j=1$, a candidate $v^i=\delta^i_j$ was checked and, as long as the final result was zero, $j$ was increased by one within a loop. Although such a method, based on brute force approach, does not seem very sophisticated, it generally leads to the correct initial guess in less than $10$ tries, so it is much more efficient than multiplication of all the projectors. The eigenvectors of (\ref{eq41}), just as other matrices present in our problem, contain many zero entries, so it is beneficial to convert them to lists of non-zero entries and indices associated with those elements. After repeating this procedure in all the subsectors of a given $p$-particle sector one obtains:
\begin{itemize}
\item \texttt{er[[]]} -- length $\binom{\mathcal{N}}{p}$ list of the basis states of the reduced Hilbert space. \texttt{er[[i,j]]} is \texttt{j}-th non-zero entry of \texttt{i}-th eigenvector.
\item \texttt{wh[[]]} -- length $\binom{\mathcal{N}}{p}$ list of lists, which provide information on the indices where the non-zero elements of the above vectors appear. \texttt{wh[[i,j]]} is the index from the\\ $2^\mathcal{N}$-dimensional \texttt{i}-th subsector which specifies the position of the non-zero entry \texttt{er[[i,j]]}.
\item \texttt{nsr[[]]} -- length $\binom{\mathcal{N}}{p}$ list of numbers. \texttt{nsr[[i]]} is the length of \texttt{er[[i]]}, i.e. the number of non-zero elements of \texttt{i}-th eigenvector.
\end{itemize}
\begin{figure}
\begin{center}
\includegraphics[width=10cm]{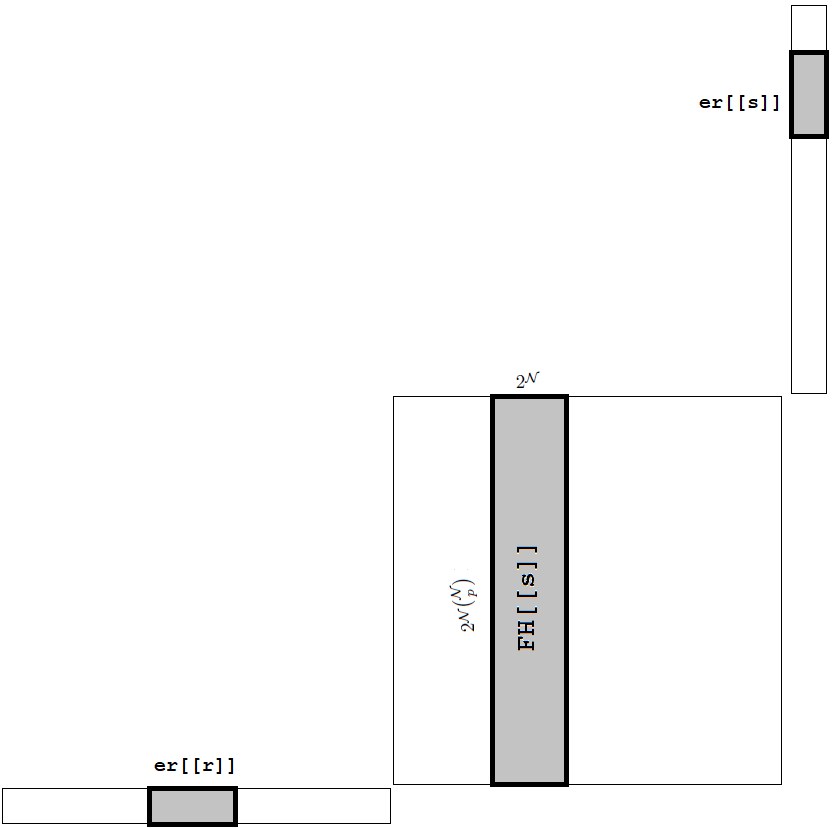}
\end{center}
\caption{Construction of the Hamiltonian in the constrained spin Hilbert space. To obtain the matrix element $\braket{r|H_S|s}$ one multiplies the Hamiltonian matrix with the row-vector \texttt{er[[r]]} and the column-vector \texttt{er[[s]]} in a standard way. Since the reduced spin Hilbert space basis vectors \texttt{er[[t]]} belong to the corresponding $t$-th subsectors, their entries are non-zero only on these subspaces (shaded in the picture) when treated like objects from the whole $p$-particle sector. Thus, it is easiest to imagine the operation (\ref{may2f}) as a multiplication of $2^{\mathcal{N}}\binom{\mathcal{N}}{p}\times 2^{\mathcal{N}}$ Hamiltonian matrix block \texttt{FH[[s]]} by the $1\times 2^{\mathcal{N}}\binom{\mathcal{N}}{p}$ matrix \texttt{er[[r]]} (extended to the whole $p$-particle sector) from the left and the $2^{\mathcal{N}}\times 1$ matrix \texttt{er[[s]]} (kept as a subsector object) from the right.}
\label{fig2}
\end{figure}
\section{Construction of the Hamiltonian}
Construction of the Hamiltonian is very similar to the procedure described in the previous subsection. In particular, all calculations are still performed within a single $p$-particle sector and, since matrices $\Gamma^i$ also have one non-zero entry per column, when link operators $S(\vec{n},\hat{e})$ act on a basis state $\ket{ii}$, the result is another basis state $\ket{jj}$. Thus again there is no need to find matrix elements between all combinations of $\ket{ii}$'s and $\bra{jj}$'s, but the proper $\bra{jj}$ can be deduced from $\ket{ii}$. However, the important difference is that subsectors are not invariant subspaces of the Hamiltonian. Thus entire Hamiltonian cannot be obtained from calculations in $2^\mathcal{N}\times2^\mathcal{N}$ blocks associated with subsectors. Instead, the Hamiltonian is constructed in $2^\mathcal{N}\binom{\mathcal{N}}{p}\times2^\mathcal{N}$ rectangular blocks \texttt{FH[[s]]}, whose columns belong to an $s$-th subsector (see Fig. \ref{fig2}). The main \texttt{Do[,}$\{$\texttt{ii,1,ns0}$\}$\texttt{]} loop of these calculations runs over all basis vectors $\ket{ii}$ of $s$-th subsector and, for every part of the Hamiltonian related to a single link, corresponding $\bra{jj}$ is assigned which yields a non-zero entry. Since the Hamiltonian causes hopping, $\bra{jj}$ is from a different subsector. Yet, apart from adaptation of rectangular blocks and doing calculations separately for every link, the rest of the procedure remains unchanged.\\
Once the eigenvectors are found and the Hamiltonian is constructed in the unconstrained space, one can use them to obtain the Hamiltonian in the reduced Hilbert space. This part requires some caution, because matrices of apparently non-matching dimensions are multiplied. The recipe to calculate \texttt{h[[r,s]]} element of the reduced Hamiltonian is:
\begin{multline}
\text{\texttt{h[[r,s]]=Sum[}}\\ \text{\texttt{Conjugate[er[[r,jr]]]*FH[[s]][[wh[[r,jr]]+(r-1)*ns0,wh[[s,ir]]]]*er[[s,ir]],}}\\ \{\text{\texttt{jr,1,nsr[[1]]}}\}\text{\texttt{,}}\{\text{\texttt{ir,1,nsr[[1]]}}\}\text{\texttt{]}},
\label{may2f}
\end{multline}
which is just the standard way to obtain matrix element $\braket{r|\text{\texttt{FH[[s]]}}|s}$. Both vector \texttt{er[[s]]} and columns of the Hamiltonian block \texttt{FH[[s]]} belong to the same \texttt{s}-th subsector. Thus, when computing their product, one can forget about the rest of the Hilbert space and treat them like objects from the \texttt{s}-th subsector. List \texttt{wh[[s]]} ensures that non-zero entries of $s$-th eigenvector are multiplied by proper elements of \texttt{FH[[s]]}. The "left part" of this matrix product is a bit less intuitive since $r$-th eigenvector is a $2^\mathcal{N}$-dimensional object from $r$-th subsector of $p$-particle sector, while \texttt{FH[[s]]} has rows from the $2^\mathcal{N}\binom{N}{p}$-dimensional $p$-particle sector. The solution to this apparent incompatibility is to understand that obviously \texttt{er[[r]]} can be extended to the entire $p$-particle sector (see Fig. \ref{fig2}), but since the "extension entries" are all zeros, it becomes clear that only rows of \texttt{FH[[s]]} from $r$-th subsector matter when it is multiplied by \texttt{er[[r]]}. This is the reason for the shift by \texttt{(r-1)*ns0}. Once the reduced Hamiltonian is found, one can find its eigenvalues and compare with the formulas obtained in the Grassmann representation, which is discussed in the next section.
\clearpage
\chapter{Fermionic Hamiltonian diagonalization}
The benefits of introducing the local bosonization scheme were discussed comprehensively in Section 1. Some tasks are easier to perform in the fermionic representation though. These include finding the eigenenergies of the system. In this section, the formulas for the eigenenergies of the system of fermions on a lattice are derived. We begin with the simplest case of 1-particle states of free fermions on a 1-dimensional lattice and later expand our scope to include many-particle sectors, 2-dimensional lattices and the interaction with the external $\mathbb{Z}_2$ field. The eigenenergy formulas obtained in the fermionic picture will be compared to the numerical results in the spin representation in Section 5, which serves as an additional verification of the bosonization proposal validity.

\section{Energy spectrum of free Hamiltonian}
\subsection{Hamiltonian in one dimension -- eigenenergies of 1-particle states}
In this section, a procedure to diagonalize Hamiltonian of a system of free fermions on a length $L$ 1-dimensional lattice

\begin{equation}
H_f=i\sum_{n=1}^L \left( \phi(n)^\dag \phi(n+1)-\phi(n+1)^\dag\phi(n)\right)
\label{r1}
\end{equation}
is shown. To find matrix elements of its first term one applies the anticommutation relation $\{\phi(m),\phi(n)^\dag\}=\delta_m^n$:
\begin{equation}
\begin{split}
\braket{l|\phi(n)^\dag\phi(n+1)|m}=&\braket{0|\phi(l)\phi(n)^\dag\phi(n+1)\phi(m)^\dag|0}=\\
&=\bra{0}\left(\delta_n^l-\phi(n)^\dag\phi(l)\right)\left(\delta_{n+1}^m-\phi(m)^\dag\phi(n+1)\right)\ket{0}=\delta_n^l\delta_{n+1}^m,
\end{split}
\label{r2}
\end{equation}
where, once the creation/annihilation operators are normally ordered, vacuum expectation value of their products is zero. Thus the only non-vanishing term is the product of Kronecker deltas. Matrix elements of the second term in Hamiltonian (\ref{r1}) are found by calculating the complex conjugate of the above equation:
\begin{equation}
\braket{l|\phi(n+1)^\dag\phi(n)|m}=\braket{m|\phi(n)^\dag\phi(n+1)|l}^*=(\delta_n^m\delta_{n+1}^l)^*=\delta_{n+1}^l\delta_n^m.
\label{r3}
\end{equation}
Hence, the matrix elements of the Hamiltonian are:
\begin{equation}
\braket{l|H_f|m}=i\sum_{n=1}^L\left(\delta_n^l\delta_{n+1}^m-\delta_{n+1}^l\delta_n^m\right)=i(\delta_{l+1}^m-\delta_{l-1}^m).
\label{r4}
\end{equation}
Therefore, in position representation, matrix $H_f$ consists of a diagonal line of $i$'s just above the main diagonal, a diagonal line of $(-i)$'s just below the main diagonal and (assuming periodic boundary conditions) two other entries $\braket{L|H_f|1}=i$ and $\braket{1|H_f|L}=-i$. To diagonalize this matrix one transforms the operators from the real space to the momentum representation, which is done with formulas:
\begin{equation*}
\phi(n)=\frac{1}{\sqrt{L}}\sum_p e^{ipn}\phi(p)
\tag{4.5a}
\label{r5a}
\end{equation*}
and
\begin{equation*}
\phi(n)^\dag=\frac{1}{\sqrt{L}}\sum_p e^{-ipn}\phi(p)^\dag,
\tag{4.5b}
\label{r5b}
\end{equation*}
\setcounter{equation}{6}
where summation takes place over a set $p\in\{2\pi/L,4\pi/L,6\pi/L,...,2\pi\}$. Formulas which yield an inverse transformation are:
\begin{equation*}
\phi(p)=\frac{1}{\sqrt{L}}\sum_{n=1}^L e^{-ipn}\phi(n)
\label{r6a}
\tag{4.6a}
\end{equation*}
and
\begin{equation*}
\phi(p)^\dag=\frac{1}{\sqrt{L}}\sum_{n=1}^L e^{ipn}\phi(n)^\dag.
\label{r6b}
\tag{4.6b}
\end{equation*}
Using these equations, it is straightforward to check that anticommutation relations in position representation imply analogous relations in momentum space: $\{\phi(p)^\dag,\phi(q)\}=\delta_q^p$ and zero in other cases. Anticommutator of a real-space annihilation operator and a momentum representation creation operator is:
\begin{equation}
\begin{split}
\{\phi(n),\phi(p)^\dag\}=\{\phi(n),\frac{1}{\sqrt{L}}\sum_me^{ipm}\phi(m)^\dag\}=\frac{1}{\sqrt{L}}\sum_me^{ipm}\{\phi(n),\phi(m)^\dag&\}=\\=\frac{1}{\sqrt{L}}\sum_m&e^{ipm}\delta_n^m=\frac{1}{\sqrt{L}}e^{ipn}.
\end{split}
\label{r7}
\end{equation}
Numerically, this result is identical with projection of a momentum state $\ket{p}$ on a position state $\bra{n}$:
\begin{equation}
\braket{n|p}=\braket{0|\phi(n)\phi(p)^\dag|0}=\braket{0|\{\phi(n),\phi(p)^\dag\}-\phi(p)^\dag\phi(n)|0}=\braket{0|\frac{1}{\sqrt{L}}e^{ipn}|0}=\frac{1}{\sqrt{L}}e^{ipn},
\label{r8}
\end{equation}
which, to no surprise, is analogous to a well-known formula for position representation of a momentum eigenfunction $\psi_p(x)=\braket{x|p}=\frac{1}{\sqrt{2\pi}}e^{ipx}$. Using equations (\ref{r4}) and (\ref{r8}) one easily obtains matrix elements of the Hamiltonian in momentum space:
\begin{equation}
\begin{split}
\braket{p|H_f|q}=\sum_{l,m=1}^L\braket{p|l}\braket{l|H_f|m}\braket{m|q}=\sum_{l,m}\frac{1}{\sqrt{L}}e^{-ipl}\cdot i&(\delta_{l+1}^m-\delta_{l-1}^m)\cdot\frac{1}{\sqrt{L}}e^{iqm}=\\
=\frac{i}{L}\sum_m\left(e^{-ip(m-1)}e^{iqm}-e^{-ip(m+1)}e^{iqm}\right)=&\frac{i}{L}\sum_m(e^{ip}-e^{-ip})e^{i(q-p)m}=\\
=&\frac{i}{L}\cdot 2i\sin p\cdot L\delta_p^q=-2\sin p\delta_p^q.
\end{split}
\label{r9}
\end{equation}
Thus, in this representation $H_f$ is diagonal and it can be written as:
\begin{equation}
H_f=\sum_p-2\sin(p)\phi(p)^\dag\phi(p).
\label{r10}
\end{equation}
Eigenenergies of 1-particle lattice excitations are evident from this form of the Hamiltonian:
\begin{equation}
E^{(1)}(p)=-2\sin p.
\label{r11}
\end{equation}
Before we proceed to generalization of this result to many-particle states, let us see that diagonalization can be done by direct substitution of transformation formulas (\ref{r5a}) and (\ref{r5b}) to the Hamiltonian:
\begin{equation}
\begin{split}
\sum_{n=1}^L&\phi(n)^\dag\phi(n+1)=\sum_n\left(\frac{1}{\sqrt{L}}\sum_qe^{-iqn}\phi(q)^\dag\right)\left(\frac{1}{\sqrt{L}}\sum_pe^{ip(n+1)}\phi(p)\right)=\\
&=\frac{1}{L}\sum_{q,p}\sum_ne^{ip}e^{i(p-q)n}\phi(q)^\dag\phi(p)=\sum_{q,p}e^{ip}\delta_q^p\phi(q)^\dag\phi(p)=\sum_p e^{ip}\phi(p)^\dag\phi(p).
\end{split}
\label{r12}
\end{equation}
Hence:
\begin{equation}
H_f=i\sum_{n=1}^L(\phi(n)^\dag\phi(n+1)-\phi(n+1)^\dag\phi(n))=i\sum_p (e^{ip}-e^{-ip})\phi(p)^\dag\phi(p)=\sum_p-2\sin(p)\phi(p)^\dag\phi(p),
\label{r13}
\end{equation}
which coincides with the result obtained previously.
\subsection{Hamiltonian in one dimension -- many-particle sectors}
The purpose of this section is to find expectation values of the Hamiltonian in $M$-particle states $\ket{p_1, p_2, ...p_M}$:
\begin{equation}
\braket{p_1...p_M|H_f|p_1...p_M}=\sum_p-2\sin(p)\braket{0|\phi(p_M)...\phi(p_1)\phi(p)^\dag\phi(p)\phi(p_1)^\dag...\phi(p_M)^\dag|0}.
\label{r14}
\end{equation}
The simplest way to achieve this is by application of Wick's theorem separately to $\phi(p)\phi(p_1)^\dag...$ $...\phi(p_M)^\dag$ and $\phi(p_M)...\phi(p_1)\phi(p)^\dag$. Let us begin with the first of these operator products. To write it as a sum of normally ordered terms, one needs to contract $\phi(p)$, which is the only annihilation operator among them, with every $\phi(p_j)^\dag$. There are no non-vanishing terms corresponding to multiple contractions, so -- apart from plus or minus signs -- this operator product is the sum of normally ordered product $(\prod_i\phi(p_i)^\dag)\phi(p)$ and $M$ terms with single contraction $\delta_{p_j}^p\prod_{i\neq j}\phi(p_i)^\dag$. To see directly, without use of Wick's theorem, how these terms emerge and to keep track of plus or minus signs, one can examine normal ordering of this product for a few lowest $M$'s. For $M=1$ it is simply an anticommutation relation:
\begin{equation}
\phi(p)\phi(p_1)^\dag=\{\phi(p),\phi(p_1)^\dag\}-\phi(p_1)^\dag\phi(p)=\delta_{p_1}^p-\phi(p_1)^\dag\phi(p).
\label{r15}
\end{equation}
To obtain the formula for $M=2$, one multiplies the previous one by $\phi(p_2)^\dag$ and applies it again to $\phi(p)\phi(p_2)^\dag$: 
\begin{multline}
\phi(p)\phi(p_1)^\dag\phi(p_2)^\dag=\left(\delta_{p_1}^p-\phi(p_1)^\dag\phi(p)\right)\phi(p_2)^\dag\stackrel{(\ref{r15})}{=}\delta_{p_1}^p\phi(p_2)^\dag-\phi(p_1)^\dag(\delta_{p_2}^p-\phi(p_2)^\dag\phi(p))=\\
=\delta_{p_1}^p\phi(p_2)^\dag-\phi(p_1)^\dag\delta_{p_2}^p+\phi(p_1)^\dag\phi(p_2)^\dag\phi(p).
\label{r16}
\end{multline}
Similarly for $M=3$:
\begin{multline}
\phi(p)\phi(p_1)^\dag\phi(p_2)^\dag\phi(p_3)^\dag=\left(\delta_{p_1}^p\phi(p_2)^\dag-\phi(p_1)^\dag\delta_{p_2}^p+\phi(p_1)^\dag\phi(p_2)^\dag\phi(p)\right)\phi(p_3)^\dag=\\
\stackrel{(\ref{r15})}{=}\delta_{p_1}^p\phi(p_2)^\dag\phi(p_3)^\dag-\phi(p_1)^\dag\delta_{p_2}^p\phi(p_3)^\dag+\phi(p_1)^\dag\phi(p_2)^\dag(\delta_{p_3}^p-\phi(p_3)^\dag\phi(p))=\\
=\delta_{p_1}^p\phi(p_2)^\dag\phi(p_3)^\dag-\phi(p_1)^\dag\delta_{p_2}^p\phi(p_3)^\dag+\phi(p_1)^\dag\phi(p_2)^\dag\delta_{p_3}^p-\phi(p_1)^\dag\phi(p_2)^\dag\phi(p_3)^\dag\phi(p).
\label{r17}
\end{multline}
At this point a general pattern should become clear. When multiplying a formula for $M$ by $\phi(p_{M+1})^\dag$ to obtain the one for $M+1$, initial $M$ terms of the sum are simply expanded by one more creation operator. On the other hand, the last term of the sum requires anticommutation of $\phi(p)$ and $\phi(p_{M+1})^\dag$. This introduces $\delta_{p_{M+1}}^p$ factor and a new normally ordered product with $\phi(p)$, which is one operator longer than before. Subsequent terms are of opposite signs. Summing up these observations and the discussion about Wick's theorem, one obtains:
\begin{equation}
\phi(p)\phi(p_1)^\dag...\phi(p_M)^\dag=\left(\sum_{j=1}^M(-1)^{j+1}\delta_{p_j}^p\prod_{i=1,\ i\neq j}^M\phi(p_i)^\dag\right)+(-1)^M\left(\prod_{i=1}^M\phi(p_i)^\dag\right)\phi(p).
\label{r18}
\end{equation}
Hermitian conjugate of this result provides normal ordering of the second operator product in (\ref{r14}):
\begin{equation}
\phi(p_M)...\phi(p_1)\phi(p)^\dag=\left(\sum_{j=1}^M(-1)^{j+1}\delta_{p_j}^p\prod_{i=M,\ i\neq j}^1\phi(p_i)\right)+(-1)^M\phi(p)^\dag\left(\prod_{i=M}^1\phi(p_i)\right).
\label{r19}
\end{equation}
Substitution of eqs. (\ref{r18}) and (\ref{r19}) to eq. (\ref{r14}) yields the final result (one neglects terms which annihilate vacuum state):
\begin{equation}
\begin{split}
\braket{p_1...p_M|H_f|p_1...p_M}&=\\
=\sum_p-2\sin p& \braket{0|\left(\sum_{j=1}^M(-1)^{j+1}\delta_{p_j}^p\prod_{i=M,\ i\neq j}^1\phi(p_i)\right)\left(\sum_{k=1}^M(-1)^{k+1}\delta_{p_k}^p\prod_{l=1,\ l\neq k}^M\phi(p_l)^\dag\right)|0}=\\
=\sum_p-2\sin p&\left(\sum_{j=1}^M(-1)^{j+1}\delta_{p_j}^p\bra{p_1,p_2,...,p_{j-1},p_{j+1},...,p_M}\right)\times\\
&\times\left(\sum_{k=1}^M(-1)^{k+1}\delta_{p_k}^p\ket{p_1,p_2,...,p_{k-1},p_{k+1},...,p_M}\right)=\\
=\sum_p-2\sin p&\sum_{j,k=1}^M(-1)^{j+k}\delta_{p_j}^p\delta_{p_k}^p\delta_k^j=\sum_p-2\sin p\sum_{k=1}^M\delta_{p_k}^p=\sum_{k=1}^M\sum_p-2\delta_{p_k}^p\sin p=\sum_{k=1}^M-2\sin p_k.
\end{split}
\label{r20}
\end{equation}
Thus, energies of $M$-particle states are:
\begin{equation}
E^{(M)}(p_1,p_2,...,p_M)=-2\sin(p_1)-2\sin(p_2)-...-2\sin(p_M).
\label{r21}
\end{equation}
Such generalization of eq. (\ref{r11}) coincides with the intuition that eigenenergy of a state of $M$ non-interacting fermions is the sum of energies of individual particles.
\section{2D Hamiltonian}
Diagonalization of the 2-dimensional Hamiltonian
\begin{equation}
H_f=i\sum_{\vec{n},\vec{e}}\left(\phi(\vec{n})^\dag\phi(\vec{n}+\vec{e})-\phi(\vec{n}+\vec{e})^\dag\phi(\vec{n})\right)
\label{r22}
\end{equation}
of free fermions on a $L_1\times L_2$ lattice is for most part analogous to the 1-dimensional case. One substitutes 2-dimensional Fourier transform formulas:
\begin{equation}
\phi(\vec{n})=\frac{1}{\sqrt{V}}\sum_{\vec{p}}e^{i\vec{p}\vec{n}}\phi(\vec{p})
\label{r23}
\end{equation}
\begin{equation}
\phi(\vec{n})^\dag=\frac{1}{\sqrt{V}}\sum_{\vec{p}}e^{-i\vec{p}\vec{n}}\phi(\vec{p})^\dag,
\label{r24}
\end{equation}
where $p_x\in\{2\pi/L_1,4\pi/L_1,6\pi/L_1,...,2\pi\}$, $p_y\in\{2\pi/L_2,4\pi/L_2,6\pi/L_2,...,2\pi\}$ and $V=L_1L_2$, to the first term of the Hamiltonian, which yields:
\begin{equation}
\begin{split}
\sum_{\vec{n},\vec{e}}&\phi(\vec{n})^\dag\phi(\vec{n}+\vec{e})=\sum_{\vec{n},\vec{e}}\left(\frac{1}{\sqrt{V}}\sum_{\vec{p}}e^{-i\vec{p}\vec{n}}\phi(\vec{p})^\dag\right)\left(\frac{1}{\sqrt{V}}\sum_{\vec{q}}e^{i\vec{q}(\vec{n}+\vec{e})}\phi(\vec{q})\right)=\\
&=\frac{1}{V}\sum_{\vec{e},\vec{p},\vec{q}}\sum_{\vec{n}}e^{i\vec{q}\vec{e}}e^{i(\vec{q}-\vec{p})\vec{n}}\phi(\vec{p})^\dag\phi(\vec{q})=\frac{1}{V}\sum_{\vec{e},\vec{p},\vec{q}}V\delta_{\vec{q}}^{\vec{p}}e^{i\vec{q}\vec{e}}\phi(\vec{p})^\dag\phi(\vec{q})=\sum_{\vec{e},\vec{p}}e^{i\vec{p}\vec{e}}\phi(\vec{p})^\dag\phi(\vec{p}).
\end{split}
\label{r25}
\end{equation}
Hence:
\begin{equation}
H_f=i\sum_{\vec{e},\vec{p}}(e^{i\vec{p}\vec{e}}-e^{-i\vec{p}\vec{e}})\phi(\vec{p})^\dag\phi(\vec{p})=\sum_{\vec{p}}(-2\sin p_x-2\sin p_y)\phi(\vec{p})^\dag\phi(\vec{p}).
\label{r26}
\end{equation}
This equation is strictly analogous to eq. (\ref{r10}) apart from two facts: (i) the numerical factor is $(-2\sin p_x-2\sin p_y)$ instead of $-2\sin p$, (ii) summation is done over 2-dimensional lattice in momentum space. The first fact can be handled by substitution $-2\sin p\to(-2\sin p_x-2\sin p_y)$, since this factor is not involved in creation/annihilation operators algebra -- as can be seen from derivation (\ref{r20}). The second difference also does not matter, because $H_f$ is diagonal in momentum space. In other words, terms like $\phi(\vec{p})^\dag\phi(\vec{p})$ are insensitive to the choice of numbering of the lattice sites or boundary conditions (contrary to $H_f$ in real space). Therefore, detailed derivation of Hamiltonian expectation values in $M$-particle states $\ket{\vec{p_1},\vec{p_2},...,\vec{p_M}}$ would be a repetition of derivation (\ref{r14})-(\ref{r20}) and the result is:
\begin{equation}
\braket{\vec{p_1}...\vec{p_M}|H_f|\vec{p_1}...\vec{p_M}}=\sum_{k=1}^M-2\sin p_{k,x}-2\sin p_{k,y}.
\label{r27}
\end{equation}
Or, equivalently:
\begin{equation}
E^{(M)}(\vec{p_1},\vec{p_2},...,\vec{p_M})=-2\sin(p_{1,x})-2\sin(p_{1,y})-2\sin(p_{2,x})-2\sin(p_{2,y})-...-2\sin(p_{M,x})-2\sin(p_{M,y}).
\label{r28}
\end{equation}
\section{Energy spectrum of Hamiltonian with external magnetic field}
\subsection{Transformation to momentum space}
To obtain a block-diagonal form of Hamiltonian (\ref{s1}), it is expressed in momentum space:
\begin{equation*}
\phi(\vec{n})=\frac{1}{\sqrt{V}}\sum_{\vec{p}}e^{i\vec{p}\vec{n}}\phi(\vec{p}),
\label{s2a}
\tag{4.29a}
\end{equation*}
\begin{equation*}
\phi(\vec{n})^\dag
=\frac{1}{\sqrt{V}}\sum_{\vec{p}}e^{-i\vec{p}\vec{n}}\phi(\vec{p})^\dag.
\label{s2b}
\tag{4.29b}
\end{equation*}
This transformation preserves the canonical anticommutation relation. Indeed, assuming that $\{\phi(\vec{p})^\dag,\phi(\vec{q})\}=\delta_{\vec{q}}^{\vec{p}}$, one obtains the relation in real space:
\setcounter{equation}{29}
\begin{multline}
\{\phi(\vec{n})^\dag,\phi(\vec{m})\}=\Big\{\frac{1}{\sqrt{V}}\sum_{\vec{p}}e^{-i\vec{p}\vec{n}}\phi(\vec{p})^\dag,
\frac{1}{\sqrt{V}}\sum_{\vec{q}}e^{i\vec{q}\vec{m}}\phi(\vec{q})\Big\}
=\\=\frac{1}{V}\sum_{\vec{p},\vec{q}}e^{i(\vec{q}\vec{m}-\vec{p}\vec{n})}\{\phi(\vec{p})^\dag,\phi(\vec{q})\}
=\frac{1}{V}\sum_{\vec{p},\vec{q}}e^{i(\vec{q}\vec{m}-\vec{p}\vec{n})}\delta^{\vec{p}}_{\vec{q}}=\frac{1}{V}\sum_{\vec{q}}e^{i\vec{q}(\vec{m}-\vec{n})}=\delta^{\vec{m}}_{\vec{n}}
\label{s3a}
\end{multline}
\setcounter{equation}{30}
Sums with respect to momentum are performed over the first Brillouin zone, i.e. domain of the momenta is:
\begin{equation}
p_{\hat{e}}=\frac{2\pi}{L_{\hat{e}}}n,\quad n=0,1,...,L_{\hat{e}}-1,\quad \hat{e}\in\{x,y\}.
\label{s3extra}
\end{equation}
Whenever momenta outside this domain appear in the equations, one identifies $\phi(\vec{p})\equiv\phi(\vec{p}\pm 2\pi\hat{e})$. To diagonalize the Hamiltonian, one begins with expression of $H_+$ in momentum space. From eqs. (\ref{s2a}) and (\ref{s2b}): 
\begin{multline}
\sum_{\vec{n}}\phi(\vec{n})^\dag\phi(\vec{n}+\hat{y})
=\sum_{\vec{n}}\Big(\frac{1}{\sqrt{V}}\sum_{\vec{p}}e^{-i\vec{p}\vec{n}}\phi(\vec{p})^\dag\Big)\Big(\frac{1}{\sqrt{V}}\sum_{\vec{q}}e^{i\vec{q}\vec{n}}e^{iq_y}\phi(\vec{q})\Big)=\\
=\frac{1}{V}\sum_{\vec{p},\vec{q}}e^{iq_y}\phi(\vec{p})^\dag\phi(\vec{q})\sum_{\vec{n}}e^{i(\vec{q}-\vec{p})\vec{n}}
=\sum_{\vec{p},\vec{q}}e^{iq_y}\phi(\vec{p})^\dag\phi(\vec{q})\delta^{\vec{p}}_{\vec{q}}=\sum_{\vec{q}}e^{iq_y}\phi(\vec{q})^\dag\phi(\vec{q}).
\label{s4a}
\end{multline}
\setcounter{equation}{32}
Hence:
\begin{multline}
H_+=i\sum_{\vec{n}}\left(\phi(\vec{n})^\dag\phi(\vec{n}+\hat{y})-\phi(\vec{n}+\hat{y})^\dag\phi(\vec{n})\right)=\sum_{\vec{q}}i(e^{iq_y}-e^{-iq_y})\phi(\vec{q})^\dag\phi(\vec{q})=\\=\sum_{\vec{q}}-2\sin q_y \phi(\vec{q})^\dag\phi(\vec{q}).
\label{s5}
\end{multline}
This is diagonal -- at no surprise due to similarity of $H_+$ and the Hamiltonian for free fermions. Very similar derivation is done for $H_-$:
\begin{multline}
\sum_{\vec{n}}(-1)^{n_y}\phi(\vec{n})^\dag\phi(\vec{n}+\hat{x})
=\sum_{\vec{n}}(-1)^{n_y}\Big(\frac{1}{\sqrt{V}}\sum_{\vec{p}}e^{-i\vec{p}\vec{n}}\phi(\vec{p})^\dag\Big)\Big(\frac{1}{\sqrt{V}}\sum_{\vec{q}}e^{i\vec{q}\vec{n}}e^{iq_x}\phi(\vec{q})\Big)=\\
=\frac{1}{V}\sum_{\vec{p},\vec{q}}e^{iq_x}\phi(\vec{p})^\dag\phi(\vec{q})\sum_{\vec{n}}(-1)^{n_y}e^{i(\vec{q}-\vec{p})\vec{n}}.
\label{s6a}
\end{multline}
The term under the sum can be written as $(-1)^{n_y}e^{i(\vec{q}-\vec{p})\vec{n}}=e^{i\pi n_y}e^{i(\vec{q}-\vec{p})\vec{n}}=e^{i(\vec{q}-\vec{p}+\pi\hat{y})\vec{n}}$. Therefore, once this sum is performed, one obtains:
\setcounter{equation}{34}
\begin{equation}
\sum_{\vec{n}}(-1)^{n_y}\phi(\vec{n})^\dag\phi(\vec{n}+\hat{x})
=\sum_{\vec{p},\vec{q}}e^{iq_x}\phi(\vec{p})^\dag\phi(\vec{q})\delta^{\vec{p}}_{\vec{q}+\pi\hat{y}}
=\sum_{\vec{q}}e^{iq_x}\phi(\vec{q}+\pi\hat{y})^\dag\phi(\vec{q})
\label{s6b}
\end{equation}
and the corresponding contribution to the Hamiltonian equals:
\begin{multline}
H_-=i\sum_{\vec{n}}(-1)^{n_y}(\phi(\vec{n})^\dag\phi(\vec{n}+\hat{x})-\phi(\vec{n}+\hat{x})^\dag\phi(\vec{n}))=\\
=i\Big(\sum_{\vec{q}}e^{iq_x}\phi(\vec{q}+\pi\hat{y})^\dag\phi(\vec{q})-\sum_{\vec{q}}e^{-iq_x}\phi(\vec{q})^\dag\phi(\vec{q}+\pi\hat{y})\Big)=\\
=i\Big(\sum_{\vec{q}}e^{iq_x}\phi(\vec{q}+\pi\hat{y})^\dag\phi(\vec{q})-\sum_{\vec{q}}e^{-iq_x}\phi(\vec{q}+\pi\hat{y})^\dag\phi(\vec{q})\Big)=\sum_{\vec{q}}-2\sin q_x \phi(\vec{q}+\pi\hat{y})^\dag\phi(\vec{q}).
\label{s7}
\end{multline}
Thus, in the end:
\begin{equation}
H=-2\sum_{\vec{q}}\left(\sin q_y\phi(\vec{q})^\dag\phi(\vec{q})+\sin q_x\phi(\vec{q}+\pi\hat{y})^\dag\phi(\vec{q})\right).
\label{s8}
\end{equation}
\subsection{Block-diagonal structure of the Hamiltonian in momentum space}
\begin{figure}
\begin{center}
\includegraphics[width=12cm]{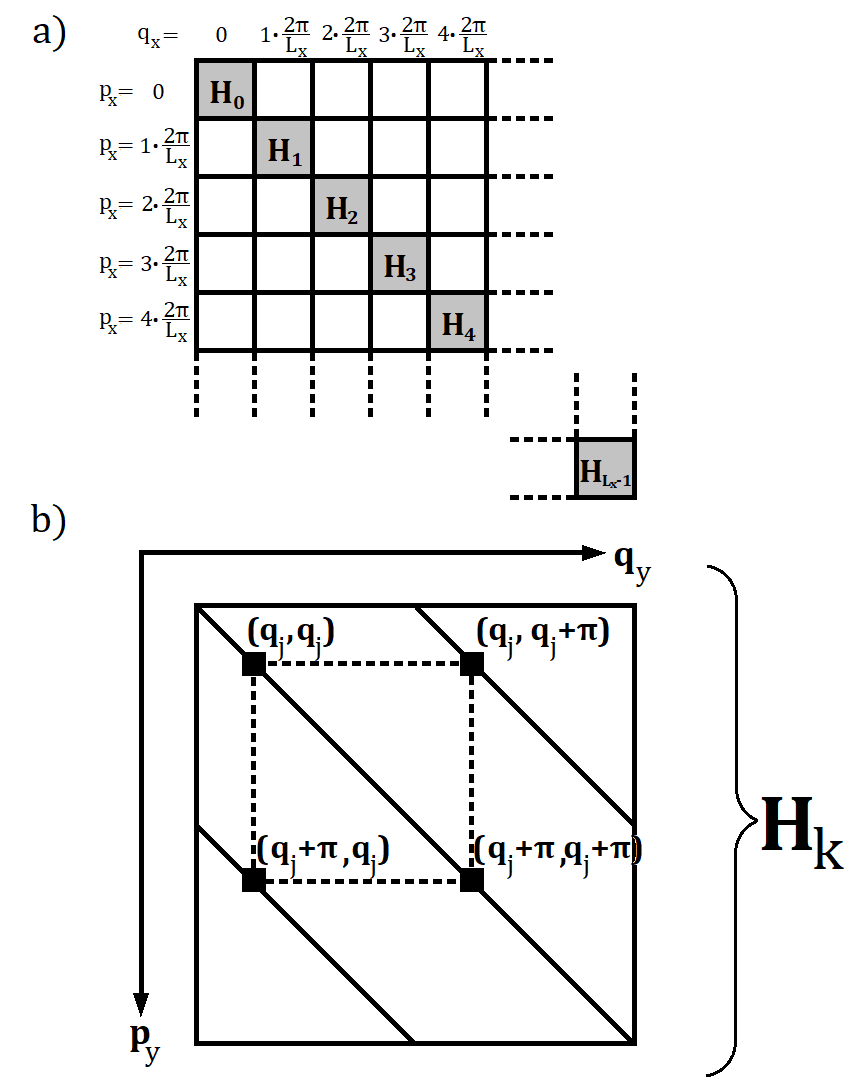}
\caption{(a) Block-diagonal structure of the Hamiltonian of fermions in a constant magnetic field. Since momentum states are ordered lexicographically, block $H_k$ is the Hamiltonian matrix reduced to a subspace of vectors with a constant x-component of momentum $2\pi k/L_x$.\\(b) Bidiagonal structure of a $L_y\times L_y$ block $H_k$. The only non-zero entries are located either at the main diagonal or at a parallel line shifted by half the size of the block, $L_y/2$. Positions at which non-zero elements occur are marked with diagonal black solid lines.}
\label{figblock}
\end{center}
\end{figure}
Matrix of Hamiltonian (\ref{s8}) expressed in momentum basis is not much different from a diagonal matrix. The first term in (\ref{s8}) is diagonal, while the second one describes hopping between states $\vec{q}$ and $\vec{q}\pm\pi\hat{y}$. Therefore, hopping occurs only between momentum states with equal x-components and whose y-components are exactly in counterphase:
\begin{equation}
\braket{\vec{p}|H|\vec{q}}\neq 0\quad \Rightarrow\quad p_x=q_x\wedge|p_y-q_y|\in\{0,\pi\}.
\end{equation}
In particular, when a momentum basis
\begin{equation}
\ket{\vec{q}_{mn}}=\Big|\frac{2\pi}{L_x}m\hat{x}+\frac{2\pi}{L_y}n\hat{y}\Big>
\label{o1}
\end{equation}
is introduced ($m=0,1,2,... ,L_x-1$, $n=0,1,2,... ,L_y-1$) and its vectors are ordered lexicographically:
\begin{equation}
\ket{\vec{q}_{m'n'}}>\ket{\vec{q}_{mn}}\quad :\Leftrightarrow\quad m'>m\vee(m'=m\wedge n'>n),
\label{o2}
\end{equation}
then the Hamiltonian matrix in this basis is block-diagonal. This follows from the fact that the basis states are ordered primarily with respect to their x-components and $\braket{\vec{p}|H|\vec{q}}=0$ when $p_x\neq q_x$ (see Fig. \ref{figblock}a). Denote those blocks on the diagonal of the Hamiltonian by $H_k$ ($0\leq k\leq L_x-1$):
\begin{equation}
(H_k)_{m'm}=\braket{\vec{q}_{km'}|H|\vec{q}_{km}}.
\label{o3}
\end{equation}
These are $L_y\times L_y$ matrices of Hamiltonian matrix elements between momentum states with a fixed x-component.\\
Furthermore, since hopping occurs only between states $\ket{\vec{q}_{mn}}$ and $\ket{\vec{q}_{mn}\pm\pi\hat{y}}=\ket{\vec{q}_{m,n\pm L_y/2}}$, blocks $H_k$ are bidiagonal (Fig. \ref{figblock}b). Such a form of these blocks greatly simplifies diagonalization.
\subsection{Eigenenergies of the Hamiltonian}
To find eigenvalues $E$ of the blocks $H_k$ one searches for the roots of the characteristic polynomial:
\begin{equation}
\text{det}(H_k-E\mathds{1})=\sum_{\sigma\in S_{L_y}}\text{sgn }\sigma\ (H_k-E\mathds{1})_{1,\sigma(1)}(H_k-E\mathds{1})_{2,\sigma(2)}...(H_k-E\mathds{1})_{L_y,\sigma(L_y)},
\label{s10}
\end{equation}
where $S_{L_y}$ is a symmetric group of permutations $\sigma:\ \{1,2,...,L_y\}\to\{1,2,...,L_y\}$. Every row/column of $H_k-E\mathds{1}$ contains only two non-zero entries, because such is the structure of $H_k$, as shown in Fig. \ref{figblock}b, and $E\mathds{1}$ is clearly diagonal. Since all non-zero entries of $H_k-E\mathds{1}$ are located either at its main diagonal or at the diagonal shifted by $L_y/2$, which is half of the size of the matrix, $(H_k-E\mathds{1})_{j,\sigma(j)}$ do not vanish only when $\sigma(j)=j$ or $\sigma(j)=j\pm L_y/2$. Thus the above sum reduces to a sum over permutations $\sigma\in S_{L_y}'\subset S_{L_y}$ such that all $(H_k-E\mathds{1})_{j,\sigma(j)}$, $1\leq j\leq L_y$, are matrix elements from either non-zero diagonal. In other words, $\sigma\in S'_{L_y}\Rightarrow \forall 1\leq j\leq L_y: \sigma(j)=j\ \vee\ \sigma(j)=j\pm L_y/2$. Taking into account also the fact that if $\sigma(j)=j$, then $\sigma(j\pm L_y/2)\neq j\Rightarrow \sigma(j\pm L_y/2)=j\pm L_y/2$ and that, similarly, $\sigma(j)=j\pm L_y/2$ implies $\sigma(j\pm L_y/2)=j$ -- one arrives finally at:
\begin{multline}
\sigma\in S_{L_y}'\ \Leftrightarrow\ \forall\ 1\leq j\leq \frac{L_y}{2}:\\ \left(\sigma(j)=j\wedge\sigma\Big(j+\frac{L_y}{2}\Big)=j+\frac{L_y}{2}\right)\vee\left(\sigma(j)=j+\frac{L_y}{2}\wedge\sigma\Big(j+\frac{L_y}{2}\Big)=j\right).
\label{s11}
\end{multline}
The first part of this disjunction corresponds to diagonal matrix elements, while the latter one to off-diagonal terms (see Fig. \ref{figblock}b for comparison). In other words, only permutations $\sigma$ which are compositions of transpositions
\begin{equation}
\begin{pmatrix}
j&j+L_y/2\\
j+L_y/2&j
\end{pmatrix},
\label{s12}
\end{equation}
$1\leq j\leq L_y/2$, yield non-zero contributions the to sum (\ref{s10}):
\begin{equation}
S_{L_y}'=\Big\{
\sigma_1\circ\sigma_2\circ...\circ\sigma_{L_y/2}\ |\ \sigma_j\in\Big\{\text{id},
\begin{pmatrix}
j&j+L_y/2\\
j+L_y/2&j
\end{pmatrix}
\Big\}
\Big\}.
\label{s13}
\end{equation}
Denote $m=m(\sigma)$ -- the number of such transpositions permutation $\sigma$ is composed of, $0\leq m\leq L_y/2$. The characteristic polynomial can be then expressed as a sum with respect to $m$:
\begin{multline}
\text{det}(H_k-E\mathds{1})=\sum_{\sigma\in S_{L_y}'}\text{sgn }\sigma\ (H_k-E\mathds{1})_{1,\sigma(1)}(H_k-E\mathds{1})_{2,\sigma(2)}...(H_k-E\mathds{1})_{L_y,\sigma(L_y)}=\\
=\sum_{m=0}^{L_y/2}(-1)^m(\sin\ q_{x,k})^{2m}\sum_{\Omega\in C(L_y/2-m)}\prod_{i\in\Omega}(\sin q_{y,i}-E)(\sin q_{y,i+L_y/2}-E),
\label{s14}
\end{multline}
where
\begin{equation}
q_{x,k}\equiv (\vec{q}_{ki})_x=\frac{2\pi}{L_x}k,\ 0\leq k\leq L_x-1;\quad q_{y,i}\equiv(\vec{q}_{ki})_y=\frac{2\pi}{L_y}i,\ 0\leq i\leq \frac{L_y}{2}-1.
\label{s15}
\end{equation}
$C(L_y/2-m)$ is a set of $(L_y/2-m)$-element subsets of $\{0,1,2,...,L_y/2-1\}$. Its elements $\Omega\in C(L_y/2-m)$ specify row/column indices of $H_k-E\mathds{1}$ which contribute diagonal matrix elements to a given term of the sum (\ref{s14}), while remaining rows/columns contribute off-diagonal elements. Relation between summation over permutations $\sigma\in S_{L_y}'$ and over $m\in[0,L_y/2]$ and $\Omega\in C(L_y/2-m)$ is therefore:
\begin{equation}
i\in\Omega\ \Leftrightarrow \sigma(i)=i\ \wedge\ \sigma(i+L_y/2)=i+L_y/2.
\label{s15ex}
\end{equation}
In the sum (\ref{s14}) terms $(\sin\ q_{x,k})^{2m}$ correspond to off-diagonal contributions to $\text{det}(H_k-E\mathds{1})$, while $(\sin q_{y,i}-E)(\sin q_{y,i+L_y/2}-E)$ are terms from the main diagonal. Note also that the domain of $i$ in eq. (\ref{s15}) is half of the whole interval $[0,L_y-1]$ once we take account for the diagonal terms $\sin q_{y,i}-E$ and $\sin q_{y,i+L_y/2}-E$ in pairs. Factor $(-2)$, which appears in front of Hamiltonian (\ref{s8}), is neglected in this part of derivation for clarity. The last factor in eq. (\ref{s14}) can be rewritten as:
\begin{multline}
(\sin q_{y,i}-E)(\sin q_{y,i+L_y/2}-E)=(\sin q_{y,i}-E)(\sin (q_{y,i}+\pi)-E)=\\=(\sin q_{y,i}-E)(-\sin q_{y,i}-E)=E^2-\sin (q_{y,i})^2.
\label{s16}
\end{multline}
Thus:
\begin{multline}
\text{det}(H_k-E\mathds{1})=\sum_{m=0}^{L_y/2}(-1)^m(\sin\ q_{x,k})^{2m}\sum_{\Omega\in C(L_y/2-m)}\prod_{i\in\Omega}(E^2-\sin (q_{y,i})^2)=\\=\prod_{i=0}^{L_y/2-1} (E^2-\sin(q_{y,i})^2-\sin(q_{x,k})^2),
\label{s17}
\end{multline}
where the last equality is easiest to prove going from the right hand side to the left by similar reasoning as binomial expansion. The roots of the characteristic polynomial are therefore:
\begin{equation}
E_{k,i}=\pm\sqrt{\sin(q_{x,k})^2+\sin(q_{y,i})^2}
\label{s18}
\end{equation}
or, skipping the $k$ and $i$ indices, and reintroducing the $(-2)$ factor:
\begin{equation}
E(q_x,q_y)=\pm2\sqrt{\sin(q_x)^2+\sin(q_y)^2},
\label{s19}
\end{equation}
where $q_{x}=0,\ 1\cdot\frac{2\pi}{L_x},\ 2\cdot\frac{2\pi}{L_{x}},\ ...\ ,\ (L_{x}-1)\cdot\frac{2\pi}{L_{x}}$ and $q_{y}=0,\ 1\cdot\frac{2\pi}{L_y},\ 2\cdot\frac{2\pi}{L_{y}},\ ...\ ,\ (\frac{L_{y}}{2}-1)\cdot\frac{2\pi}{L_{y}}$.\\
In $M$-particle case one obtains:
\begin{equation}
E^{(M)}(\vec{q_1},\vec{q_2},...,\vec{q_M})=\sum_{k=1}^M \pm2\sqrt{\sin(q_{k,x})^2+\sin(q_{k,y})^2},
\label{may2a}
\end{equation}
where the choices of $\pm$ signs are independent for all the terms of the sum.
\clearpage

\chapter{Results}

\section{Methods}

The procedure to construct the constraints and the Hamiltonian in the spin representation, which was described in Section 3 alongside with the implementation in Wolfram Mathematica, is used to verify numerically the theoretical predictions for small lattice sizes.\\
To get a closer insight into mutual relations of the constraints, a step-by-step examination of the spin Hilbert space dimension reduction is performed.
The set of constraints imposed on the spin space is expanded from a single constraint to the full set of $\mathcal{N}+2$ constraints. After each new constraint is appended, the dimension of the partly-constrained spin space is computed. One expects that action of all independent constraints reduces the dimension of a subsector from $2^\mathcal{N}$ to $1$, i.e. to the size of a fermionic subsector. It is also predicted that, in a generic case, a single constraint reduces the space dimension by a factor two. This prediction is based on symmetry between projection on $P_s=+1$ and $P_s=-1$, whose images are subspaces of equal sizes. Furthermore, a few different orders to append the projection operators are applied and studied to discover relations between them and figure out which ones are independent.

These results are compared to conclusions from Section 2.3. Moreover, the energetic spectrum is found by diagonalization of the Hamiltonian in the spin representation. Comparison of the eigenenergies with predictions from formulas (\ref{r28}) and (\ref{may2a}), which are obtained in the fermionic representation, provides yet another confirmation of equivalence between these two approaches.\\
In order to study how dimension of the spin Hilbert space is reduced by the constraints, one computes traces of products of the constraints. Indeed, the trace of a product $\Sigma$ of constraints:
\begin{equation}
\Sigma=\prod_{\Sigma_i\in\Lambda}\Sigma_i\quad,\text{where}\qquad \Lambda\subset\{\Sigma_{ij}|1\leq i\leq L_x,1\leq j\leq L_y\}\cup\{\Sigma_x,\Sigma_y\},
\label{s20new}
\end{equation}
which itself is a projector (see eq. (\ref{eq31s5})), yields:
\begin{equation}
\text{Tr }\Sigma=\text{Tr}(\mathcal{U}\Sigma_{\text{diag}}\mathcal{U}^{-1})=\text{Tr}(\mathcal{U}^{-1}\mathcal{U}\Sigma_{\text{diag}})=\text{Tr }\Sigma_{\text{diag}},
\label{s21new}
\end{equation}
where $\Sigma_{\text{diag}}$ is a diagonalized form of the operator $\Sigma$. The latter quantity is the dimension of the image of the projection operator $\Sigma$, because it is a number of its eigenvectors associated with eigenvalue $1$.

Consider the case when the boundary conditions are chosen correctly, i.e. in such a manner that the condition (\ref{eq34}) is satisfied. Since one expects that the constrained Hilbert space of a subsector is one-dimensional, the expected trace of the product of all the plaquette projectors and two perpendicular Polyakov line projection operators is one. When the opposite boundary conditions are applied, antiperiodic when periodic are needed to fulfill eq. (\ref{eq34}) or vice versa, the trace of all the constraints equals zero. The constraints have no solution in this case and the image of their composition is $0$-dimensional. Therefore, the basic information obtained from examination of the traces is the verification of the formula (\ref{eq34}) and determination which boundary conditions are the proper choice in each $p$-particle sector.

The traces of products of projection operators are arranged in tables with respect to increasing number of the constraints (from zero to $\mathcal{N}+2$). Lattice sizes $3\times3$, $4\times3$ and $4\times4$ are studied and the results for them are presented in the forthcoming sections.\\
An independent constraint reduces the spin Hilbert space dimension by a factor two, which is indicated by a corresponding trace reduction. On the other hand, when a constraint is appended to the product of the projectors and the resulting space reduction is other than two, this new operator is dependent on the previous ones.\\
Relative dependencies between the constraints are inferred from study of various arrangements to impose them. If, for example, the last plaquette operator appended to the product provides no dimensional reduction whether it acts directly after all the other plaquette projectors, or after all of them and a Polyakov line projector, or at the very end -- then this plaquette projection operator is dependent on all the other plaquettes, but not on the Polyakov lines.

The equivalence between the spin matrices and Grassmann variables descriptions of the system is further tested by comparison of the eigenenergies obtained from numerical calculations and predictions from analytic formulas. The eigenenergies are found from diagonalization of the reduced Hamiltonian in the spin representation and then compared to formulas (\ref{r28}) or (\ref{may2a}). In addition, these data are supplemented by plots of energy levels and their degeneracy.
\section{Odd-odd lattice -- results for a $3\times3$ lattice}
\begin{table}[h!]
\begin{center}
\begin{tabular}{|p{3.98cm}|p{3.98cm}|}
\hline
Tr $\mathds{1}$&512\\ \hline
Tr $\Sigma_{11}$&256\\ \hline
Tr $\Sigma_{11}\Sigma_{12}$&128\\ \hline
Tr $\Sigma_{11}...\Sigma_{13}$&64\\ \hline
Tr $\Sigma_{11}...\Sigma_{21}$&32\\ \hline
Tr $\Sigma_{11}...\Sigma_{22}$&16\\ \hline
Tr $\Sigma_{11}...\Sigma_{23}$&8\\ \hline
Tr $\Sigma_{11}...\Sigma_{31}$&4\\ \hline
Tr $\Sigma_{11}...\Sigma_{32}$&2\\ \hline
Tr $\Sigma_{11}...\Sigma_{33}$&2\\ \hline
\end{tabular}
\\
\begin{tabular}{|p{2.7cm}|p{0.8cm}||p{2.7cm}|p{0.8cm}|}
\hline
Tr $\Sigma_{11}...\Sigma_{33}\Sigma_{x}$&1&Tr $\Sigma_{11}...\Sigma_{33}\Sigma_{y}$&1\\
\hline
Tr $\Sigma_{11}...\Sigma_{y}$&1 (0)&Tr $\Sigma_{11}...\Sigma_{x}$&1 (0)\\
\hline
\end{tabular}
\caption{Hilbert space reduction for all sectors $0\leq p\leq 9$ on a $3\times 3$ lattice. The last two rows of the table correspond to two distinct cases, when either the operator $\Sigma_x$ is included in the product of projection operators before $\Sigma_y$ (left part) or inversely (right part). The results are obtained for both types of boundary conditions -- periodic ($\epsilon_x=\epsilon_y=1$) and antiperiodic ($\epsilon_x=-1$, $\epsilon_y=1$). When the result depends on the choice of boundary conditions, traces of projection operators products for incorrect boundary conditions -- i.e. antiperiodic for odd $p$ or periodic for even $p$ -- are written in parentheses.}
\label{tab1}
\end{center}
\end{table}
\begin{figure}[h!]
\centering
\includegraphics{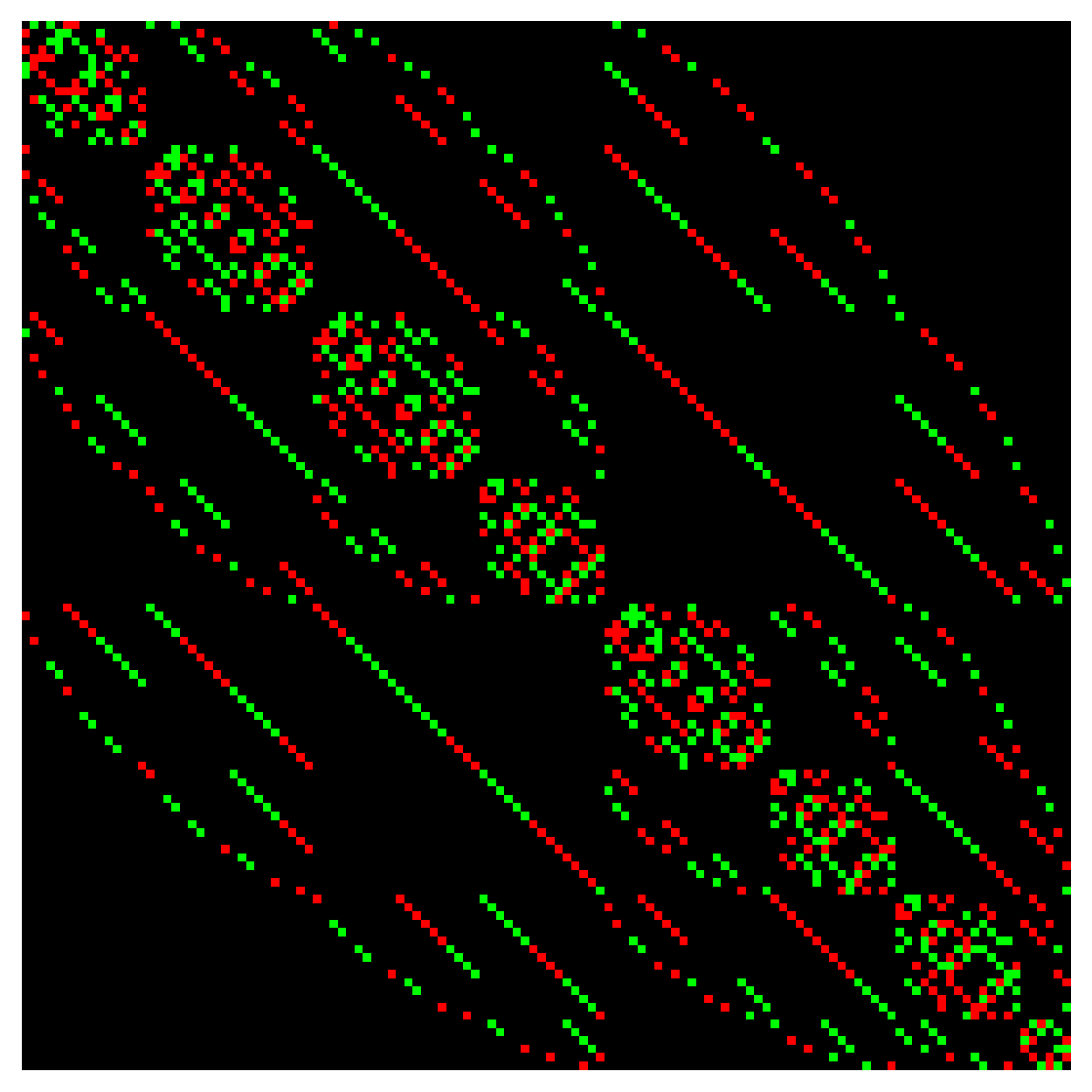}
\caption{The reduced Hamiltonian (\ref{may2f}) in the spin representation for the system of free fermions on a $3\times3$ lattice in $p=4$ sector. The green squares represent the matrix elements equal $i$, the red ones stand for $-i$ entries, while the black background depicts zeroes.}
\label{may10fig2}
\end{figure}
\begin{table}[h!]
\begin{center}
\begin{tabular}{|c||c|c|c|c|c|c|c|c|}
\hline
Eigenenergy&$p=1$&$p=2$&$p=3$&$p=4$&$p=5$&$p=6$&$p=7$&$p=8$\\ \hline
$-4\sqrt{3}$&0&0&1&3&3&1&0&0\\ \hline
$-3\sqrt{3}$&0&2&6&8&8&6&2&0\\ \hline
$-2\sqrt{3}$&1&4&10&17&17&10&4&1\\ \hline
$-\sqrt{3}$&2&8&16&22&22&16&8&2\\ \hline
$0$&3&8&18&26&26&18&8&3\\ \hline
$\sqrt{3}$&2&8&16&22&22&16&8&2\\ \hline
$2\sqrt{3}$&1&4&10&17&17&10&4&1\\ \hline
$3\sqrt{3}$&0&2&6&8&8&6&2&0\\ \hline
$4\sqrt{3}$&0&0&1&3&3&1&0&0\\ \hline
\end{tabular}
\caption{Spectrum of the Hamiltonian for all sectors $1\leq p\leq8$ on a $3\times 3$ lattice. The columns 2--9 contain degeneracies of corresponding eigenenergies.}
\label{tab4}
\end{center}
\end{table}
The results for a $3\times3$ lattice showing how the spin Hilbert space is reduced by application of the constraints are presented in Tab. \ref{tab1}. These calculations were performed in all $p$-particle sectors ($0\leq p\leq 9$) and for both periodic and antiperiodic boundary conditions. When the result depends on the choice of boundary conditions, the numbers obtained for incorrect boundary conditions are written in parentheses. The last two rows of the table describe space reduction in two distinct cases when either the operator $\Sigma_x$ is included in the product of projection operators before $\Sigma_y$ (the two columns to the left) or inversely (the two columns to the right).\\
A number of observations can be deduced from these data. First, they confirm prediction (\ref{eq34}) about the right choice of boundary conditions to reduce the dimension of the spin Hilbert space from $2^{\mathcal{N}}$ to $1$, but not to zero. Periodic boundary conditions are the proper choice for odd $p$ sectors, while antiperiodic ($\epsilon_x=-1$, $\epsilon_y=1$) boundary conditions lead to the correct space reduction in even sectors. This coincides with eq. (\ref{eq34}) when $L_x=3$, $L_y=3$ and $\eta=-1$.\\
A step-by-step examination of space reduction and calculations with different order of the projection operators enable analysis of their mutual relations. Eight plaquette projectors provide independent constraints. Each of them reduces the Hilbert space dimension by a factor 2. However, the last plaquette provides no additional space reduction. Therefore, it is dependent on the other plaquette projection operators and the constraint associated with this operator is redundant. The Polyakov line projector which is applied first -- whether it is $\Sigma_x$ or $\Sigma_y$ -- gives an independent constraint. Yet, the last one contributes to a redundant constraint when the boundary conditions are correct, and provides an inconsistent constraint otherwise. This leads to a conclusion that the two Polyakov lines are independent of the plaquettes, but are dependent on each other.\\
The reduced Hamiltonian (\ref{may2f}) in the spin representation is shown in Fig. \ref{may10fig2} for $p=4$ sector and in \emph{Appendix D} for the remaining $1\leq p\leq 8$ sectors. In these figures, the green squares represent the matrix elements equal $i$, the red ones stand for $-i$ entries and the black background depicts zeroes. Even in the reduced spin space, the matrices are still quite sparse. They are denser close to the main diagonal, which corresponds to the action of the Hamiltonian hopping terms between states with similar positioning of the particles on the lattice. Other structures, like off-diagonal parallel lines of non-zero entries are also observed. Energies of the system of fermions on the $3\times 3$ lattice, obtained from diagonalization of the reduced Hamiltonian, are presented in Tab. \ref{tab4}. The first column of the table lists all eigenenergies which occur in any $p$-particle sector, $1\leq p\leq 8$. Remaining columns contain degeneracies of corresponding eigenenergies in subsequent sectors. To further test equivalence between the spin matrices and the Grassmann variables descriptions of the system, one compares energy spectrum from Tab. \ref{tab4} to the analytic formula for eigenenergies (\ref{r28}), which was derived in fermionic representation. The results agree in all sectors.

\section{Even-odd lattice -- results for a $4\times3$ lattice}
\begin{table}[h!]
\begin{center}
\begin{tabular}{|p{6.73cm}|p{6.73cm}|}
\hline
Tr $\mathds{1}$&4096\\ \hline
Tr $\Sigma_{11}$&2048\\ \hline
Tr $\Sigma_{11}\Sigma_{21}$&1024\\ \hline
Tr $\Sigma_{11}...\Sigma_{31}$&512\\ \hline
Tr $\Sigma_{11}...\Sigma_{41}$&256\\ \hline
Tr $\Sigma_{11}...\Sigma_{12}$&128\\ \hline
Tr $\Sigma_{11}...\Sigma_{22}$&64\\ \hline
Tr $\Sigma_{11}...\Sigma_{32}$&32\\ \hline
Tr $\Sigma_{11}...\Sigma_{42}$&16\\ \hline
Tr $\Sigma_{11}...\Sigma_{13}$&8\\ \hline
Tr $\Sigma_{11}...\Sigma_{23}$&4\\ \hline
Tr $\Sigma_{11}...\Sigma_{33}$&2\\ \hline
\end{tabular}
\\
\begin{tabular}{|p{2cm}|p{0.65cm}||p{2cm}|p{0.65cm}||p{2cm}|p{0.65cm}||p{2cm}|p{0.65cm}|}
\hline
Tr $\Sigma_{11}...\Sigma_{y}$&1&Tr $\Sigma_{11}...\Sigma_{y}$&1&Tr $\Sigma_{11}...\Sigma_{43}$&2&Tr $\Sigma_{11}...\Sigma_{x}$&2(0)\\
\hline
Tr $\Sigma_{11}...\Sigma_{43}$&1&Tr $\Sigma_{11}...\Sigma_{x}$&1(0)&Tr $\Sigma_{11}...\Sigma_{x}$&2(0)&Tr $\Sigma_{11}...\Sigma_{43}$&2(0)\\
\hline
Tr $\Sigma_{11}...\Sigma_{x}$&1(0)&Tr $\Sigma_{11}...\Sigma_{43}$&1(0)&Tr $\Sigma_{11}...\Sigma_{y}$&1(0)&Tr $\Sigma_{11}...\Sigma_{y}$&1(0)\\
\hline
\end{tabular}
\caption{Hilbert space reduction on a $4\times 3$ lattice. The last three rows of the table correspond to four distinct cases, which differ by the order the operators $\Sigma_{43}$, $\Sigma_x$ and $\Sigma_y$ are included in the product of projection operators. The results are obtained for both types of boundary conditions -- periodic ($\epsilon_x=\epsilon_y=1$) and antiperiodic ($\epsilon_x=-1$, $\epsilon_y=1$). When the result depends on the choice of boundary conditions, traces of projection operators products for incorrect boundary conditions -- i.e. antiperiodic for odd $p$ or periodic for even $p$ -- are written in parentheses.}
\label{tab2}
\end{center}
\end{table}
\begin{figure}[h!]
\centering
\begin{subfigure}{0.4\textwidth}
\centering
\includegraphics[width=\textwidth]{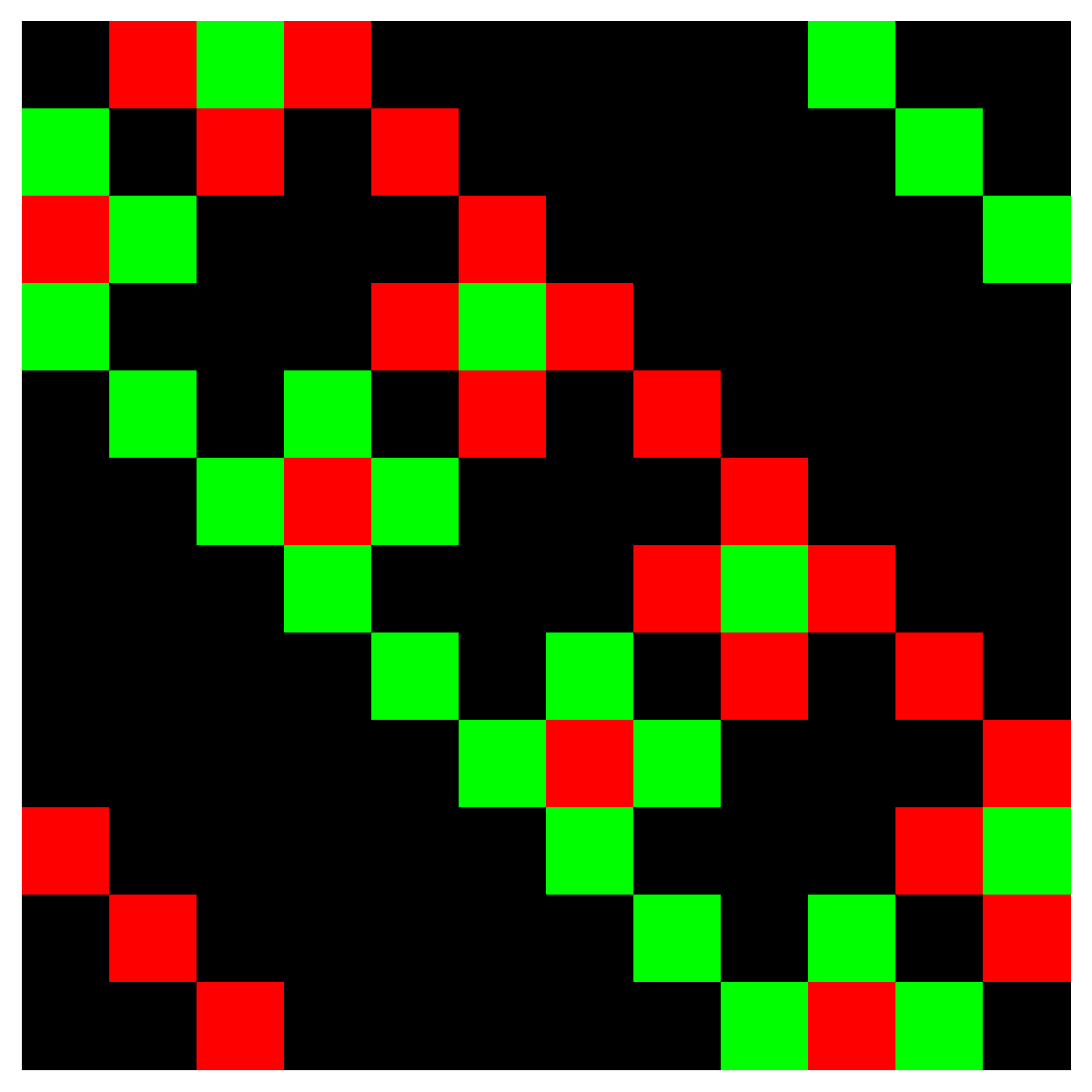}
\caption{}
\label{may10f10}
\end{subfigure}
\hfill
\begin{subfigure}{0.4\textwidth}
\centering
\includegraphics[width=\textwidth]{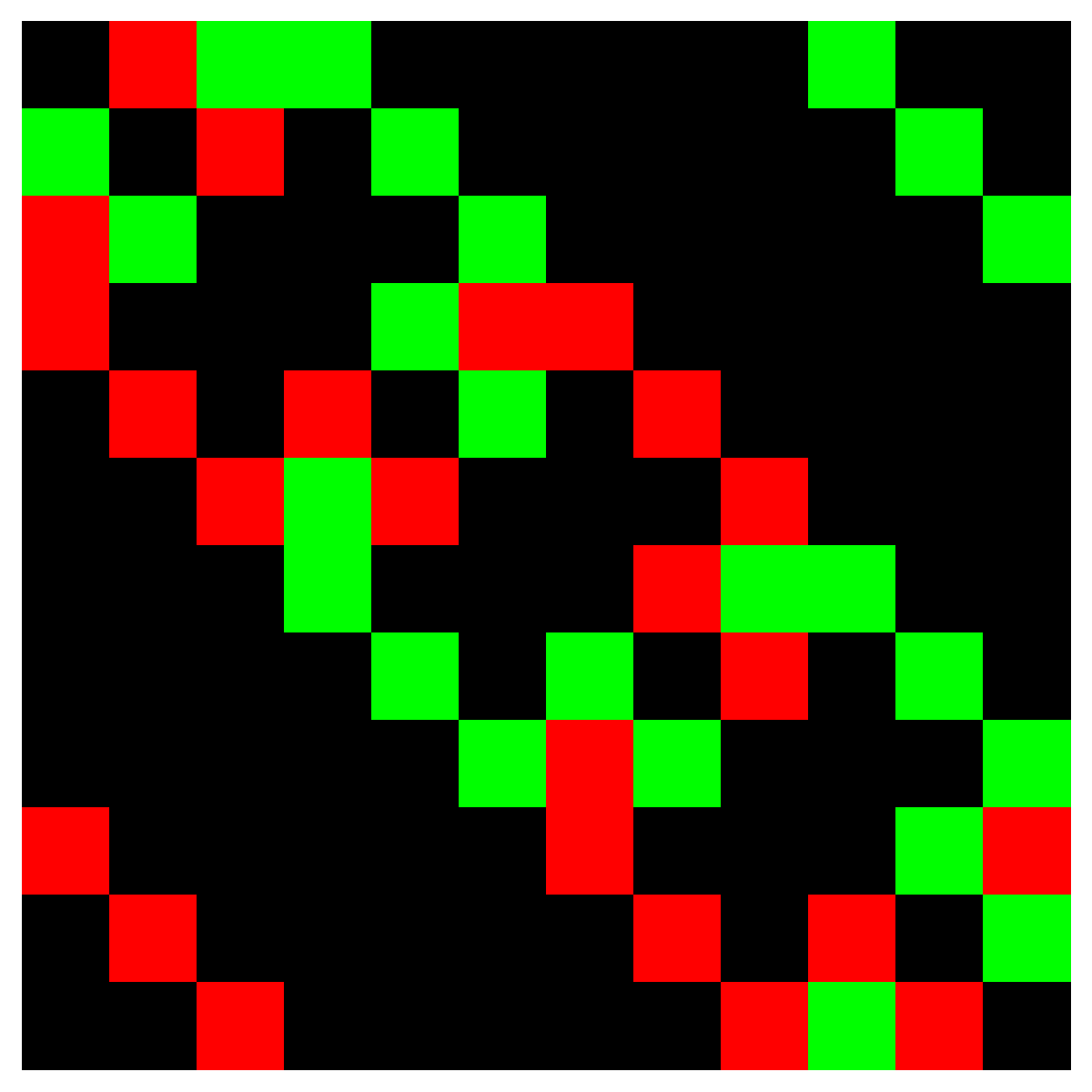}
\caption{}
\label{may10f11}
\end{subfigure}
\caption{The reduced Hamiltonian (\ref{may2f}) in the spin representation for the system of fermions on a $4\times3$ lattice in $p=1$ sector: (a) for free fermions, (b) in the constant magnetic field. The green squares represent the matrix elements equal $i$, the red ones stand for $-i$ entries, while the black background depicts zeroes.}
\label{may10fig4}
\end{figure}
\begin{table}
\begin{center}
\begin{tabular}{|l||*{3}{c|}}\hline
\backslashbox{$\alpha$}{$\beta$}
&\makebox[1em]{\textbf{-1}}&\makebox[1em]{\textbf{0}}&\makebox[1em]{\textbf{1}}\\\hline\hline
\textbf{-1} &1 & 1 & 1\\\hline
\textbf{0} &2 & 2 & 2\\\hline
\textbf{1} &1 & 1 & 1\\\hline
\end{tabular}
\caption{Histogram of eigenenergies for a system of free fermions on a $4\times 3$ lattice (in $p=1$ sector). The number on the intersection of $\alpha$ row and $\beta$ column is the degeneracy of the eigenenergy $E_{\alpha,\beta}=2\alpha+\sqrt{3}\beta$.}
\label{tab7}
\end{center}
\end{table}

\begin{table}
\begin{center}
\begin{tabular}{|l||*{5}{c|}}\hline
\backslashbox{$\alpha$}{$\beta$}
&\makebox[1em]{\textbf{-2}}&\makebox[1em]{\textbf{-1}}&\makebox[1em]{\textbf{0}}&\makebox[1em]{\textbf{1}}&\makebox[1em]{\textbf{2}}\\\hline\hline
\textbf{-2} &0 & 1 & 1 & 1 & 0\\\hline
\textbf{-1} &2 & 4 & 6 & 4 & 2\\\hline
\textbf{0} &2 & 6 & 8 & 6 & 2\\\hline
\textbf{1} &2 & 4 & 6 & 4 & 2\\\hline
\textbf{2} &0 & 1 & 1 & 1 & 0\\\hline
\end{tabular}
\caption{Histogram of eigenenergies for a system of free fermions on a $4\times 3$ lattice (in $p=2$ sector). The number on the intersection of $\alpha$ row and $\beta$ column is the degeneracy of the eigenenergy $E_{\alpha,\beta}=2\alpha+\sqrt{3}\beta$.}
\label{tab8}
\end{center}
\end{table}

\begin{figure}[h!]
\begin{center}
\includegraphics[width=12cm]{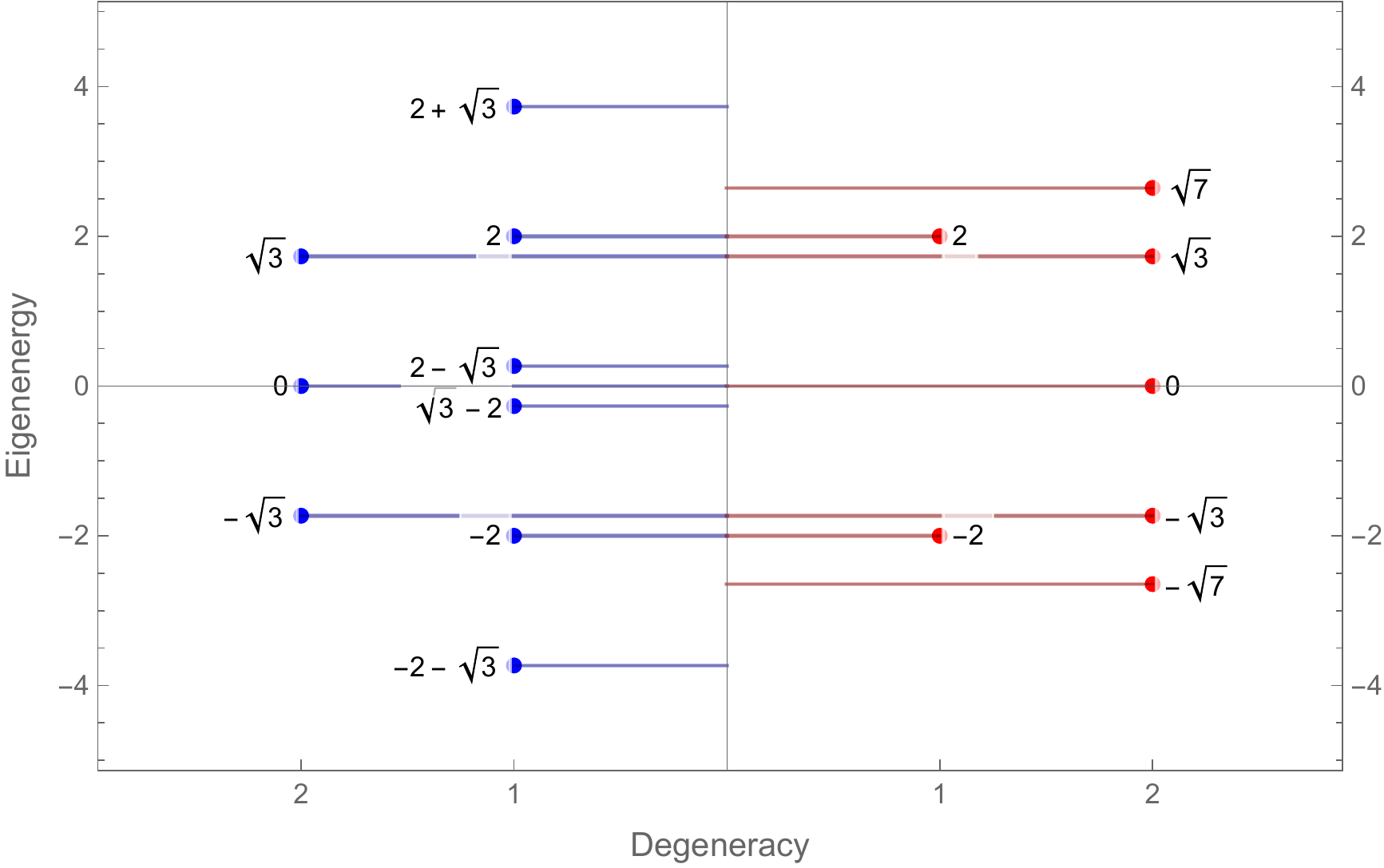}
\caption{
Comparison of the energy levels of a system of fermions on a $4\times 3$ lattice in $p=1$ sector for free fermions (left) and in the constant magnetic field (right). The degeneracies of the eigenenergies are indicated by the lengths of the corresponding lines.
}
\label{fig4}
\end{center}
\end{figure}

\begin{figure}[h!]
\begin{center}
\includegraphics[width=12cm]{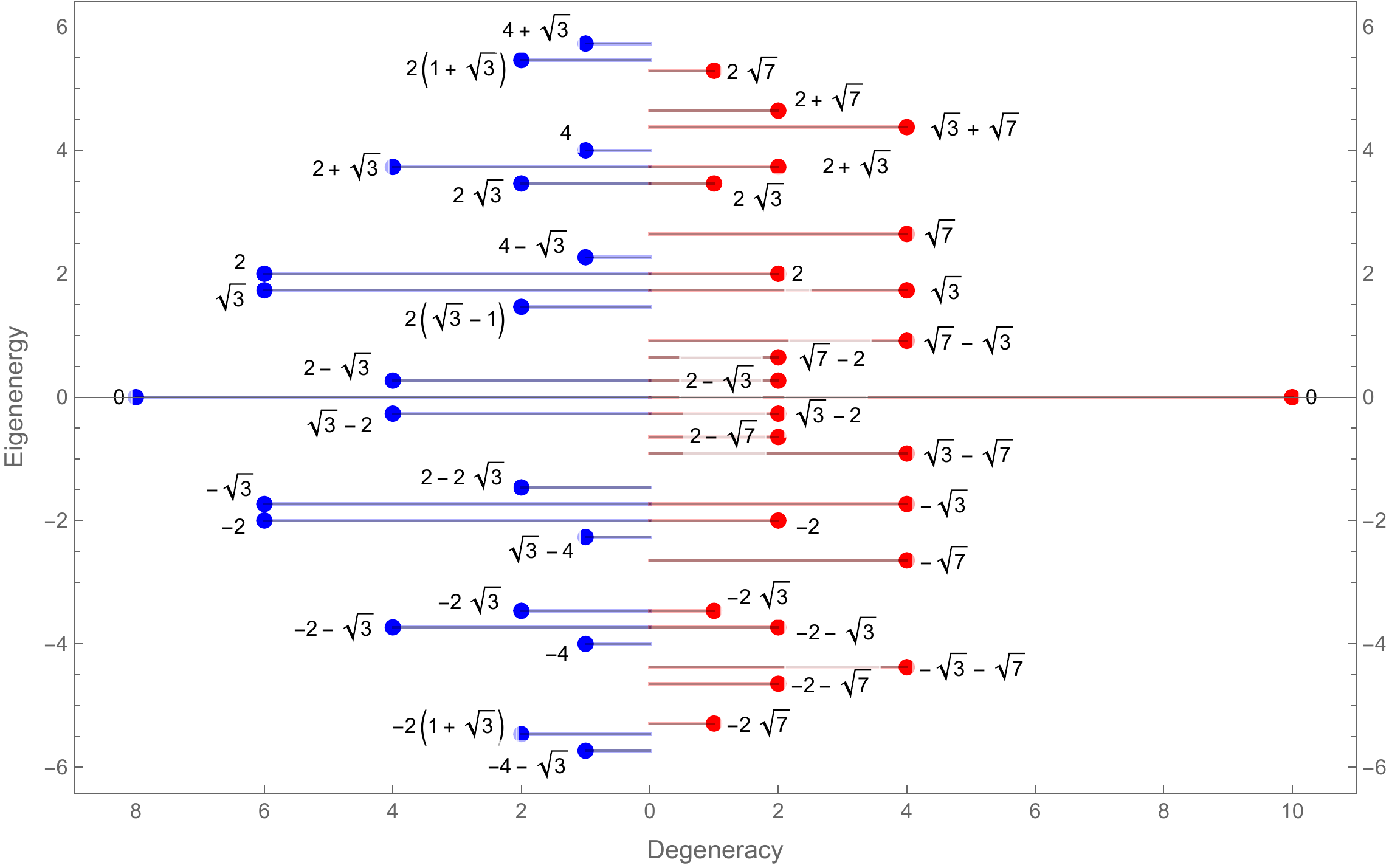}
\caption{
Comparison of the energy levels of a system of fermions on a $4\times 3$ lattice in $p=2$ sector for free fermions (left) and in the constant magnetic field (right). The degeneracies of the eigenenergies are indicated by the lengths of the corresponding lines.}
\label{fig5}
\end{center}
\end{figure}

\begin{figure}
\centering
\includegraphics[width=12cm]{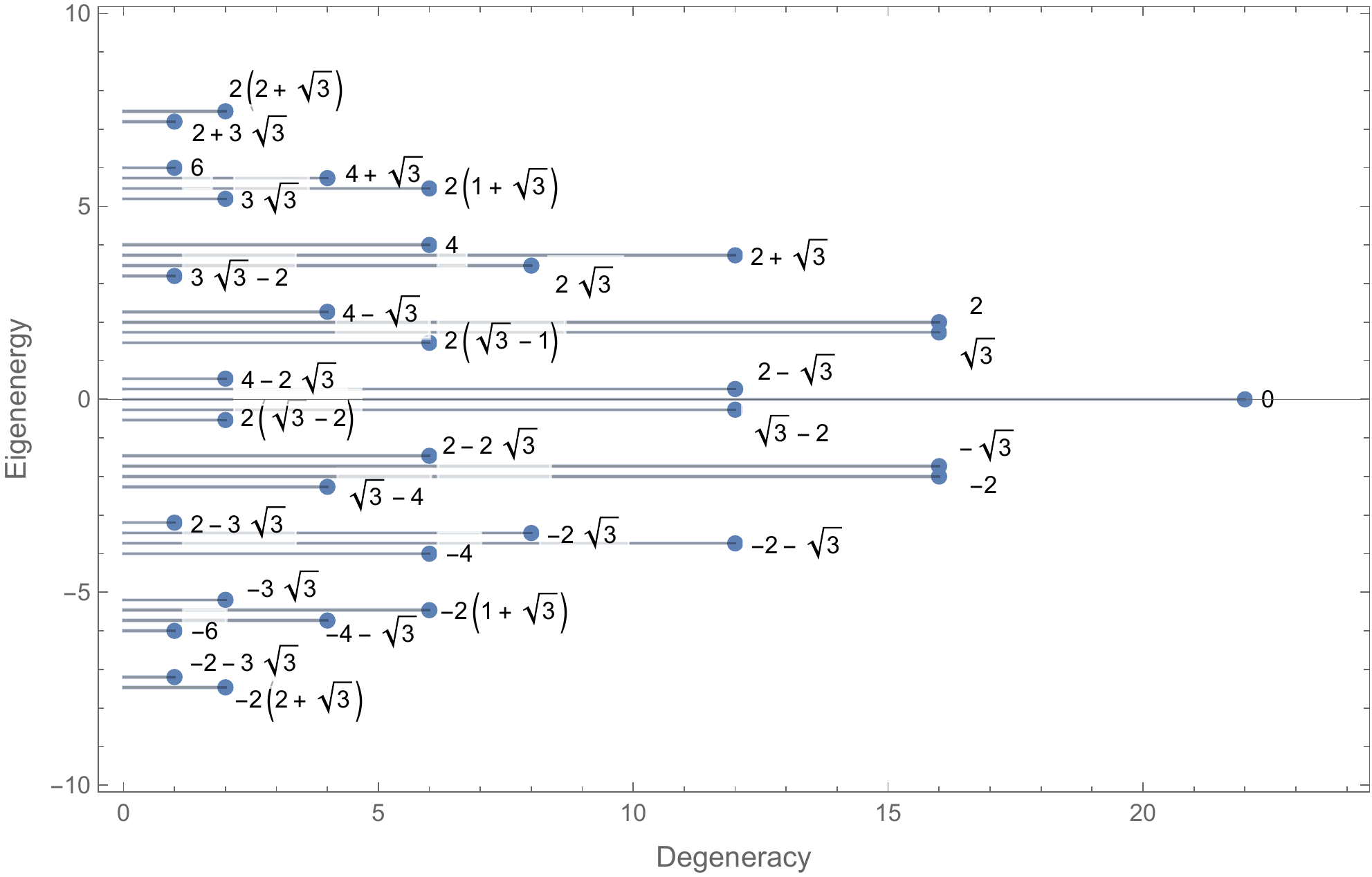}
\caption{The energy levels of a free fermions system on a $4\times 3$ lattice in $p=3$ sector.}
\label{figmay15}
\end{figure}

\begin{table}
\begin{center}
\begin{tabular}{|l||*{7}{c|}}\hline
\backslashbox{$\alpha$}{$\beta$}
&\makebox[1em]{\textbf{-3}}&\makebox[1em]{\textbf{-2}}&\makebox[1em]{\textbf{-1}}&\makebox[1em]{\textbf{0}}&\makebox[1em]{\textbf{1}}&\makebox[1em]{\textbf{2}}&\makebox[1em]{\textbf{3}}\\\hline\hline
\textbf{-3} &0 & 0 & 0 & 1 & 0 & 0 & 0\\\hline
\textbf{-2} &0 & 2 & 4 & 6 & 4 & 2 & 0\\\hline
\textbf{-1} &1 & 6 & 12 & 16 & 12 & 6 & 1\\\hline
\textbf{0} &2 & 8 & 16 & 22 & 16 & 8 & 2\\\hline
\textbf{1} &1 & 6 & 12 & 16 & 12 & 6 & 1\\\hline
\textbf{2} &0 & 2 & 4 & 6 & 4 & 2 & 0\\\hline
\textbf{3} &0 & 0 & 0 & 1 & 0 & 0 & 0\\\hline
\end{tabular}
\caption{Histogram of eigenenergies for a system of free fermions on a $4\times 3$ lattice (in $p=3$ sector). The number on the intersection of $\alpha$ row and $\beta$ column is the degeneracy of the eigenenergy $E_{\alpha,\beta}=2\alpha+\sqrt{3}\beta$.}
\label{tab9}
\end{center}
\end{table}
The results of Hilbert space reduction for a $4\times 3$ lattice are presented in Tab. \ref{tab2}. Similarly to the previous case, periodic boundary conditions are the correct option in odd $p$-particle sectors and to obtain the right space reduction in even sectors one has to use antiperiodic ($\epsilon_x=-1$, $\epsilon_y=1$) boundary conditions. The results for incorrect boundary conditions are written in parentheses. The last three rows of the table correspond to different orders the last plaquette projector $\Sigma_{43}$ and the projection operators associated with Polyakov lines are applied.\\
Eleven plaquette projection operators provide independent constraints, while the last plaquette depends on them. It contributes to a redundant constraint even when it acts directly after $\Sigma_{33}$. The operator $\Sigma_y$ does not depend on the other projectors. It reduces the Hilbert space dimension by a factor 2 even when applied as the last one. On the contrary, the other Polyakov line projector $\Sigma_x$ is dependent on the other projection operators. When the boundary conditions are correct, it provides no additional constraint whether it acts before both the last plaquette and $\Sigma_y$, after any of them or both. Furthermore, for the wrong boundary conditions, the constraint from this Polyakov line is inconsistent with the other constraints and reduces the space dimension to zero.\\
The reduced Hamiltonian of the system of free fermions on a $4\times 3$ lattice is presented in Fig. \ref{may10f10}. The eigenenergies are computed by diagonalization of this Hamiltonian. The results are shown in Tab. \ref{tab7} for sector $p=1$, in Tab. \ref{tab8} for sector $p=2$ and in Tab. \ref{tab9} for $p=3$. Since the eigenenergies are linear combinations with integer coefficients of $2$ and $\sqrt{3}$ in this case, degeneracies of the eigenenergies are presented in tables, whose rows and columns are labeled by those coefficients. The number on the intersection of $\alpha$ row and $\beta$ column is the degeneracy of the eigenenergy $E_{\alpha,\beta}=2\alpha+\sqrt{3}\beta$. The energy levels are also depicted in Fig. \ref{fig4} ($p=1$), Fig. \ref{fig5} ($p=2$) and Fig. \ref{figmay15} ($p=3$). Positions of these lines along energy axis inform about eigenenergies of those states, while their lengths along horizontal axis provides information on degeneracies of corresponding eigenenergies.\\
As a lattice with at least one even dimension, the $4\times 3$ lattice is a good starting candidate to test the predictions about the system of fermions in an external constant magnetic field. Its reduced Hamiltonian in $p=1$ sector is shown in Fig. \ref{may10f11}. The energy levels of fermions in the magnetic field found from diagonalization in spin representation are compared to the eigenenergies in the free fermions case in Fig. \ref{fig4} for sector $p=1$ and in Fig. \ref{fig5} for $p=2$. Both for the free system and in the magnetic field, the energy spectra from Tabs. \ref{tab7}-\ref{tab9} and Figs. \ref{fig4}-\ref{figmay15} agree with fermionic analytic formulas (\ref{r28}) and (\ref{may2a}).
\clearpage
\section{Even-even lattice -- results for a $4\times 4$ lattice}
\begin{table}[h!]
\begin{center}
\begin{tabular}{|p{3.98cm}|p{3.98cm}|}
\hline
Tr $\mathds{1}$&65536\\ \hline
Tr $\Sigma_{11}$&32768\\ \hline
Tr $\Sigma_{11}\Sigma_{21}$&16384\\ \hline
Tr $\Sigma_{11}...\Sigma_{31}$&8192\\ \hline
Tr $\Sigma_{11}...\Sigma_{41}$&4096\\ \hline
Tr $\Sigma_{11}...\Sigma_{12}$&2048\\ \hline
Tr $\Sigma_{11}...\Sigma_{22}$&1024\\ \hline
Tr $\Sigma_{11}...\Sigma_{32}$&512\\ \hline
Tr $\Sigma_{11}...\Sigma_{42}$&256\\ \hline
Tr $\Sigma_{11}...\Sigma_{13}$&128\\ \hline
Tr $\Sigma_{11}...\Sigma_{23}$&64\\ \hline
Tr $\Sigma_{11}...\Sigma_{33}$&32\\ \hline
Tr $\Sigma_{11}...\Sigma_{43}$&16\\ \hline
Tr $\Sigma_{11}...\Sigma_{14}$&8\\ \hline
Tr $\Sigma_{11}...\Sigma_{24}$&4\\ \hline
\end{tabular}
\\
\begin{tabular}{|p{2.7cm}|p{0.8cm}||p{2.7cm}|p{0.8cm}|}
\hline
Tr $\Sigma_{11}..\Sigma_{24}\Sigma_{34}$&4 (0)&Tr $\Sigma_{11}...\Sigma_{24}\Sigma_{x}$&2\\
\hline
Tr $\Sigma_{11}...\Sigma_{44}$&4 (0)&Tr $\Sigma_{11}...\Sigma_{y}$&1\\
\hline
Tr $\Sigma_{11}...\Sigma_{x}$&2 (0)&Tr $\Sigma_{11}...\Sigma_{34}$&1 (0)\\
\hline
Tr $\Sigma_{11}...\Sigma_{y}$&1 (0)&Tr $\Sigma_{11}...\Sigma_{44}$&1 (0)\\
\hline
\end{tabular}
\caption{Hilbert space reduction on a $4\times 4$ lattice. The last four rows of the table correspond to two distinct cases, which differ by the order the operators $\Sigma_{34}$, $\Sigma_{44}$, $\Sigma_x$ and $\Sigma_y$ are included in the product of projection operators. In sectors with odd $p$ the condition (\ref{eq34}) is not met neither for periodic or antiperiodic boundary conditions. Traces of projection operators products for odd $p$ are written in parentheses when they differ from the results for even $p$.}
\label{tab3}
\end{center}
\end{table}
It is a straightforward observation from eq. (\ref{eq34}) that, for an even-even lattice, Hilbert space reduction cannot be "fixed" in the sectors of wrong parity by switching from periodic boundary conditions to antiperiodic. Indeed, when $L_x=L_y=4$, $(-\epsilon_y^\prime/\epsilon_y)^{L_x}=(-\epsilon_y^\prime/\epsilon_y)^{4}$ and $(-\epsilon_x^\prime/\epsilon_x)^{L_y}=(-\epsilon_x^\prime/\epsilon_x)^{4}$ are both equal to $1$ whether periodic or antiperiodic boundary conditions are selected. Taking into account also that $\eta^{L_xL_y}=\eta^{16}=1$, the RHS of eq. (\ref{eq34}) equals 1 independently on the choice of boundary conditions. The condition (\ref{eq34}) is therefore satisfied only when $p$ is even -- only even $p$ sectors are accessible in the case of a $4\times 4$ lattice. The results showing how the spin space is reduced by application of the constraints are shown in Tab. \ref{tab3}. The numbers obtained for odd sectors are written in parentheses when they differ from results in even sectors. The last four rows of the table correspond to cases when either the last two plaquette projectors $\Sigma_{34}$ and $\Sigma_{44}$ act before the Polyakov line projectors or the order of these pairs of operators is inversed. In both cases, Polyakov lines provide independent constraints, while the last two plaquettes are dependent on the fourteen remaining plaquettes. In even sectors, the plaquette projectors $\Sigma_{34}$ and $\Sigma_{44}$ contribute to redundant constraints, which is not affected by the order these two plaquettes and the Polyakov lines are applied. In odd sectors, when the condition (\ref{eq34}) is not met, the Hilbert space dimension is reduced to zero once $\Sigma_{34}$ or $\Sigma_{44}$ is applied.
\clearpage

\chapter{Summary}

The introduction of the basis in the spin representation along with the construction of the projector operators and the Hamiltonian in this basis, which were presented in Section 3, show how the transformation proposed in this work can be applied to obtain observables in the spin picture. Explicit construction of the constraints in the basis allows one to solve them and, once the basis vectors of the reduced spin Hilbert space are found, the spin Hamiltonian is expressed in this basis and diagonalized. This goal -- finding the exact solution of the constraints -- is the main motivation and achievement of this work. Application of the algorithms and programs proposed in this thesis enables this task, which used to be unaccesible due to its complexity. This step makes the prescription \cite{w,n15} complete and suitable for actual physics problems. It also provides a direct check of the equivalence between the fermionic and spin descriptions by comparison of the eigenenergies obtained in both representations, which was carried out in Section 5.
\vspace{0.5cm}

At the beginning of this thesis, the proposal of a transformation from fermionic operators to spin matrices was introduced. The description of the system of fermions occupying the sites of a rectangular lattice in the Grassmann and spin representations was presented in Section~2. In particular, the scheme to assign fermionic/spin operators to lattice sites was explained. The necessary quantities in both representations were constructed and discussed -- including the Clifford variables (\ref{eq15}), (\ref{eq20}) and the link operators (\ref{eq16}), (\ref{eq21}). The algebra of the link operators in the fermionic picture was determined (\ref{eq16x})-(\ref{eq16z}), (\ref{eq21x})-(\ref{eq21z}) and the equivalent spin description was found by the requirement to preserve this algebra. The set of spin matrices, which provide the equivalent description, was established and the Hamiltonian and the particle number operator were expressed in both representations. In one dimension, it was checked in Section 2.1 that this proposal yields the same result as the well-known Jordan-Wigner transformation \cite{jw}, while not being restricted to $d=1$, but having a straightforward generalization to $d>1$.
\vspace{0.5cm}

The simple fermions studied in this work have 2-dimensional Hilbert spaces while in $d=2$ one employs 4-dimensional generalized Euclidean Dirac matrices for the equivalent description in the spin picture. Therefore, the proposed transformation from the Grassmann to spin variables results in an increased number of degrees of freedom -- from $2^{\mathcal{N}}$ to $4^{\mathcal{N}}$, where $\mathcal{N}$ is the total number of lattice sites. Therefore, the full product space, obtained as a tensor product of spin spaces associated with individual lattice sites, needs to be constrained in order to obtain the physical Hilbert space in the spin representation. Once these constraints are imposed, the two descriptions are indeed equivalent. Since finding the reduced spin Hilbert space constitutes a substantial stage of the whole procedure, the constraints were paid particular attention. \vspace{0.2cm}\\
First, the full set of necessary constraints, including both the plaquette operators $P_{f,s}(\vec{n})$ (\ref{eq25}) and the Polyakov lines $\mathcal{L}_{x,y}(n_{y,x})$ (\ref{eq33}), was determined in both representations in Section 2.3. The requirement that the identity $P_f(\vec{n})=1$, which is true in the Grassmann representation (\ref{eq26}), is satisfied by its spin counterpart $P_s(\vec{n})$, amounts to the set of $\mathcal{N}$ constraints, which commute among themselves (\ref{eq31s1}). Seemingly a set of $\mathcal{N}$ constraints, which reduce the spin space dimension by a factor two each, would be exactly the number of constraints necessary to reduce the dimension of the space from $4^{\mathcal{N}}$ to $2^{\mathcal{N}}$. Yet, not all of them are independent. Because the product of all spin-picture plaquettes is identically the unity operator, only $\mathcal{N}-1$ plaquettes are independent. Furthermore, when both lattice dimensions $L_x$ and $L_y$ are even, there are only $\mathcal{N}-2$ independent plaquettes \cite{brww}. Thus, the set of plaquette constraints is insufficient to fully reduce the spin space and it was expanded by the Polyakov line constraints. Since the product of any two parallel Polyakov lines and all the plaquettes in the rectangle spanned by those lines is identically the unity operator, there are at most two independent Polyakov lines -- "horizontal" $\mathcal{L}_x$ and "vertical" $\mathcal{L}_y$. They can be chosen arbitrarily from the families $\mathcal{L}_x(n_y)$ or $\mathcal{L}_y(n_x)$ of equivalent choices and all other Polyakov lines can be obtained by multiplication of $\mathcal{L}_x$ or $\mathcal{L}_y$ by plaquettes. The two Polyakov lines provide independent constraints on even$\times$even lattices, while they are related when either $L_x$ or $L_y$ is odd (see the commentary after eq. (\ref{extra3}) and Appendix C). It is easily seen that in both cases the set of plaquette constraints and Polyakov line constraints provides the necessary number of $\mathcal{N}$ independent conditions, which are used to project the full spin space onto its physical subspace.\vspace{0.2cm}\\
The proper boundary conditions for the fermionic operators and the spin-like variables on finite-size lattices were established in Section 2.1.2 for 1-dimensional lattices and in Appendix C in the 2-dimensional case. In $d=1$, the boundary conditions for the fermionic variables and for the spin matrices $\sigma^1$, $\sigma^2$ are the same (both periodic or both antiperiodic) when the total number of particles on the lattice $p$ is odd, while they must be taken opposite in the even $p$ sectors. In $d=2$, the correct boundary conditions, i.e. the ones which result in a non-trivial solution of the constraints, are determined by the condition $(-1)^p=\eta^{L_xL_y}(-\epsilon_y^\prime/\epsilon_y)^{L_x}(-\epsilon_x^\prime/\epsilon_x)^{L_y}$ -- where $L_x$ and $L_y$ are the dimensions of the rectangular lattice, $\epsilon_x^\prime$ and $\epsilon_y^\prime$ are the boundary condition coefficients in the spin representation in directions $x$ and $y$ respectively, similarly $\epsilon_x$, $\epsilon_y$ are the boundary conditions for the Grassmann variables and $\eta=-1$.
\vspace{0.5cm}

Apart from the free fermions case, the method to introduce the interaction with the external $\mathbb{Z}_2$ field $U(\vec{n},\vec{n}+\hat{e})$ was also proposed in Section 2.4. In the fermionic picture, the field $U$ is assigned to the links of the lattice as shown in the Hamiltonian (\ref{s01}). Since the field attributes an additional factor $\prod_{l\in C(\vec{n})}U(l)$ to the fermionic plaquette operator $P_f(\vec{n})$ associated with the plaquette $C(\vec{n})$, the constraints $P_s(\vec{n})=1$ need to be replaced by $P_s(\vec{n})=\prod_{l\in C(\vec{n})}U(l)\equiv B(\vec{n})$ in order to project the full spin space onto the physical, reduced Hilbert space and obtain the equivalent spin description. $B(\vec{n})$ is understood as a $\mathbb{Z}_2$ magnetic field, which resides on the lattice faces. Similarly to the constraints (\ref{intr31d}), which arise in the Chen-Kapustin-Radicevic bosonization proposal \cite{n5}, the constraints $P_s(\vec{n})=B(\vec{n})$ can be interpreted as a lattice version of a modified Gauss law for the bosonic system. Particular attention was paid to the case of a constant magnetic field ($B(\vec{n})=-1$), for which the Hamiltonians were derived and their eigenvalues were found, and shown to agree, in both fermionic and spin pictures.
\vspace{0.5cm}

To approach the challenging task to solve the constraints and find the exact form of the basis vectors of the reduced spin Hilbert space, a suitable basis was proposed along with a number of observations and ideas which improve the efficiency of the programs used in their study. The crucial observation follows from the fact that the Hamiltonian commutes with the particle number operator and thus the $p$-particle sectors are invariant subspaces of the Hamiltonian. The Hamiltonian is therefore block-diagonal in bases arranged primarily according to the particle number or, in other words, this allows to solve the problem separately in every $p$-particle sector. Furthermore, there exist in addition "fixed coordinates" subsectors -- the subspaces of the $p$-particle sectors which correspond to states with fixed positions of the particles. They are invariant subspaces of the constraints, which enables to solve the constraints independently in every subsector of fixed coordinates of $p$ particles. Consequently, the dimension of matrices involved in the computations are yet further reduced. Considering these facts, the basis (\ref{eq38}) was chosen for the calculations in the spin representation. Its basis vectors are tensor products of the eigenvectors of $\Gamma^5$ matrices associated with all the lattice sites and they are arranged primarily by the particle number $p$, and secondly by the subsectors. The basis spin states were labeled with $L_x\times L_y$ matrices $(s_{ij})$ containing labels $\{1,2,3,4\}$ of four eigenvectors of $\Gamma^5(n_{ij})$ -- see eqs. (\ref{eq38})-(\ref{eq39}). A detailed scheme to implement this basis in Wolfram Mathematica programs was explained in Section 3.1 and the listing is presented in Appendix B.
\vspace{0.5cm}

The plaquette and Polyakov line projection operators were constructed in this basis independently within subsectors. Such an approach reduces the dimensions of the projection operator matrices from $2^{\mathcal{N}}\binom{\mathcal{N}}{p}\times 2^{\mathcal{N}}\binom{\mathcal{N}}{p}$ to $2^{\mathcal{N}}\times 2^{\mathcal{N}}$. Yet, working with such still huge matrices requires a thoughtful approach -- the methods and tricks used to decrease the computational complexity of the programs as much as possible are briefly reminded in this paragraph. Because imposing all the constraints reduces the spin space dimension by a factor $2^{\mathcal{N}}$, the subsectors are projected onto 1-dimensional spaces by the projection operators. This fact greatly simplifies finding the basis vectors of the reduced spin Hilbert space as, since there is only one eigenvector of the product of all the projection operators (\ref{s20new}) per subsector, it can be obtained by methods less difficult than a straightforward diagonalization of eq. (\ref{s20new}) -- e.g. by acting with this operator on simple test vectors, as discussed in Section 3.2. Another simplification arises from the fact that the plaquettes and Polyakov lines acting on a basis state $\ket{ii}$ yield another basis state $\ket{jj}$. In other words, the matrices of the constraints $\braket{jj|P_s,\ \mathcal{L}_{x,y}|ii}$ have only one non-zero element per column in this basis. To take advantage of this factor in the proposed algorithm to construct the constraints matrices, the method to determine the proper $\ket{jj}$ for every $\ket{ii}$ was designed, which allows one to skip the vanishing matrix elements. Also, due to the fact that these matrices consist mostly of vanishing elements, built-in Wolfram Mathematica functions and methods for sparse matrices were used in the programs to further improve their efficiency. On the other hand, the "fixed coordinate" subsectors are not invariant subspaces of the Hamiltonian. Thus, in Section 3.3, it was constructed in rectangular $2^{\mathcal{N}}\binom{\mathcal{N}}{p}\times 2^{\mathcal{N}}$ blocks, which were then used to obtain the $\binom{\mathcal{N}}{p}\times\binom{\mathcal{N}}{p}$ "reduced" Hamiltonian -- i.e. the Hamiltonian expressed in the basis of the $\binom{\mathcal{N}}{p}$-dimensional reduced spin Hilbert space describing the $p$-particle sector. After that, the reduced Hamiltonian is diagonalized to find the eigenenergies. Apart from the fact that calculations in the full $p$-particle sector are needed, these computations are quite similar to the ones associated with the constraints though, so other simplifications mentioned above still apply.
\vspace{0.5cm}

The analytic formulas for the eigenenergies were derived in the Grassmann representation in Section 4 to compare their predictions with the numerical results obtained in the spin representation. In the free fermions case, the Fourier transform to the momentum space makes the Hamiltonian diagonal, which yields the formulas (\ref{r21}) and (\ref{r28}). These equations are linear in sines of momenta. For fermions in a constant magnetic field, the Hamiltonian in the momentum space is not diagonal, yet the Fourier transform takes it to the "almost diagonal" form (\ref{s8}). It was shown in Section 4.3.2 that this Hamiltonian is block-diagonal in a lexicographically ordered (\ref{o2}) momentum basis (\ref{o1}) (see Fig. \ref{figblock}). The final formula (\ref{s19}) was derived by analysis of the characteristic polynomial. In this case, the energy-momentum dispersion relation behaves as $\sqrt{\sin(q_x)^2+\sin(q_y)^2}$.
\vspace{0.5cm}

The constraints were constructed in the basis as discussed above and analyzed with the Wolfram Mathematica programs for lattice sizes $3\times3$, $4\times3$ and $4\times4$. The results of the computations are presented and discussed in Section 5. To understand how the projection operators reduce the spin space and determine the relations between them, they were imposed in various orders and the number of imposed constraints was increased gradually from $1$ to $\mathcal{N}+2$. The dimension of the image of these projection operators sets, i.e. the dimension of the partially reduced spin space, was determined by calculating the trace of the projection operators products (\ref{s21new}). The prediction that a single plaquette or Polyakov line projection operator reduces the dimension of the spin space by a factor two was confirmed. When different orders of applying the constraints were studied, it was concluded that the constraints which do not reduce further the spin space dimension are dependent on the preceding ones. These observations confirm the predictions from \cite{brww} on the number of independent plaquette constraints and Polyakov lines, which are listed in the end of Section 2.3. Examination, how the spin Hilbert space is reduced by the constraints, also verifies the relation (\ref{eq34}), which specifies the boundary conditions used on the lattice. In every subsector, the single basis vector of the reduced spin Hilbert space was obtained by acting with the product of $\mathcal{N}$ independent projection operators on test vectors. Once the full basis was established, the reduced spin Hamiltonian was constructed and diagonalized. It was verified that the eigenenergies obtained in the spin picture agree with the formulas from the fermionic representation (\ref{r28}) and (\ref{s19}), which provides an additional check of the equivalence between the two descriptions.
\vspace{0.5cm}

We would like to finish this section by providing a couple of suggestions how the topic covered in this work could be further studied. One idea to continue this research, which is an important and natural next step, is to find an analytic formula for the basis states of the reduced spin Hilbert space:
\begin{equation}
\ket{r}=\sum_s r_sv_s=\sum_s r_s\bigotimes_{1\leq i\leq L_x,\ 1\leq j\leq L_y}e_{s_{ij}},
\label{sum1}
\end{equation}
where $\ket{r}$ is a basis vector of the reduced spin Hilbert space associated with some subsector and the sum goes over the spin configurations matching this subsector, which are labeled by the matrices $(s_{ij})$ (see eq. (\ref{eq38})). $v_s$ are basis vectors of the full spin space (before the reduction), $e_{s_{ij}}$ are the eigenvectors of $\Gamma^5$ matrices assigned to the corresponding lattice sites and the coefficients $r_s$ are functions of $r$ and $s$, which one aims to determine. The two major tasks encountered in the numerical computation of the eigenenergies in the spin picture are to obtain the basis vectors of the reduced spin Hilbert space and to express the Hamiltonian in this basis. Having an exact formula like (\ref{sum1}), one does not need to approach the first of these steps numerically, while also the latter one could be significantly simplified -- e.g. in situations when the functions $r_s$ are simple, a closed form expression for the Hamiltonian matrix elements $\braket{t|H_s|r}$ might be found.
One way to approach this problem is to study a system of fermions/spins on a $2\times 2$ lattice. For such a miniature lattice, the dimension of the unconstrained spin space is $4^4=256$, while the dimension of the subsectors is $2^4=16$.
The algebraic expressions are much simpler in this case compared to the larger lattices and the matrices have sizes suitable to view their whole structure at once -- which gives the best chances to observe some regularities and obtain the analytic formula.\\
Another interesting area of research is the application of the transformation studied in this work to physics problems to which the Jordan-Wigner transformation is a vital step. Due to the limitations of the Jordan-Wigner transformation such problems are restricted to $d=1$ spatial dimension. Thus, switching from the Grassmann to spin variables in the way described in this work provides an opportunity to extend their scope. This class of problems includes study of discretized field theories. A good example is the tensor network approach to find the phase structure of the massive Thirring model \cite{thirring}:
\begin{equation}
S_{\text{Th}}[\psi,\bar{\psi}]=\int d^2x\left[ \bar{\psi}i\gamma^\mu\partial_\mu\psi-m\bar{\psi}\psi-\frac{g}{2}(\bar{\psi}\gamma_\mu\psi)(\bar{\psi}\gamma^\mu\psi) \right],
\label{sum2}
\end{equation}
which is proposed in \cite{cichy}. In this paper, the model (\ref{sum2}) is expressed in Hamiltonian formulation and then discretized to obtain a model of fermions on the lattice. After that, the Jordan-Wigner transformation is employed to formulate the problem as an equivalent quantum spin chain model, which is handled with the tensor network methods. Similarly, study of the Hamiltonians occurring in the quantum computing models, mentioned in Section 1.2.4, might benefit from application of the $d>1$ dimensional transformation from Grassmann to spin-like variables.\\
Last but not least, the approach to express the projection operators in the basis to solve the constraints, which was discussed in Section 3, might be reproduced for other bosonization schemes. A recommended choice to begin with is the versatile Chen-Kapustin-Radi\v{c}evi\'{c} proposal \cite{n5}, discussed in Section 1.2.3.1. In other words, one might try to apply the techniques learned in the case of the constraints (\ref{may2}) for the purpose of the conditions (\ref{intr31d}).

\clearpage
\chapter*{Acknowledgments}
\thispagestyle{empty}
I would like to thank 
prof. Jacek Wosiek and dr hab. Piotr Korcyl for all the invaluable help with creation of this thesis and sharing their expert knowledge with me. 

\noindent
I thank my colleagues from the Institute of Physics, Arek Bochniak and Błażej Ruba for numerous discussions.\\

\noindent
This work was supported by the NCN grant: UMO2016/21/B/ST2/01492.
\newpage
\begin{appendices}
\chapter{Derivation of Pauli matrix identities}
Using Pauli matrices commutation relations:\begin{equation}
\sigma^+\sigma^-=\frac{1}{4}(\sigma^1+i\sigma^2)(\sigma^1-i\sigma^2)=\frac{1}{4}\left((\sigma^1)^2+(\sigma^2)^2-i[\sigma^1,\sigma^2]\right)=\frac{1}{4}(2+2\sigma^3)=\frac{1+\sigma^3}{2},
\label{apa1}
\end{equation}
\begin{equation}
\sigma^-\sigma^+=\frac{1}{4}(\sigma^1-i\sigma^2)(\sigma^1+i\sigma^2)=\frac{1}{4}\left((\sigma^1)^2+(\sigma^2)^2+i[\sigma^1,\sigma^2]\right)=\frac{1-\sigma^3}{2}.
\label{apa2}
\end{equation}
$\sigma^3\sigma^\pm=-\sigma^\pm\sigma^3$, because different Pauli matrices anticommute while $\sigma^\pm$ are linear combinations of $\sigma^1$ and $\sigma^2$. Therefore:
\begin{equation}
\sigma^3\sigma^-=-\sigma^-\sigma^3=-\frac{1}{2}\left(\sigma^1-i\sigma^2\right)\sigma^3=\frac{1}{2}\left(i\sigma^2\sigma^3-\sigma^1\sigma^3\right)=\frac{1}{2}\left(-\sigma^1+i\sigma^2\right)=-\sigma^-.
\label{apa3}
\end{equation}
Similarly:
\begin{equation}
\sigma^3\sigma^+=-\sigma^+\sigma^3=-\frac{1}{2}\left(\sigma^1+i\sigma^2\right)\sigma^3=-\frac{1}{2}\left(-i\sigma^2-\sigma^1\right)=\sigma^+.
\label{apa4}
\end{equation}
Operators (\ref{apa1}) and (\ref{apa2}) are projection operators, indeed:
\begin{equation}
\left(\frac{1+\sigma^3}{2}\right)^2=\frac{1+1+2\sigma^3}{4}=\frac{1+\sigma^3}{2},
\label{apa5}
\end{equation}
\begin{equation}
\left(\frac{1-\sigma^3}{2}\right)^2=\frac{1+1-2\sigma^3}{4}=\frac{1-\sigma^3}{2}.
\label{apa6}
\end{equation}
This fact is useful in derivation of the following formula:
\begin{multline}
e^{i\pi\sigma^+\sigma^-}=e^{i\pi\frac{1+\sigma^3}{2}}=\left(1+\sum_{k=1}^{\infty}\frac{(i\pi)^k}{k!}\left(\frac{1+\sigma^3}{2}\right)^k\right)=\left(1+\sum_{k=1}^{\infty}\frac{(i\pi)^k}{k!}\frac{1+\sigma^3}{2}\right)=\\
=\left(1+(e^{i\pi}-1)\frac{1+\sigma^3}{2}\right)=1-(1+\sigma^3)=-\sigma^3.
\label{apa7}
\end{multline}
Thus, in particular:
\begin{equation}
\sigma^\pm e^{i\pi\sigma^+\sigma^-}=-\sigma^\pm\sigma^3=\sigma^3\sigma^\pm=-e^{i\pi\sigma^+\sigma^-}\sigma^\pm.
\label{apa8}
\end{equation}
Under Jordan-Wigner transformation the anticommutator of creation and annihilation fermionic operators transforms to:
\begin{multline}
\{\phi(n)^\dag,\phi(m)\}\rightarrow\{\sigma^+(n)\prod_{j=1}^{n-1}e^{-i\pi\sigma^+(j)\sigma^-(j)},\left(\prod_{k=1}^{m-1}e^{i\pi\sigma^+(k)\sigma^-(k)}\right)\sigma^-(m)\}=\\
=\sigma^+(n)\left(\prod_{j=1}^{n-1}e^{-i\pi\sigma^+(j)\sigma^-(j)}\right)\left(\prod_{k=1}^{m-1}e^{i\pi\sigma^+(k)\sigma^-(k)}\right)\sigma^-(m)+\\
+\left(\prod_{k=1}^{m-1}e^{i\pi\sigma^+(k)\sigma^-(k)}\right)\sigma^-(m)\sigma^+(n)\left(\prod_{j=1}^{n-1}e^{-i\pi\sigma^+(j)\sigma^-(j)}\right).
\label{apa9}
\end{multline}
Operator $\sigma^+(n)$ commutes with the product of operators to its left because all operators $\sigma^\pm(j)$ which occur in this product are associated with other lattice sites, $j\neq n$. Similarly, $\sigma^-(m)$ commutes with the product to its left. Therefore:
\begin{multline}
\{\phi(n)^\dag,\phi(m)\}\rightarrow \sigma^+(n)\left(\prod_{j=1}^{n-1}e^{-i\pi\sigma^+(j)\sigma^-(j)}\right)\left(\prod_{k=1}^{m-1}e^{i\pi\sigma^+(k)\sigma^-(k)}\right)\sigma^-(m)+\\
+\sigma^-(m)\left(\prod_{k=1}^{m-1}e^{i\pi\sigma^+(k)\sigma^-(k)}\right)\left(\prod_{j=1}^{n-1}e^{-i\pi\sigma^+(j)\sigma^-(j)}\right)\sigma^+(n)=\\
=\sigma^+(n)\Omega(m,n)\sigma^-(m)+\sigma^-(m)\Omega(m,n)\sigma^+(n),
\label{apa10}
\end{multline}
where $\Omega(m,n)$ is a product of operators $e^{\pm i\pi\sigma^+(l)\sigma^-(l)}$ for $l$ between $m$ and $n$:
\begin{equation}
\Omega(m,n)=\prod_{l=\min(m,n)}^{\max(m,n)-1}e^{\text{sgn}(m-n)i\pi\sigma^+(l)\sigma^-(l)}.
\label{apa11}
\end{equation}
In particular
\begin{equation}
\Omega(m,n)=1,
\label{apa12}
\end{equation}
thus
\begin{equation}
\{\phi(m)^\dag,\phi(m)\}\rightarrow\sigma^+(m)\sigma^-(m)+\sigma^-(m)\sigma^+(m)=\frac{1+\sigma^3(m)}{2}+\frac{1-\sigma^3(m)}{2}=1.
\label{apa13}
\end{equation}
When $m\neq n$, exactly one is true -- either term with $l=m$ occurs in the product (\ref{apa11}) or the one with $l=n$. Therefore, using also eq. (\ref{apa8}), either $\Omega(m,n)\sigma^-(m)=+\sigma^-(m)\Omega(m,n)$ and $\Omega(m,n)\sigma^+(n)=-\sigma^+(n)\Omega(m,n)$ or $\Omega(m,n)\sigma^-(m)=-\sigma^-(m)\Omega(m,n)$ while $\Omega(m,n)\sigma^+(n)=+\sigma^+(n)\Omega(m,n)$. Thus, for $m\neq n$, eq. (\ref{apa10} becomes):
\begin{equation}
\{\phi(n)^\dag,\phi(m)\}\rightarrow\pm\left(\sigma^+(n)\sigma^-(m)-\sigma^-(m)\sigma^+(n)\right)\Omega(m,n)=0.
\label{apa14}
\end{equation}
Eqs. (\ref{apa13}) and (\ref{apa14}) finally yield the desired result:
\begin{equation}
\{\phi(n)^\dag,\phi(m)\}\rightarrow \{\sigma^+(n)\prod_{j=1}^{n-1}e^{-i\pi\sigma^+(j)\sigma^-(j)},\left(\prod_{k=1}^{m-1}e^{i\pi\sigma^+(k)\sigma^-(k)}\right)\sigma^-(m)\}=\delta_{mn}.
\end{equation}
\clearpage

\chapter{
Construction of bases and constraints in Wolfram Mathematica
}
\begin{lstlisting}[frame=single, numbers=left, numberstyle=\scriptsize,breaklines=true,basicstyle=\ttfamily,firstnumber=1]
L=3;
Nc=L^2;
ns0=2^Nc;

SetBasis[sector_]:=Module[{kk,in,ix,di,listpocc,ii,l,k,nosectors,nssec,perms},
nosectors=Binomial[Nc,sector];
nssec=ns0*nosectors;
in=Table[0,{Nc}];
ix=Table[0,{Nc}];
di=Table[0,{Nc}];
stp=Table[{},{nssec}];
JJ=Table[0,{2},{2},{2},
           {2},{2},{2},
           {2},{2},{2},{nosectors}];
perms=Permutations@Join[Table[a,{Nc-sector}],Table[b,{sector}]];
pos=Table[Flatten@Position[perms[[i]],b],{i,1,Length[perms]}];
ii=0;
Do[{
Do[{in[[rr]]=2;ix[[rr]]=3;di[[rr]]=1},{rr,1,Nc}];
listpocc=pos[[pocc]];
Do[{in[[listpocc[[rr]]]]=1;ix[[listpocc[[rr]]]]=4;di[[listpocc[[rr]]]]=3},{rr,1,sector}];
Do[{ii=ii+1;stp[[ii]]={{k[1],k[2],k[3]},
                       {k[4],k[5],k[6]},
                       {k[7],k[8],k[9]}};
Do[{l[kk]=k[kk]-1},{kk,1,Nc}];
Do[l[listpocc[[rr]]]=(k[listpocc[[rr]]]-1)/3+1,{rr,1,sector}];
JJ[[l[1],l[2],l[3],l[4],l[5],l[6],l[7],l[8],l[9],pocc]]=ii;   
   },
{k[1],in[[1]],ix[[1]],di[[1]]},{k[2],in[[2]],ix[[2]],di[[2]]},
{k[3],in[[3]],ix[[3]],di[[3]]},{k[4],in[[4]],ix[[4]],di[[4]]},
{k[5],in[[5]],ix[[5]],di[[5]]},{k[6],in[[6]],ix[[6]],di[[6]]},
{k[7],in[[7]],ix[[7]],di[[7]]},{k[8],in[[8]],ix[[8]],di[[8]]},
{k[9],in[[9]],ix[[9]],di[[9]]}
]}
,{pocc,1,nosectors}];

FromState[s_] :=
(*for a given state s it returns a list listpocc with indices of occupied sites and a linear index jj (in the whole p-particle subspace) of the state s*)
  Module[{tsts, listpocc, jj, l, pocc},
   tsts = Flatten[s];
   listpocc=Sort@Flatten@Join[Position[tsts,1],Position[tsts,4]];
   Do[l[kk] = tsts[[kk]] - 1, {kk, 1, Nc}];
   Do[l[listpocc[[rr]]]=(tsts[[listpocc[[rr]]]-1)/3+1,{rr,1, 
     Length[listpocc]}];
   pocc = Position[pos, listpocc][[1, 1]];
   jj = JJ[[l[1],l[2],l[3],l[4],l[5],l[6],l[7],l[8],l[9],pocc]];
   Return[{listpocc, jj}]
   ];

GetEv[r_] := Module[{stR, tt, i, j, st, jj, jnd, ind},
   stR = Table[stp[[ii]], {ii, ns0*(r - 1) + 1, ns0*r}];   
   JG32 = {1, 2, 3, 4}; MG32 = {-I, I, -I, I};
   JG43 = {4, 3, 2, 1}; MG43 = {-1, -1, 1, 1};
   JG14 = {1, 2, 3, 4}; MG14 = {-I, -I, I, I};
   JG12 = {4, 3, 2, 1}; MG12 = {-1, 1, -1, 1};   
   JG31 = {4, 3, 2, 1}; MG31 = {-I, -I, -I, -I};
   JG13 = {4, 3, 2, 1}; MG13 = {I, I, I, I};
   JG42 = {4, 3, 2, 1}; MG42 = {-I, I, I, -I};
   JG24 = {4, 3, 2, 1}; MG24 = {I, -I, -I, I};
   PMS11 = SparseArray[{{1, 1} -> 0}, ns0];
   PMS12 = SparseArray[{{1, 1} -> 0}, ns0];
   PMS13 = SparseArray[{{1, 1} -> 0}, ns0];
   PMS21 = SparseArray[{{1, 1} -> 0}, ns0];
   PMS22 = SparseArray[{{1, 1} -> 0}, ns0];
   PMS23 = SparseArray[{{1, 1} -> 0}, ns0];
   PMS31 = SparseArray[{{1, 1} -> 0}, ns0];
   PMS32 = SparseArray[{{1, 1} -> 0}, ns0];
   PMS33 = SparseArray[{{1, 1} -> 0}, ns0];
   LSX = SparseArray[{{1, 1} -> 0}, ns0];
   LSY = SparseArray[{{1, 1} -> 0}, ns0];
   JId = {1, 2, 3, 4};
   Do[{  st = stR[[ii]];
(*this loop is used to construct plaquette located at (1,1) in spin representation*)
     i[1, 3]=st[[1, 3]]; i[2, 3]=st[[2, 3]]; i[3, 3]=st[[3, 3]];
     i[1, 2]=st[[1, 2]]; i[2, 2]=st[[2, 2]]; i[3, 2]=st[[3, 2]];
     i[1, 1]=st[[1, 1]]; i[2, 1]=st[[2, 1]]; i[3, 1]=st[[3, 1]];
j[1,3]=JId[[i[1,3]]];j[2,3]= JId[[i[2,3]]];j[3,3]=JId[[i[3,3]]];  
j[1,2]=JG14[[i[1,2]]];j[2,2]=JG43[[i[2,2]]];j[3,2]=JId[[i[3,2]]];  
j[1,1]=JG12[[i[1,1]]];j[2,1]=JG32[[i[2,1]]];j[3,1]=JId[[i[3,1]]];        
     tt = Table[j[k1, k2], {k1, 1, L}, {k2, 1, L}];
     jj = FromState[tt][[2]] - (r - 1)*ns0;          
     PMS11[[jj, ii]] = 
     MG12[[i[1,1]]] MG32[[i[2,1]]] MG43[[i[2,2]]] MG14[[i[1,2]]]; 
     }, {ii, 1, ns0}];

(*we skip analogous loops for remaining 8 plaquettes in this listing for brevity, but keep the loops for Polyakov lines:*)

   Do[{  st = stR[[ii]];     
       i[1,3]= st[[1,3]]; i[2,3]= st[[2,3]]; i[3,3]= st[[3,3]];
       i[1,2]= st[[1,2]]; i[2,2] = st[[2,2]]; i[3,2] = st[[3,2]];
       i[1,1]= st[[1,1]]; i[2,1] = st[[2,1]]; i[3,1] = st[[3,1]];     
 j[1,3]=JId[[i[1,3]]];j[2,3]=JId[[i[2,3]]];j[3,3]=JId[[i[3,3]]];  
 j[1,2]=JId[[i[1,2]]];j[2,2]=JId[[i[2,2]]];j[3,2]=JId[[i[3,2]]];  
 j[1,1]=JG13[[i[1,1]]];j[2,1]=JG31[[i[2,1]]];j[3,1]=JG31[[i[3,1]]];     
     tt = Table[j[k1, k2], {k1, 1, L}, {k2, 1, L}];
     jj = FromState[tt][[2]] - (r - 1)*ns0;      
     LSX[[jj,ii]]=I*MG13[[i[1,1]]] MG31[[i[2,1]]] MG31[[i[3,1]]];      
     }, {ii, 1, ns0}];
   
   Do[{  st = stR[[ii]];     
       i[1,3]= st[[1,3]];i[2,3]= st[[2,3]];i[3,3]= st[[3,3]];
       i[1,2]= st[[1,2]];i[2,2]= st[[2,2]];i[3,2]= st[[3,2]];
       i[1,1]= st[[1,1]];i[2,1]= st[[2,1]];i[3,1]= st[[3,1]];     
 j[1,3]=JG42[[i[1,3]]];j[2,3]=JId[[i[2,3]]];j[3,3]=JId[[i[3,3]]];
 j[1,2]=JG42[[i[1,2]]];j[2,2]=JId[[i[2,2]]];j[3,2]=JId[[i[3,2]]]; 
 j[1,1]=JG24[[i[1,1]]];j[2,1]=JId[[i[2,1]]];j[3,1]=JId[[i[3,1]]];     
     tt = Table[j[k1, k2], {k1, 1, L}, {k2, 1, L}];
     jj = FromState[tt][[2]] - (r - 1)*ns0;      
     LSY[[jj,ii]]=I* MG24[[i[1,1]]] MG42[[i[1,2]]] MG42[[i[1,3]]];      
     }, {ii, 1, ns0}];
   Jd = SparseArray[{{1, 1} -> 0}, ns0];
   Do[Jd[[ii, ii]] = 1, {ii, 1, ns0}];
   Q = Table[{}, {Nc + 2}];
   Q[[1]] = (Jd + PMS11)/2;
   Q[[2]] = (Jd + PMS21)/2;
   Q[[3]] = (Jd + PMS31)/2;
   Q[[4]] = (Jd + PMS12)/2;
   Q[[5]] = (Jd + PMS22)/2;
   Q[[6]] = (Jd + PMS32)/2;
   Q[[7]] = (Jd + PMS13)/2;
   Q[[8]] = (Jd + PMS23)/2;
   Q[[9]] = (Jd + PMS33)/2;
   Q[[10]] = (Jd + LSX)/2;
   Q[[11]] = (Jd + LSY)/2;
   ind = 1;
   While[ind <= ns0,
    v0 = Table[0, {ns0}]; v0[[ind]] = 1;
    jnd = 1;
    While[jnd <= 11,
     v0 = Q[[jnd]].v0;
     If[Sum[Abs[v0[[ff]]], {ff, 1, ns0}] == 0, Break[]];
     jnd++
     ];
    If[jnd == 12, Break[]];
    ind++
    ] ; (*the output that matters most is v0 - the eigenvector and ind - index of the proper basis vector to obtain the eigenvector by projections*)
   Print["ind=", ind]
   ];
\end{lstlisting}

\clearpage

\chapter{Derivation of relations between the constraints
}

Since the total number of plaquettes plus one horizontal $\mathcal{L}_x$ and one vertical $\mathcal{L}_y$ Polyakov lines is $\mathcal{N}+2$, while the required number of independent constraints is $\mathcal{N}$, one expects some of them are related to the other constraints. These relations are briefly listed in the end of Section 2.3. In this section, their origin is discussed more thoroughly. The first identity, which enables expressing one plaquette in terms of the rest, is:
\begin{equation}
\prod_{\vec{n}}P_s(\vec{n})=1.
\label{cceq1}
\end{equation}
It is a straightforward consequence of the fact that $(\Gamma^k)^2=1$ and applies to any dimensions of the rectangular lattice. Therefore, there is always no more than $\mathcal{N}-1$ independent plaquette constraints in $d=2$. To understand the relation of $\mathcal{L}_x$ and $\mathcal{L}_y$, let us remind eq. (\ref{april3}), which can be written as:
\begin{equation}
\Gamma^5(\vec{n})=\eta(1-2N_s(\vec{n}))=\eta(-1)^{N_s(\vec{n})}.
\label{cceq1b}
\end{equation}
From the definitions of the Polyakov lines (\ref{eq33}), the boundary conditions (\ref{eq19half}) and the above equation one obtains the formula, which links the product of all the Polyakov lines to the spin boundary conditions, the lattice dimensions and the particle number operator $N_s$:
\begin{equation}
\prod_{n_y=1}^{L_y}\mathcal{L}_x(n_y)\prod_{n_x=1}^{L_x}\mathcal{L}_y(n_x)=(-\epsilon_x^\prime)^{L_y}(-\epsilon_y^\prime)^{L_x}\prod_{\vec{n}}-\Gamma^5(\vec{n})=(-\epsilon_x^\prime)^{L_y}(-\epsilon_y^\prime)^{L_x}(-\eta)^{L_x L_y}(-1)^{N_s}.
\label{cceq2}
\end{equation}
The boundary conditions coefficients enter this equation because the last link operator in every Polyakov line connects the sites on the opposite borders of the lattice, i.e. it includes a term $\Gamma^k(L_e+1)=\epsilon_e^\prime\Gamma^k(1)$. The minus signs in the middle equation arise due to anticommutations necessary to combine the Gamma matrices from the Polyakov lines into $\Gamma^1\Gamma^2\Gamma^3\Gamma^4=\Gamma^5$. The latter equality results directly from eq. (\ref{cceq1b}). Hence, within a $p$-particle sector:
\begin{equation}
\prod_{n_y=1}^{L_y}\mathcal{L}_x(n_y)\prod_{n_x=1}^{L_x}\mathcal{L}_y(n_x)=(-\epsilon_x^\prime)^{L_y}(-\epsilon_y^\prime)^{L_x}(-\eta)^{L_x L_y}(-1)^{p}.
\label{cceq2half}
\end{equation}
As observed in Section 2.3, all Polyakov lines oriented along $x$-axis can be obtained by multiplication of an arbitrary $x$-oriented Polyakov line $\mathcal{L}_x$ by appropriate plaquettes. Analogous fact occurs for $y$-direction and a chosen Polyakov line $\mathcal{L}_y$. Therefore, on the subspace determined by a choice of the values of the plaquette operators, the left hand side of the above equation is proportional to $\mathcal{L}_x^{L_y}\mathcal{L}_y^{L_x}$. When $L_x$ ($L_y$) is odd and $L_y$ ($L_x$) is even, the sign of $\mathcal{L}_y$ ($\mathcal{L}_x$) is thus fixed by eq. (\ref{cceq2half}). When both $L_x$ and $L_y$ are odd, the values of the two Polyakov lines are not fixed by this equation, but they are related to each other. In both cases the Polyakov lines provide only a single independent constraint. Only on even$\times$even lattices $\mathcal{L}_x$ and $\mathcal{L}_y$ provide two independent constraints.

Apart from the observation that there is only one independent Polyakov line constraint when $L_x$ or $L_y$ is odd, the equation (\ref{cceq2half}) also provides a way to set up the correct boundary conditions. Using the formulas (\ref{wombat3}) and (\ref{wombat4}), one easily finds the spin picture equivalent of the product (\ref{cceq2half}):
\begin{equation}
\prod_{n_y=1}^{L_y}\mathcal{L}_x^f(n_y)\prod_{n_x=1}^{L_x}\mathcal{L}_y^f(n_x)=(i^{L_x}\epsilon_x)^{L_y}(i^{L_y}\epsilon_y)^{L_x}=(-1)^{L_xL_y}\epsilon_x^{L_y}\epsilon_y^{L_x}.
\label{cceq5}
\end{equation}
Equating the right hand sides of the formulas in the two representations one obtains the desired equation involving the boundary conditions: 
\begin{equation}
(-1)^p=\eta^{L_xL_y}\left(-\frac{\epsilon_x^\prime}{\epsilon_x}\right)^{L_y}\left(-\frac{\epsilon_y^\prime}{\epsilon_y}\right)^{L_x}.
\label{cceq3}
\end{equation}

The identities (\ref{cceq1}) and (\ref{cceq2}) indicate that on odd$\times$odd and even$\times$odd lattices there is one independent Polyakov line and $\mathcal{N}-1$ independent plaquettes. The last remaining fact from Section 2.3 about relations between the constraints, which is yet to be proven, is therefore the statement that only $\mathcal{N}-2$ plaquettes are independent on even$\times$even lattices. On lattices with both dimensions even the plaquettes can be arranged into two sets characterized by the parity of $n_x+n_y$, where $\vec{n}$ is the location of the left-lower corner of the plaquette, and every spin variable $\Gamma^k(\vec{m})$ occurs in each set exactly once. Such a partition is simple to imagine as a chessboard pattern. Thus:
\begin{equation}
\prod_{\vec{n}\in \Omega}P_s(\vec{n})=(-1)^{N_s},
\label{cceq4}
\end{equation}
where $\Omega$ is the set which consists only of plaquettes with $n_x+n_y$ even or only odd. This identity follows directly from eq. (\ref{cceq1b}) and $\Gamma^4\Gamma^3\Gamma^1\Gamma^2=\Gamma^1\Gamma^4\Gamma^3\Gamma^2=-\Gamma^5$. The two equations (\ref{cceq4}) and eq. (\ref{cceq1}) form a set of three identities, out of which two are independent -- thus limiting the number of independent plaquettes on even$\times$even lattices to $\mathcal{N}-2$.
\clearpage

\chapter{Hamiltonian matrices in $3\times3$ case}
\begin{figure}[h!]
\centering
\includegraphics[width=10.5cm]{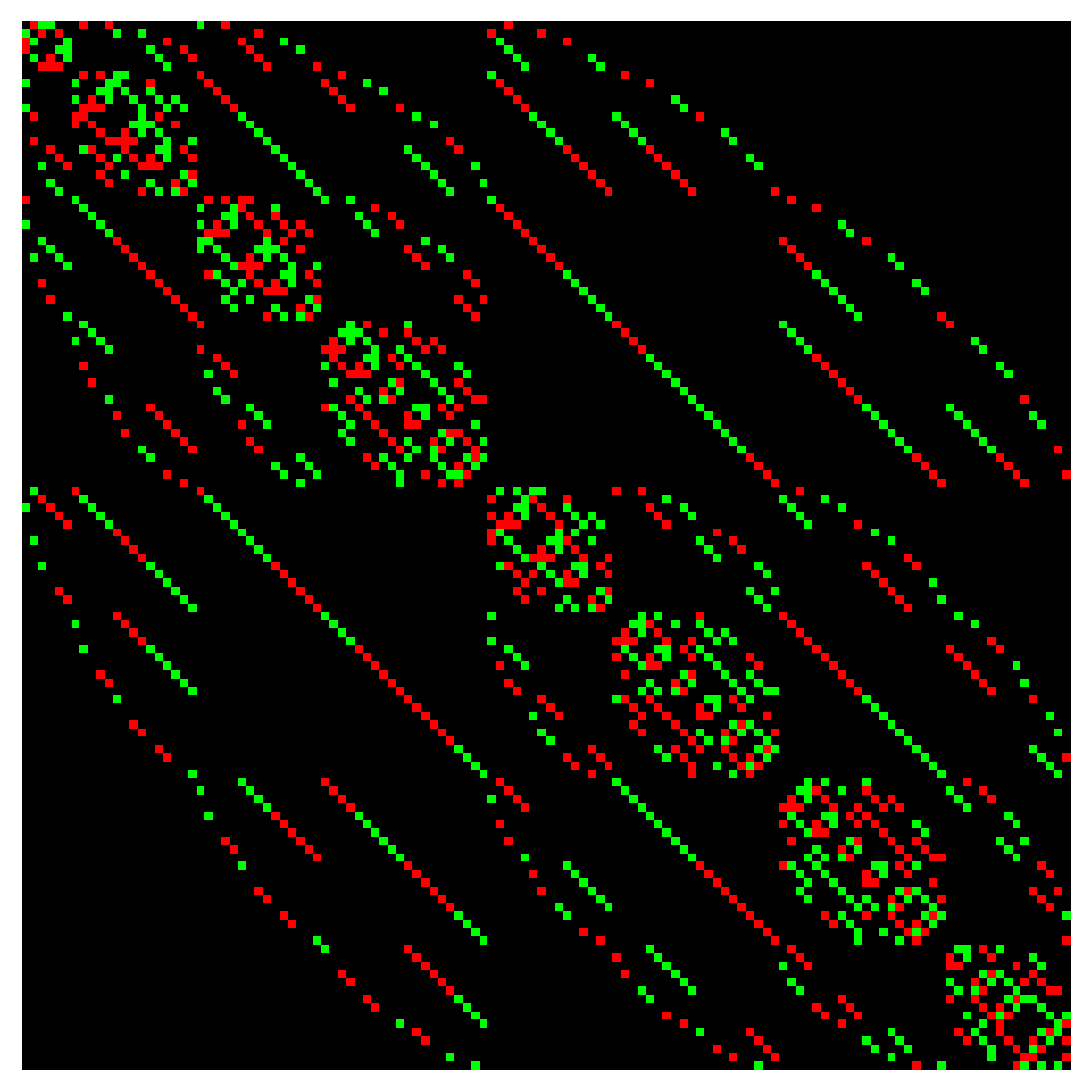}
\caption{The reduced Hamiltonian (\ref{may2f}) in the spin representation for the system of free fermions on a $3\times3$ lattice in $p=5$ sector. The green squares represent the matrix elements equal $i$, the red ones stand for $-i$ entries, while the black background depicts zeroes.}
\label{may10fig3}
\end{figure}

\begin{figure}
\centering
\begin{subfigure}[b]{0.3\textwidth}
\centering
\includegraphics[width=\textwidth]{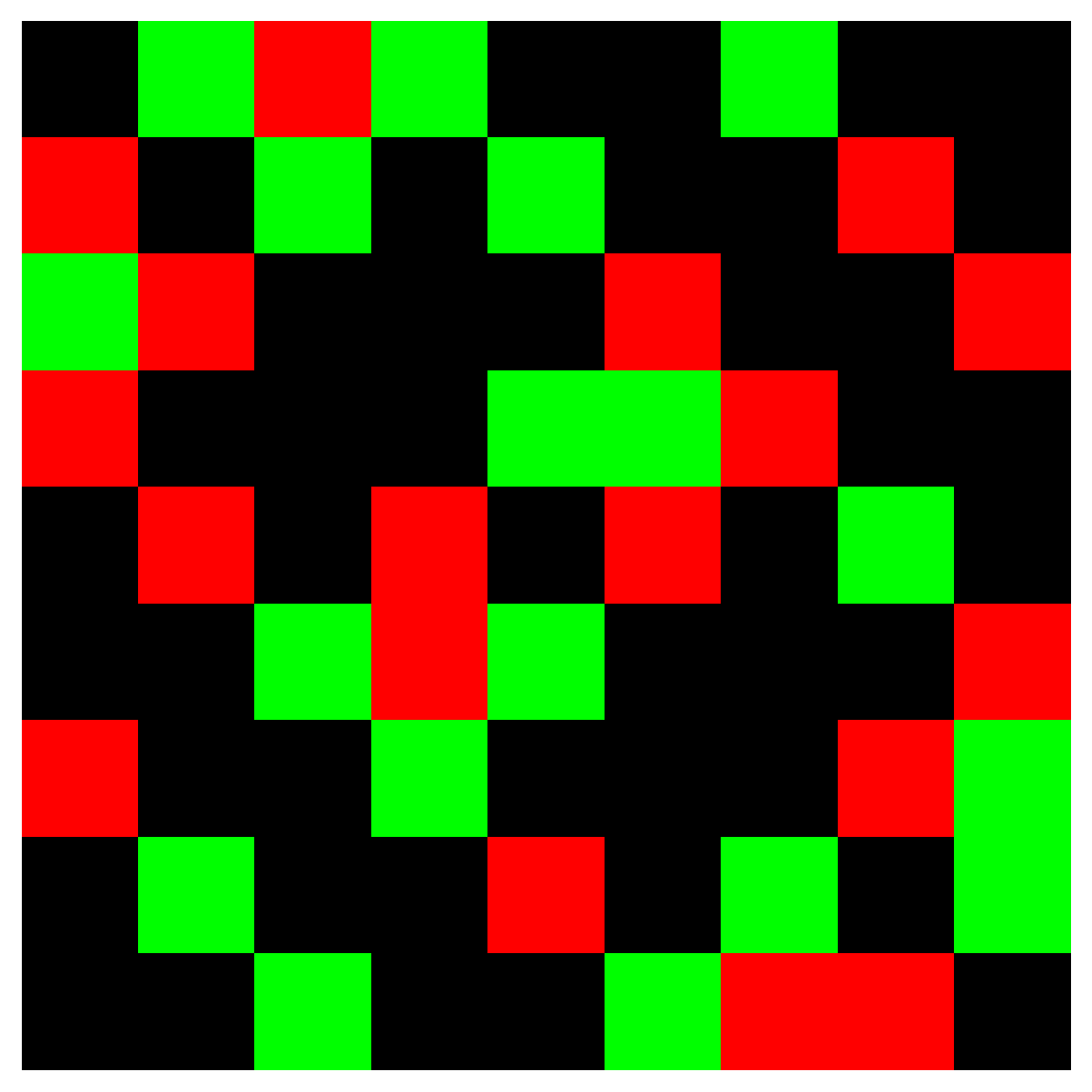}
\caption{$p=1$}
\label{may10f1}
\end{subfigure}
\hfill
\begin{subfigure}[b]{0.3\textwidth}
\centering
\includegraphics[width=\textwidth]{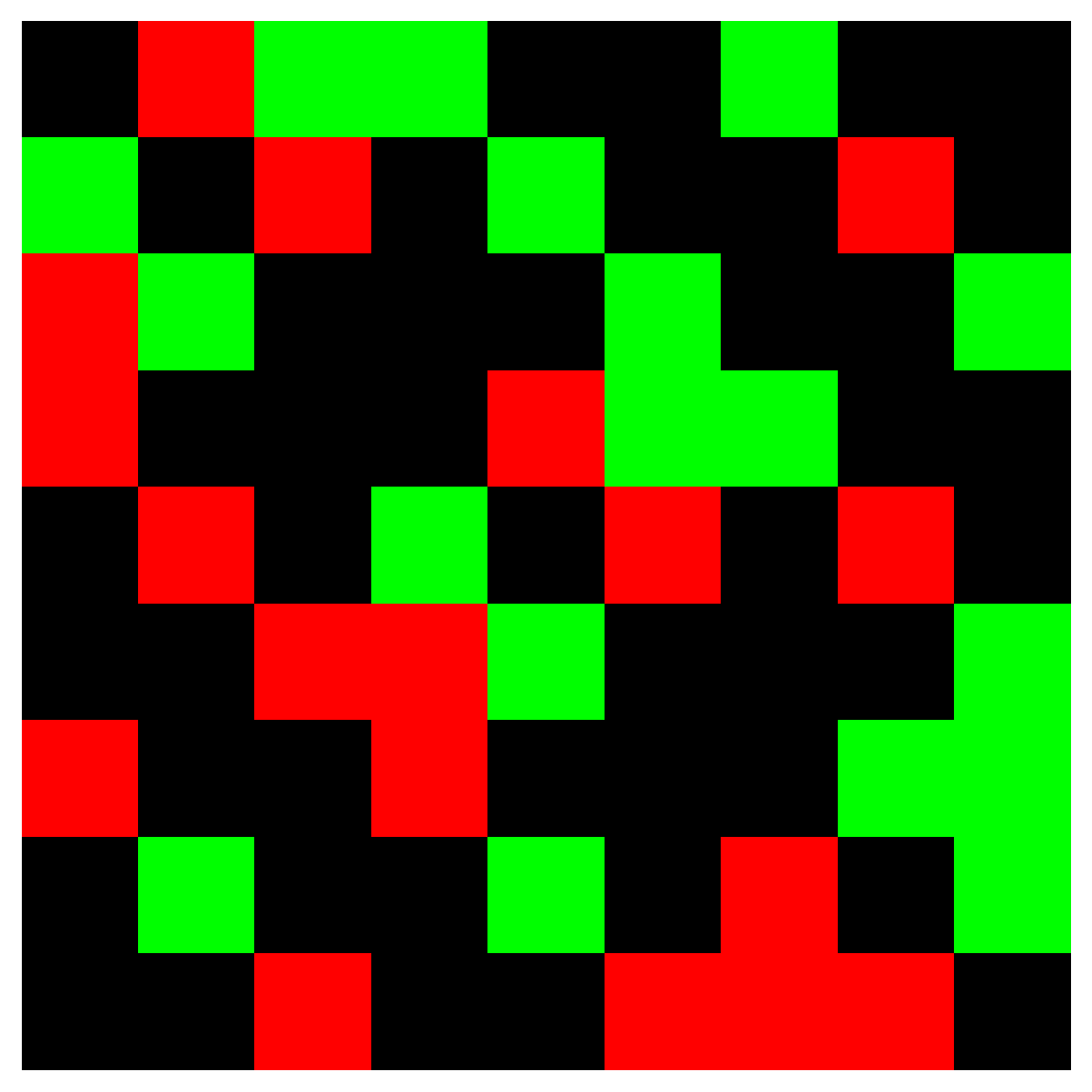}
\caption{$p=8$}
\label{may10f2}
\end{subfigure}
\begin{subfigure}[b]{0.4\textwidth}
\centering
\includegraphics[width=\textwidth]{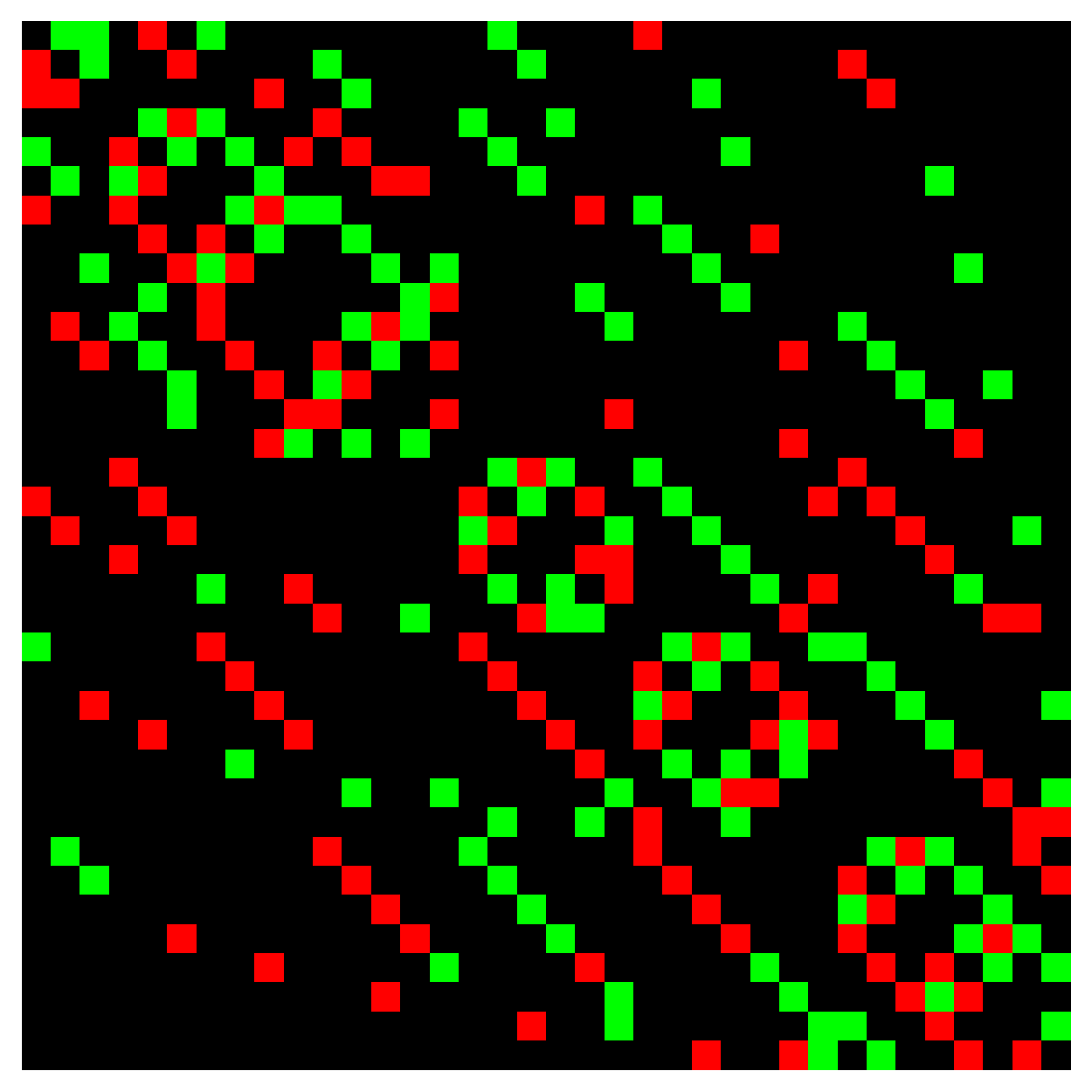}
\caption{$p=2$}
\label{may10f3}
\end{subfigure}
\hfill
\begin{subfigure}[b]{0.4\textwidth}
\centering
\includegraphics[width=\textwidth]{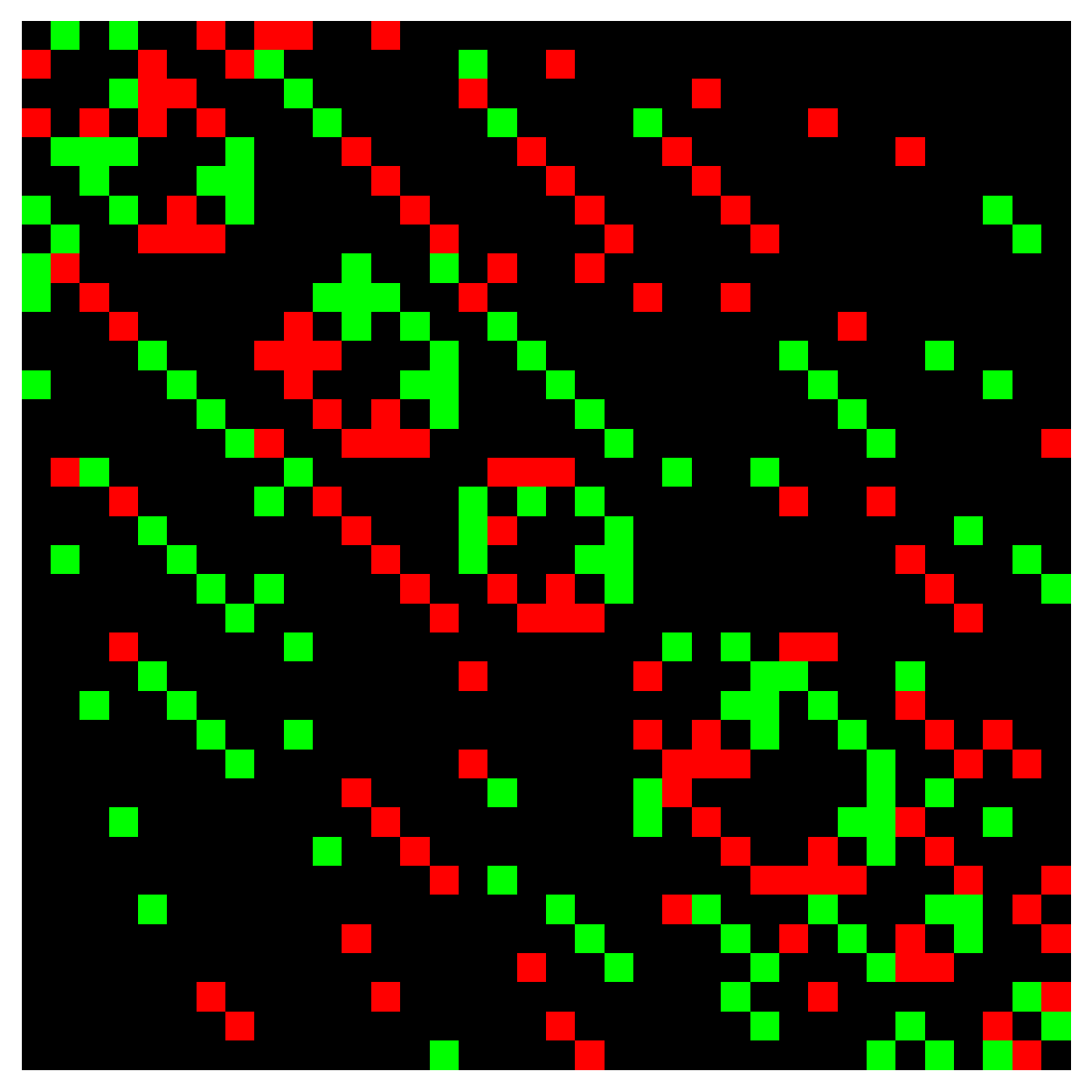}
\caption{$p=7$}
\label{may10f4}
\end{subfigure}
\begin{subfigure}[b]{0.49\textwidth}
\centering
\includegraphics[width=\textwidth]{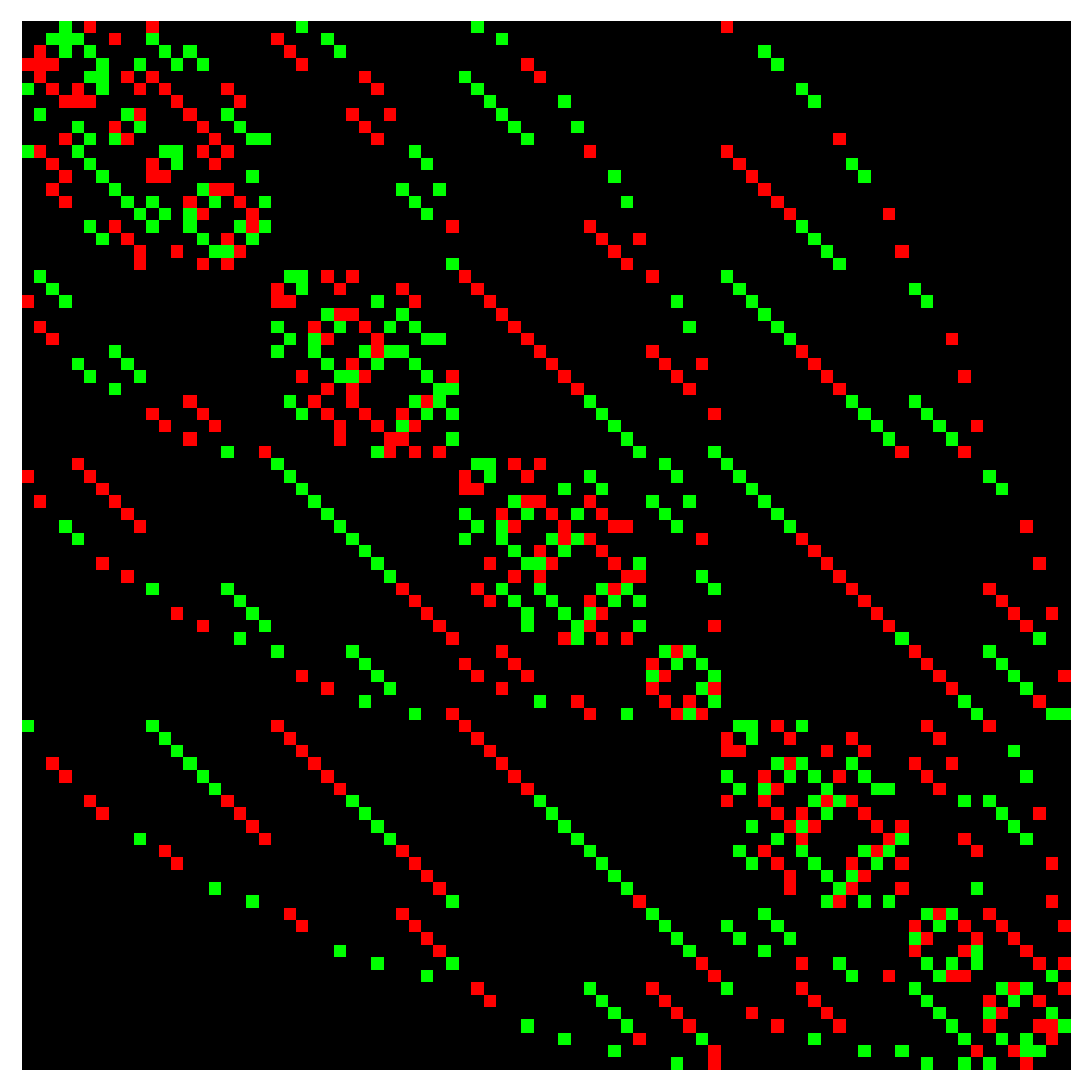}
\caption{$p=3$}
\label{may10f5}
\end{subfigure}
\hfill
\begin{subfigure}[b]{0.49\textwidth}
\centering
\includegraphics[width=\textwidth]{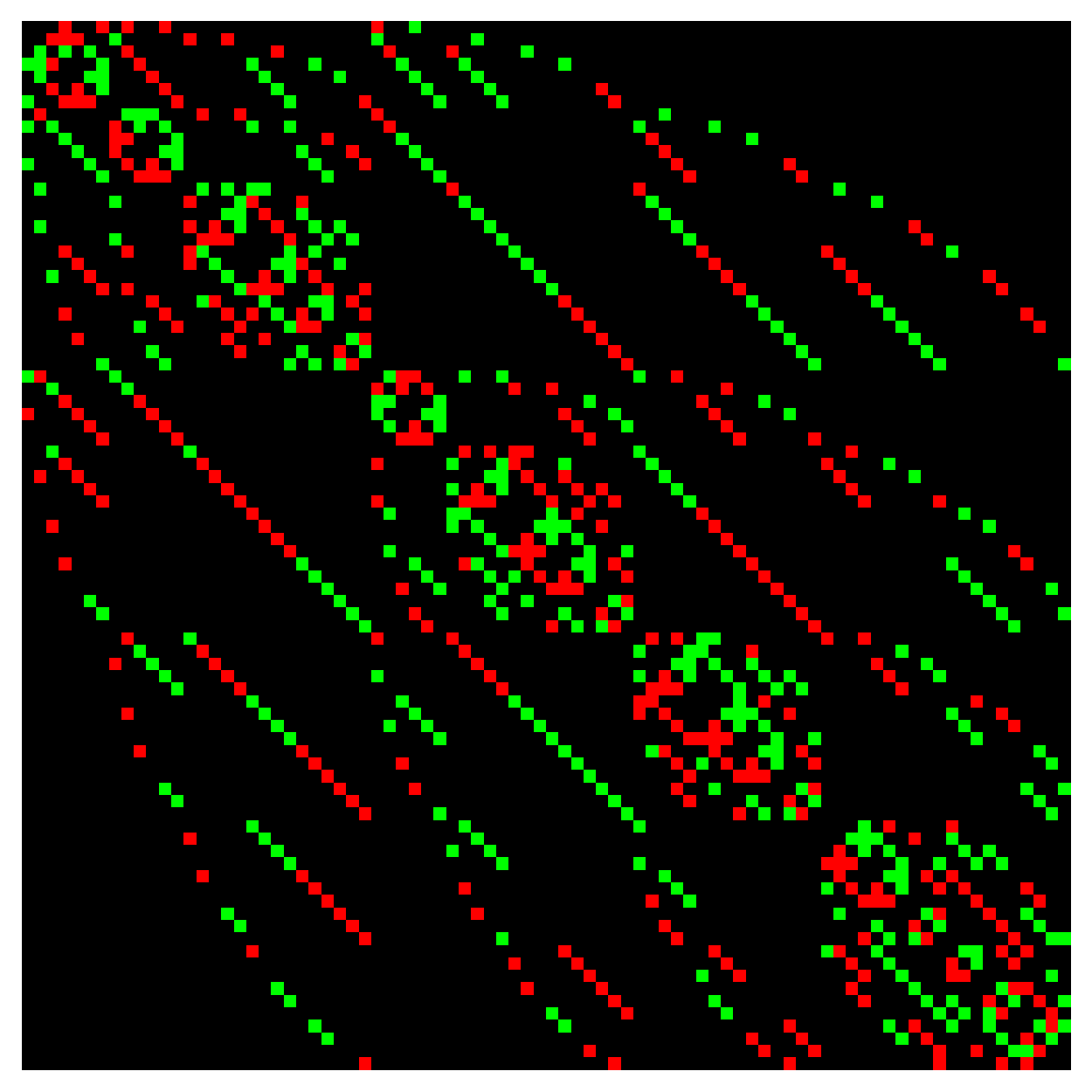}
\caption{$p=6$}
\label{may10f6}
\end{subfigure}
\caption{The reduced Hamiltonian (\ref{may2f}) in the spin representation for the system of free fermions on a $3\times3$ lattice in sectors $p=1,2,3,6,7,8$. The green squares represent the matrix elements equal $i$, the red ones stand for $-i$ entries, while the black background depicts zeroes.}
\label{may10fig1}
\end{figure}

\end{appendices}

\end{document}